\documentclass[twocolumn]{aastex63}

\submitjournal{AJ}

\shorttitle{MPs along the red HB of GCs}
\shortauthors{Dondoglio et al.}

\begin{document}

\title{Multiple stellar populations along the red Horizontal Branch and Red Clump of Globular Clusters.}

\correspondingauthor{Emanuele Dondoglio}
\email{emanuele.dondoglio@studenti.unipd.it}

\author[0000-0003-1713-0082]{E.\,Dondoglio} 
\affiliation{Dipartimento di Fisica e Astronomia ``Galileo Galilei'', Universit\`{a} di Padova, Vicolo dell'Osservatorio 3, I-35122, Padua, Italy}

\author[0000-0001-7506-930X]{A.\,P.\,Milone}
\affiliation{Dipartimento di Fisica e Astronomia ``Galileo Galilei'', Universit\`{a} di Padova, Vicolo dell'Osservatorio 3, I-35122, Padua, Italy}
\affiliation{Istituto Nazionale di Astrofisica - Osservatorio Astronomico di Padova, Vicolo dell'Osservatorio 5, IT-35122, Padua, Italy}

\author[0000-0002-1128-098X]{E.\,P.\,Lagioia}
\affiliation{Dipartimento di Fisica e Astronomia ``Galileo Galilei'', Universit\`{a} di Padova, Vicolo dell'Osservatorio 3, I-35122, Padua, Italy}

\author[0000-0002-1276-5487]{A.\,F.\.Marino}
\affiliation{Dipartimento di Fisica e Astronomia ``Galileo Galilei'', Universit\`{a} di Padova, Vicolo dell'Osservatorio 3, I-35122, Padua, Italy}
\affiliation{Istituto Nazionale di Astrofisica - Osservatorio Astronomico di Padova, Vicolo dell'Osservatorio 5, IT-35122, Padua, Italy}
\affiliation{Centro di Ateneo di Studi e Attivit\`{a} Spaziali ``Giuseppe Colombo'' - CISAS, via Venezia 15, I-35131, Padua, Italy}

\author[0000-0002-1128-098X]{M.\,Tailo}
\affiliation{Dipartimento di Fisica e Astronomia ``Galileo Galilei'', Universit\`{a} di Padova, Vicolo dell'Osservatorio 3, I-35122, Padua, Italy}

\author[0000-0002-7690-7683]{G.\,Cordoni}
\affiliation{Dipartimento di Fisica e Astronomia ``Galileo Galilei'', Universit\`{a} di Padova, Vicolo dell'Osservatorio 3, I-35122, Padua, Italy}

\author[0000-0002-7690-7683]{S.\,Jang}
\affiliation{Dipartimento di Fisica e Astronomia ``Galileo Galilei'', Universit\`{a} di Padova, Vicolo dell'Osservatorio 3, I-35122, Padua, Italy}
\affiliation{Center for Galaxy Evolution Research and Department of Astronomy, Yonsei University, Seoul 03722, Korea}

\author[0000-0002-7690-7683]{M.\,Carlos}
\affiliation{Departamento de Astronomia, IAG, Universidade de S\~{a}o Paulo, Rua do Mat\~{a}o 1226, 05509-900 S\~{a}o Paulo, Brazil}

\begin{abstract}
We exploit multi-band {\it Hubble Space Telescope} photometry to investigate multiple populations (MPs) along the red horizontal branches (HBs) and red clumps of fourteen metal-rich Globular Clusters (GCs), including twelve Milky Way GCs and the Magellanic Cloud GCs NGC\,1978 and NGC\,416.

Based on appropriate two-color diagrams we find that the fraction of 1G stars in Galactic
GCs correlates with cluster mass, confirming previous results based on red-giant branch (RGB) stars. Magellanic-Cloud GCs show higher fractions of 1G stars than Galactic GCs with similar masses, thus suggesting that the environment affects the MP phenomenon. 
We compared and combined our population fractions based on HB with previous estimates from MS and RGB, and we used ground-based UBVI photometry (available for NGC\,104, NGC\,5927, NGC\,6366, NGC\,6838) to extend the investigation over a wide field of view. 
All studied GCs are consistent with flat distributions of 1G and 2G stars within $\sim$1 arcmin from the cluster center except NGC\,416, where the 2G is more centrally concentrated. 2G stars of NGC\,104 and NGC\,5927 are more centrally-concentrated than the 1G, whereas the distribution is flat for NGC\,6366 and NGC\,6838.

We discover that most of the analyzed GCs exhibit extended sequences of 1G stars along the red HB, not consistent with a simple population. The comparison between appropriate synthetic and observed CMDs reveals that these extended distributions are consistent with either star-to-star variation in helium or with an internal metallicity spread, recalling the inhomogeneity of 1G stars along the ChMs.
\end{abstract}


\section{Introduction}\label{sec:intro}

Multi-band {\it Hubble Space Telescope} ({\it HST}) photometry revealed that the color-magnitude diagram (CMD) of most Globular clusters (GCs) is composed of two main stellar populations that can be identified along the main evolutionary stages \citep[e.g.][]{milone2012a}. 
One population, dubbed first generation (1G), is composed of stars with the same chemical composition of halo field stars with similar metallicity, while the remaining stars are enhanced in helium, nitrogen and sodium and depleted in carbon and oxygen and are named second-generation \citep[2G, e.g.][]{kraft1994a, carretta2009a, marino2019a}.

Studies based on pseudo two-color diagrams called chromosome maps (ChMs) from homogeneous {\it HST} photometry of 59 Milky-Way GCs revealed that the fraction of 1G stars and the internal variations of helium and nitrogen depend on cluster mass, thus indicating that the complexity of the multiple-population phenomenon is associated with the GC mass \citep{milone2017a, zennaro2019a}.  The fraction of 1G stars of some Magellanic-Clouds GCs can be even higher than that observed in Galactic GCs, which suggests that the properties of the multiple populations depend on the host galaxy \citep[][]{lagioia2019a, milone2020a}. 
Intriguingly, 1G stars define extended sequences along the ChMs of many GCs, which may imply that they host stars with different chemical compositions \citep[][]{milone2015a, milone2017a}. Star-to-star variations in metallicity and helium are possible responsible for the color extension of 1G stars of some GCs but a comprehensive solution is still missing \citep[e.g.][]{milone2015a, dantona2016a, marino2019b, tailo2019b}. 

These studies are typically based either on red-giant branch (RGB) or main-sequence (MS) stars. 
Although the horizontal branch (HB) is rarely used to derive accurate determinations of the relative numbers of 1G and 2G stars, it would provide  unique information on multiple populations in GCs.

Early evidence of stellar populations with different chemical composition along the red HB of GCs is provided by \citet{norris1982a}, who combined photometry and spectroscopy of fourteen stars of NGC\,104 (47\,Tuc) and detected bimodal CN distribution and C-N anticorrelation.
 A similar behaviour has been observed also in NGC\,6838 by \citet{smith1989a}.
More recently, high-resolution spectroscopy has provided direct evidence that in GCs with intermediate metallicities ([Fe/H]$\sim -1.0$--$-1.4$) Na-poor and O-rich stars populate the reddest part of the HB, while Na-rich and O-poor stars are mostly located on the blue HB \citep[e.g.][]{marino2011a, marino2014a}.

In some GCs with intermediate metallicities, the red HB can host multiple populations with different light-element abundances that are merely mixed in the photometric diagrams \citep[e.g.][]{gratton2011a, marino2014a}.
On the contrary, distinct sequences of 1G and 2G stars can be detected along the red HB of metal rich GCs with [Fe/H]$\gtrsim -1.0$, by means of appropriate photometric diagrams. 
\citet{milone2012a} have shown that 1G red-HB stars have redder  $m_{\rm F275W}-m_{\rm F336W}$ colors than 2G red-HB stars with the same $m_{\rm F336W}-m_{\rm F438W}$. As a consequence, the $m_{\rm F275W}-m_{\rm F336W}$ vs.\,$m_{\rm F336W}-m_{\rm F438W}$ two-color diagram is a powerful tool to identify stellar populations along the red HB of metal-rich GCs.
Photometric diagrams obtained from appropriate combination of $U$, $B$, $I$ magnitudes \citep[][e.g.]{marino2008a, sbordone2011a} or from the so-called JWL indices \citep[][]{lee2017a, lee2018a, lee2019a} are exquisite tools to identify multiple populations among giant stars by using ground-based telescopes and have allowed to detect split red HBs in some GCs, including 47\,Tuc and NGC\,6838 \citep[e.g.][]{milone2012a, monelli2013a, lee2020a, cordoni2020a}.
In this work, we exploit multi-band {\it HST} photometry of the Large Magellanic Cloud (LMC) GC NGC\,1978, of the GC NGC\,416 in the Small Magellanic Cloud (SMC) and of twelve Galactic GCs, namely NGC\,104, NGC\,5927, NGC\,6304, NGC\,6352, NGC\,6366, NGC\,6388, NGC\,6441, NGC\,6496, NGC\,6624, NGC\,6637, NGC\,6652 and NGC\,6838.

The paper is organized as follows. In Section~\ref{sec:data} we describe the dataset and data reduction. The photometric diagrams of  red HB stars are presented in Section~\ref{sec:MPs} where we investigate multiple populations along the red HB and derive the fraction of 1G stars. Section~\ref{sec:rela} investigates the relations between the fraction of 1G stars and the main parameters of the host clusters. In Section~\ref{sec:6388} we present the discovery of extended HB sequences for 1G stars of several GCs, while Section~\ref{sec:RD} is focused on the radial distributions of multiple populations.
 Finally, summary and discussions are provided in Section~\ref{sec:conclusions}.

\section{Data and data reduction}\label{sec:data}
To investigate the stellar populations along the HB in Galactic GCs we used the photometric catalogues by \citet{anderson2008a}, \citet{milone2012a, milone2017a} and \citet{milone2018a}, which provide astrometry and differential-reddening corrected photometry in the F275W, F336W and F438W bands of the Ultraviolet and Visual Channel of the Wide Field Camera 3 (UVIS/WFC3) on board of {\it HST} and in the F606W and F814W bands of the Wide Field Channel of the Advanced Camera for Survey (ACS/WFC).  
To study the stellar populations in the GCs NGC\,1978 and NGC\,416, we derived stellar positions and multi-band photometry by using the archive {\it HST} images summarized in Table~\ref{dataset}. 
 Stellar magnitudes and positions are measured with the computer program KS2, which is developed by Jay Anderson and is an evolution of the program kitchen\_sync \citep{anderson2008a}. 
 
\begin{table*}
\caption{Summary of the data of NGC\,416 and NGC\,1978 used in this work.}
\small
\begin{tabular}{cccccc}
\label{dataset}
\\
\hline\hline
     \\
     DATE & N$\times$ EXPTIME & FILTER & INSTRUMENT & PROGRAM & PI\\
    \hline \hline
         &  & NGC\,416 & & \\ \\ 
     2019 Jun 18 & 1500s+1512s+2$\times$1529s+2$\times$1525s & F275W & UVIS/WFC3 & 15630 & N. Bastian\\
     2019 Jul 31 & 1530s+1500s+2$\times$1533s+2$\times$1534s & F275W & UVIS/WFC3 & 15630 & N. Bastian\\
     2019 Aug 05 & 2$\times$1500s+1512s+2$\times$1515s+1523s & F275W & UVIS/WFC3 & 15630 & N. Bastian\\
     2016 Jun 16 & 700s+1160s+1200s & F336W & UVIS/WFC3 & 14069 & N. Bastian\\
     2016 Jun 16 & 500s+800s+1650s+1655s & F343N & UVIS/WFC3 & 14069 & N. Bastian\\
     2016 Jun 16 & 75s+150s+440s+460s & F438W & UVIS/WFC3 & 14069 & N. Bastian\\
     2005 Nov 22 & 2$\times$20s & F555W & ACS/WFC & 10396 & J. Gallagher\\
     2006 Mar 08 & 2$\times$20s+4$\times$496s & F555W & ACS/WFC & 10396 & J. Gallagher\\
     2005 Nov 22 & 2$\times$10s+4$\times$474s & F814W & ACS/WFC & 10396 & J. Gallagher\\
     2006 Mar 08 & 2$\times$10s+4$\times$474s & F814W & ACS/WFC & 10396 & J. Gallagher\\
     \\
              &  & NGC\,1978 & & \\ 
                                 \\
     2019 Sep 17 & 2$\times$1493s+2$\times$1498s+2$\times$1500s+2$\times$1499s & F275W & UVIS/WFC3 & 15630 & N. Bastian\\
      & 1501s+1502s+1495s+1492s &  &  &  & \\
     2011 Aug 15 & 380s+460s & F336W & UVIS/WFC3 & 12257 & L. Girardi\\
     2016 Sep 25 & 660s+740s & F336W & UVIS/WFC3 & 14069 & N. Bastian\\
     2016 Sep 25 & 425s+450s+500s+2$\times$800s+1000s & F343N & UVIS/WFC3 & 14069 & N. Bastian\\
     2016 Sep 25 & 75s+120s+420s+460s+650s+750s & F438W & UVIS/WFC3 & 14069 & N. Bastian\\
     2003 Oct 07 & 300s & F555W & ACS/WFC & 9891 & G. Gilmore\\
     2011 Aug 15 & 60s+300s+680s & F555W & ACS/WFC & 12257 & L. Girardi\\
     2003 Oct 07 & 200s & F814W & ACS/WFC & 9891 & G. Gilmore\\
\hline\hline
\end{tabular}\\
\label{tab:data}
\end{table*}

We used different methods to derive magnitudes and positions of bright and faint stars.
The bright stars have enough flux to derive accurate magnitudes and positions from each individual exposure. Hence they are measured by fitting the appropriate point-spread function (PSF) model for position and flux in each exposure independently. The best estimates of magnitude and positions of each star correspond to the average of the various measurements. 
 To derive photometry of faint stars we combined the information from the various exposures. Specifically, we calculated the average position of each star from all the exposures, and then we fitted each exposure’s pixels with the PSF, solving only for the flux. We refer to the paper by \citet{sabbi2016a} for details on the KS2 program.
 
 We used the various diagnostics provided by KS2, to select high-quality stars that are relatively isolated and well fitted by the PSF model as in \citet{milone2009a} and \citet{bedin2009a}. 
  We calibrated the photometry into the Vega system as in \citet{bedin2005a} and by using the zero points provided by the Space Telescope Science Institute webpage\footnote{ \url{http://www.stsci.edu/hst/instrumentation/wfc3/calibration} 
   and
  \url{http://www.stsci.edu/hst/acs/analysis/zeropoints} for WFC3/UVIS and ACS/WFC photometry, respectively.}. 
Stellar positions are corrected for geometric distortion by using the solutions provided by \citet{bellini2009a} and \citet{bellini2011a}. 

In addition to {\it HST} data, we exploit the astrometric and photometric catalogs of NGC\,104, NGC\,5927, NGC\,6366 and NGC\,6838 by \citet{stetson2019a}, which provide positions and $U, B, V, I$ magnitudes of stars over wide field of views. Details on the data and the data analysis are provided by \citet{stetson2005a}, \citet[][]{stetson2019a} and references therein. 

To investigate multiple populations from ground-based photometry, we combined photometry from \citet{stetson2019a} and stellar proper motions from GAIA DR2 \citep{gaia2018a}. We selected a sample of cluster members with accurate astro-photometric measurements by using the recipe by \citet{cordoni2018a, cordoni2020a}. 
In a nutshell, we first identified stars with accurate proper motions measurements, by using both the \texttt{astrometric\_gof\_al} (\texttt{As\_gof\_al}) and the Renormalized Unit Weight Error (RUWE) parameters \citep[][]{lindegren2018a}. We then selected cluster members from the proper motion vector-point diagram \citep[see][for details]{cordoni2018a}.  

Both {\it HST} and ground-based photometry has been corrected for the effect of spatial reddening variation by using the method and the computer programs by \citet[][see their Section~3.1]{milone2012b}.

\begin{figure*} 
\begin{center} 
  \includegraphics[height=10cm,trim={1.0cm 5cm 6cm 4cm}]{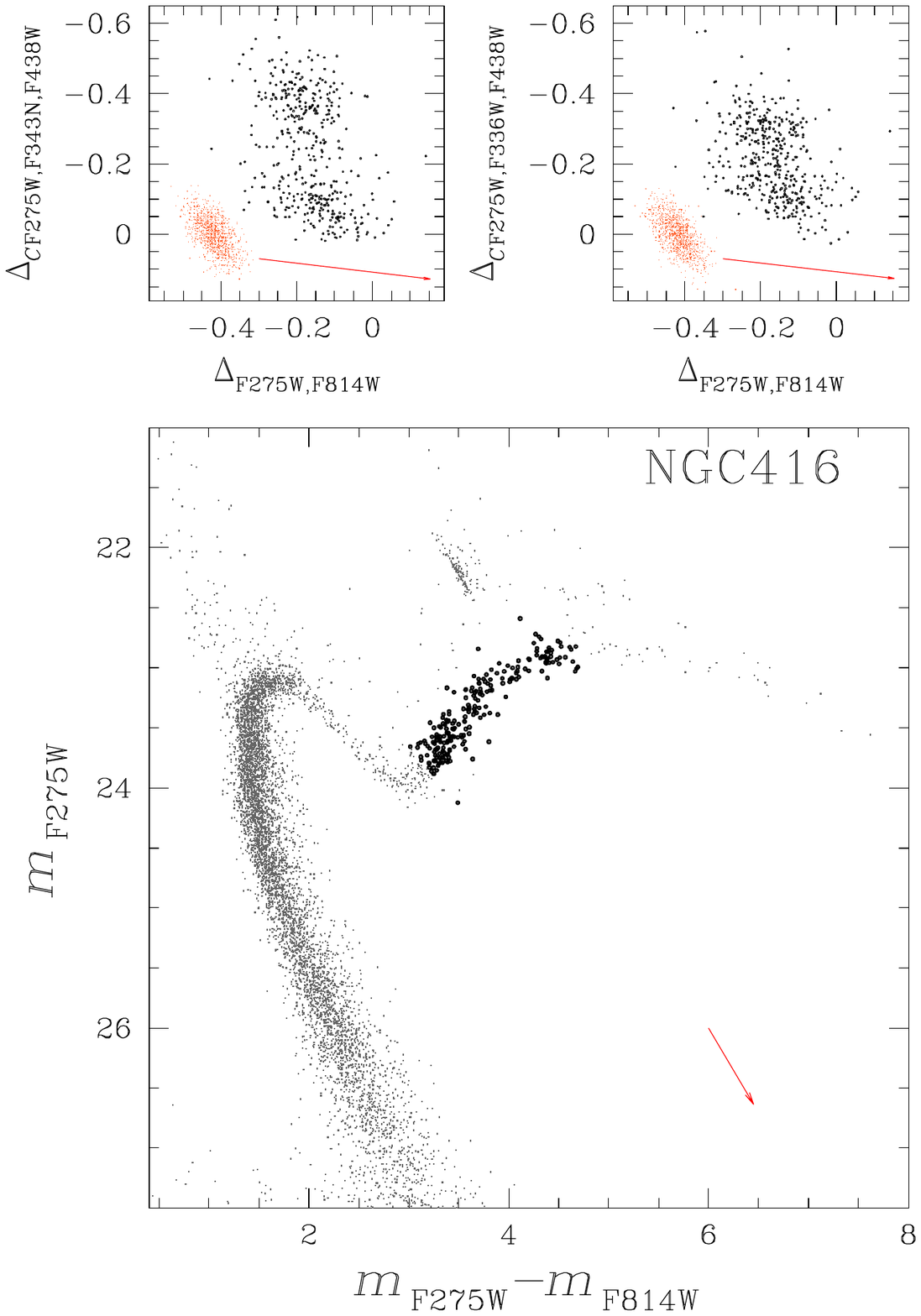}\includegraphics[height=10cm,,trim={1.0cm 5cm 6cm 4cm}]{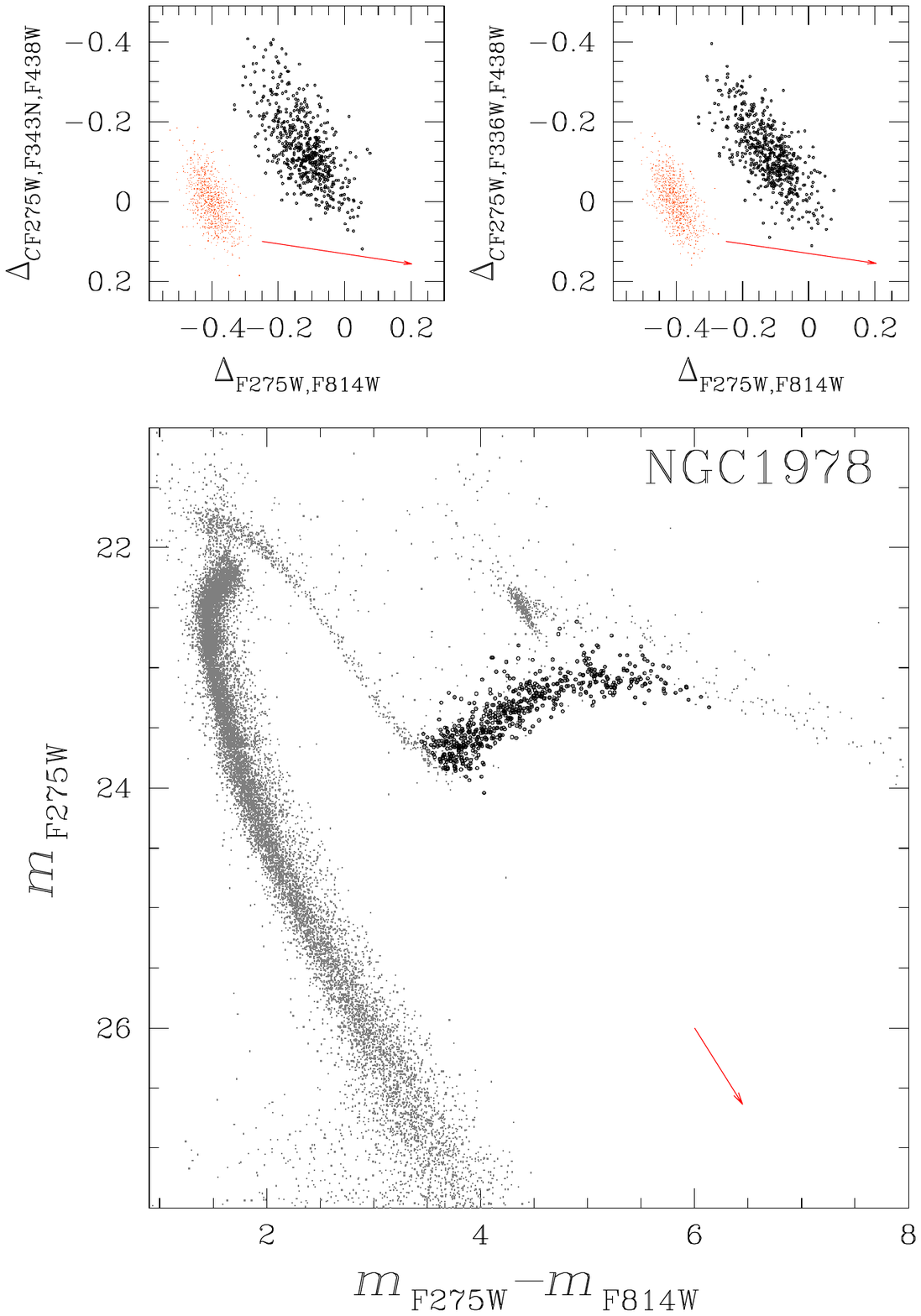}
  \caption{\textit{Bottom panels.} $m_{\rm F275W}$ vs.\,$m_{\rm F275W}-m_{\rm F814W}$ CMD of stars in NGC\,416 (left) and NGC\,1978 (right). \textit{Upper panels.} $\Delta_{\rm{\it C}F275W,F343N,F438W}$ vs.\,$\Delta_{\rm F275W,F814W}$ and $\Delta_{\rm{\it C}F275W,F336W,F438W}$ vs.\,$\Delta_{\rm F275W,F814W}$ ChMs of RGB stars marked with black dots in the bottom panels. Orange points mark the distribution of observational errors, including errors on differential reddening. Red arrows indicate the reddening vectors and correspond to a reddening variation $\Delta E(B-V) = 0.1$.\\}
 \label{fig:CMDNGC1978} 
 \end{center} 
\end{figure*} 

As an example of the photometry derived in this paper, the bottom panels of Figure~\ref{fig:CMDNGC1978} show the $m_{\rm F275W}$ vs.\,$m_{\rm F275W}-m_{\rm F814W}$ CMD of NGC\,416 and NGC\,1978 that we used to study multiple populations along both the RGB and the HB. In particular, to characterize stellar populations along the RGB we derived the $\Delta_{\rm{\it C}F275W,F343N,F438W}$ vs.\,$\Delta_{\rm F275W,F814W}$ and $\Delta_{\rm{\it C}F275W,F336W,F438W}$ vs.\,$\Delta_{\rm F275W,F814W}$ ChMs of RGB stars plotted in the upper panels of Figure~\ref{fig:CMDNGC1978}. 
The orange points represent observational errors, including errors on differential reddening, and correspond to the ChMs expected from a simple population \citep[see][for details]{milone2017a, milone2020a}.  
 Clearly, the fact that the observed pseudo-color distributions are wider than those expected from observational errors alone corroborates the evidence that NGC\,1978 and NGC\,416 hosts multiple stellar populations \citep{niederhofer2017a, martocchia2018a, martocchia2018b, lagioia2019b, milone2020a}.

\begin{table*}
\caption{This table lists the average reddening $E(B-V)$ of each cluster and the random mean scatter of reddening in the field of view, the core radius, the half light radius, the concentration and maximum radial distance of our {\it HST} observations. We also list the maximum radius of  the field of view of ground-based photometry of NGC\,104, NGC\,5927, NGC\,6366 and NGC\,6838. }
\begin{tabular}{ccccccc}
\\
\hline\hline
     \\
    CLUSTER & $E(B-V)$ & r.m.s. & $r_{\rm c}$ & $r_{\rm hl}$ & $c$ & $r_{\rm FoV}$ \\
            & [mag]    & [mag] & [arcmin] & [arcmin] & & [arcmin] \\  
    \\
    \hline
    \\
     NGC\,0104 & 0.04$^{a}$ & 0.004$^{f}$ & 0.36$^{a}$ & 3.17$^{a}$ & 2.07$^{a}$ & 1.66-24.57 \\
     NGC\,5927 & 0.45$^{a}$ & 0.017$^{f}$ & 0.42$^{a}$ & 1.10$^{a}$ & 1.60$^{a}$ &  0.95-5.96 \\
     NGC\,6304 & 0.54$^{a}$ & 0.012$^{f}$ & 0.21$^{a}$ & 1.42$^{a}$ & 1.80$^{a}$ &  0.93 \\
     NGC\,6352 & 0.22$^{a}$ & 0.017$^{f}$ & 0.83$^{a}$ & 2.05$^{a}$ & 1.10$^{a}$ &  0.91 \\
     NGC\,6366 & 0.71$^{a}$ & 0.019$^{f}$ & 2.17$^{a}$ & 2.92$^{a}$ & 0.74$^{a}$ &  0.90-7.99 \\
     NGC\,6388 & 0.37$^{a}$ & 0.012$^{f}$ & 0.12$^{a}$ & 0.52$^{a}$ & 1.75$^{a}$ &  1.00 \\
     NGC\,6441 & 0.47$^{a}$ & 0.020$^{f}$ & 0.13$^{a}$ & 0.57$^{a}$ & 1.75$^{a}$ &  0.92 \\
     NGC\,6496 & 0.15$^{a}$ & 0.014$^{f}$ & 0.95$^{a}$ & 1.02$^{a}$ & 0.70$^{a}$ &  0.89 \\
     NGC\,6624 & 0.28$^{a}$ & 0.008$^{f}$ & 0.06$^{a}$ & 0.82$^{a}$ & 2.50$^{a}$ &  0.89 \\
     NGC\,6637 & 0.18$^{a}$ & 0.007$^{f}$ & 0.33$^{a}$ & 0.84$^{a}$ & 1.38$^{a}$ &  0.94 \\
     NGC\,6652 & 0.09$^{a}$ & 0.005$^{f}$ & 0.10$^{a}$ & 0.48$^{a}$ & 1.80$^{a}$ &  0.91 \\
     NGC\,6838 & 0.25$^{a}$ & 0.012$^{f}$ & 0.63$^{a}$ & 1.67$^{a}$ & 1.15$^{a}$ &  0.97-8.94 \\
     NGC\,1978 & 0.08$^{b}$ & 0.003$^{g}$ & 0.30$^{c}$ & 0.67$^{c}$ & 1.26$^{e}$ &  0.55 \\
     NGC\,0416 & 0.05$^{b}$ & 0.020$^{g}$ & 0.17$^{d}$ & 0.25$^{d}$ & 1.02$^{h}$ &  0.55 \\
\hline\hline
\end{tabular}\\
\footnotesize{{\it{Note.}} References: $^{a}$\citet[][2010 edition]{harris1996a}; $^{b}$\citet{chantereau2019a}; $^{c}$\citet{mclaughlin2005a}; $^{d}$\citet{fischer1992}; $^{e}$\citet{mateo1987a}; $^{f}$\citet{milone2012a, milone2017a}; $^{g}$\citet{milone2020a}; $^{h}$\citet{glatt2009a}.}
\label{tab:param}
\end{table*}

\section{Multiple populations along the red HB}\label{sec:MPs}

As discussed in Section~\ref{sec:intro}, the $m_{\rm F275W}-m_{\rm F336W}$ vs.\,$m_{\rm F336W}-m_{\rm F438W}$ two-color diagram is  an efficient tool to identify multiple stellar populations along the MS, sub-giant branch, and RGB of GCs. The reason is that the amount of stellar flux in the F275W, F336W, and F438W bandpasses of {\it HST} depends on the strengths of the OH, NH and CN molecules, which are comprised in the wavelength  range of these filters, and are different for 1G and 2G stars with similar luminosities \citep[e.g.][]{milone2012a, milone2013a}.

For the same reason, 1G and 2G stars define distinct sequences of red-HB stars in the $m_{\rm F275W}-m_{\rm F336W}$ vs.\,$m_{\rm F336W}-m_{\rm F438W}$ plane, and the red-HB split is more evident in metal-rich GCs with [Fe/H]$\gtrsim -1.0$. Indeed, similarly to RGB, SGB and MS stars, the fluxes of their relatively cold red-HB stars are strongly affected by the abundances of C, N, and O. 

In the following subsection, we exploit photometry in F275W, F336W, and F438W of twelve Galactic GCs with [Fe/H]$\gtrsim -0.8$, namely NGC\,104 (47\,Tuc), NGC\,5927, MGC\,6304, NGC\,6352, NGC\,6366, NGC\,6388, NGC\,6441, NGC\,6496, NGC\,6624, NGC\,6637,  NGC\,6652 and NGC\,6838 to investigate multiple populations along the red HB. Section~\ref{sec:mcpop} is dedicated to the LMC GC NGC\,1978 and the SMC GC NGC\,416.

Table~\ref{tab:param} provides relevant quantities for all studied clusters. These comprise the average reddening in the analyzed FoV, $E(B-V)$, the random mean scatter (r.m.s) of reddening, the core radius ($r_{\rm c}$), half-light radius ($r_{\rm hl}$) and concentration ($c$). We also indicate the maximum radius covered by {it HST} and ground-based observations, $r_{\rm FoV}$.

\subsection{Disentangling first- and second-generation stars along the HB of Galactic GCs}
\begin{figure*}
 \begin{centering} 
  \includegraphics[height=7.cm, clip]{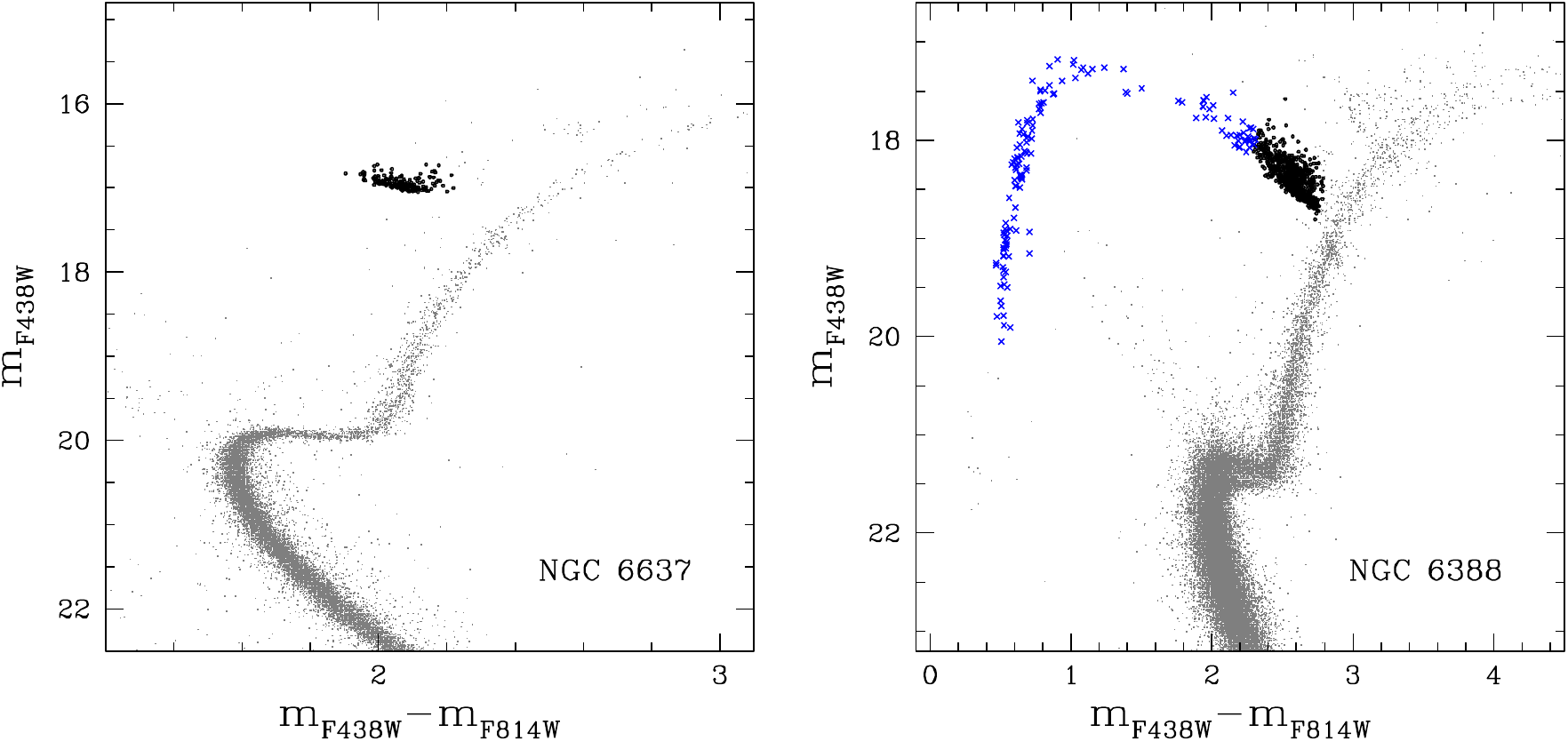}
  \caption{$m_{\rm F438W}$ vs.\,$m_{\rm F438W}-m_{\rm F814W}$ CMDs corrected for differential reddening of NGC\,6388 (left) and NGC\,6637 (right). Red-HB stars are colored black, while blue-HB stars of NGC\,6388 are represented with blue crosses.\\}
 \label{fig:sel} 
\end{centering} 
\end{figure*}

All clusters with [Fe/H]$\gtrsim -0.8$ studied in this paper exhibit the red HB alone. Remarkable exceptions are provided by NGC\,6388 and NGC\,6441, which also show a blue HB \citep[e.g.][]{rich1997a}. 
The procedure to identify 1G and 2G stars along the HB of Galactic GCs is illustrated in Figure~\ref{fig:sel} for NGC\,6637, which exhibits the red HB alone, and for NGC\,6388, whose HB is populated on both sides of the RR-Lyrae instability strip. 


\begin{figure*} 
\begin{center} 
  \includegraphics[height=13cm,clip]{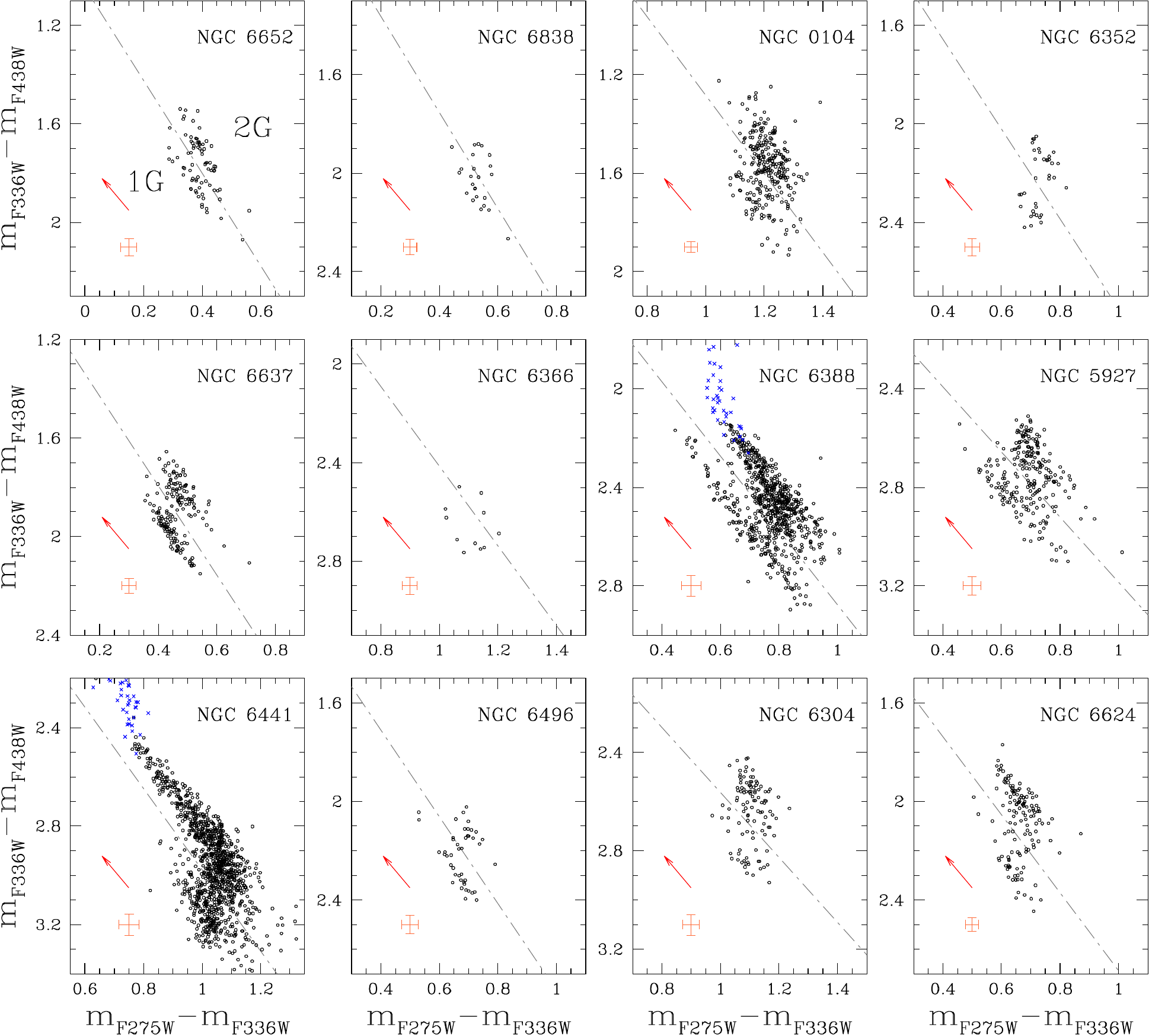}
  \caption{Collection of $m_{\rm F275W}-m_{\rm F336W}$ vs.\,$m_{\rm F336W}-m_{\rm F438W}$ differential-reddening corrected two-color diagram diagrams for the red-HB stars (black dots) of the studied Galactic GCs sorted by metallicity, from the most metal poor, to the most metal rich. Blue HB stars are represented with blue crosses. Grey dot-dashed lines separate the bulk of 1G stars from the 2G. For the sake of comparison, all the panels span the same interval of $m_{\rm F275W}-m_{\rm F336W}$ and $m_{\rm F336W}-m_{\rm F438W}$. Observational error is represented with orange bars. The reddening vectors corresponding to a reddening variation $\Delta E(B-V) = 0.1$ are represented with red arrows.\\}
 \label{fig:2col}
 \end{center} 
\end{figure*}

 We used the differential-reddening corrected $m_{\rm F438W}$ vs.\,$m_{\rm F438W}-m_{\rm F814W}$ CMDs plotted in Figure~\ref{fig:sel} to identify by eye the sample of red HB stars that we represented with black dots. Blue-HB stars of NGC\,6388 are plotted with blue crosses.
   Once these stars have been selected, they have been plotted in the $m_{\rm F275W}-m_{\rm F336W}$ vs.\,$m_{\rm F336W}-m_{\rm F438W}$ two-color diagrams shown in Figure~\ref{fig:2col}. 
  
Previous work, based on the synergy of photometry, spectroscopy and theoretical models, has provided empirical evidence that 1G stars populate the red-HB sequence with the bluest $m_{\rm F336W}-m_{\rm F438W}$ and reddest $m_{\rm F275W}-m_{\rm F336W}$ colors, while 2G stars exhibit redder $m_{\rm F336W}-m_{\rm F438W}$ and $m_{\rm F275W}-m_{\rm F336W}$ colors than 1G stars \citep[e.g.][]{milone2012a}. 
As discussed at the beginning of this section, the physical interpretation is straightforward. Indeed, 2G stars are enhanced in nitrogen and depleted in oxygen and carbon with respect to the 1G. Since the F336W filters includes strong NH molecular bands, 2G stars exhibit fainter F336W magnitudes than 1G stars with similar structure. Similarly, 2G stars have brighter F275W fluxes than the 1G due to the lower strengths of the OH bands. The F438W band is affected by the CH G band, which makes 2G stars brighter than the 1G, and by the CN band at $\sim$4200\AA, which reduces the F438W flux of 2G stars. The combined effects of these molecular bands can make 2G red-HB stars slightly brighter F438W magnitudes than the 1G. As a consequence, 1G stars have bluer $m_{\rm F336W}-m_{\rm F438W}$ and redder $m_{\rm F275W}-m_{\rm F336W}$ colors than the bulk of 2G stars.

The dashed gray lines shown in Figure~\ref{fig:2col} are derived by eye with the criterion of following the fiducial lines of the HB sequence formed by the majority of 2G stars. 
Driven by the evidence that 1G and 2G stars lie in distinct areas of the two-color diagrams, these lines have been shifted towards bluer $m_{\rm F275W}-m_{\rm F336W}$ colors to separate the majority of 1G stars from the remaining red-HB stars.

  
  This collection of diagrams plotted in in Figure~\ref{fig:2col} reveals that all the Galactic clusters of our sample show multiple populations along the red HB and the morphology of 1G and 2G stars dramatically changes from one cluster to another. In some clusters, like NGC\,6838 and NGC\,6352, both 1G and 2G stars span a small interval of less then 0.2 mag in $m_{\rm F275W}-m_{\rm F336W}$, whereas the corresponding color extension of the 1G and 2G sequences of NGC\,6388 is wider than $\sim$0.6 mag. In NGC\,6441, the $m_{\rm F275W}-m_{\rm F336W}$ extension of 2G red-HB stars is more than two times wider than the 1G color extension.
  
The number of sub-populations also shows a high degree of variety. NGC\,6637 and NGC\,6352 exhibit two distinct groups of 1G and 2G stars alone, whereas NGC\,6388 and NGC\,6441, in addition to the sequences populated by the bulk of 1G and 2G stars, show a sub-population of 2G stars with intermediate $m_{\rm F336W}-m_{\rm F438W}$ colors.

\begin{figure*} 
\begin{center} 
  \includegraphics[height=10cm,clip]{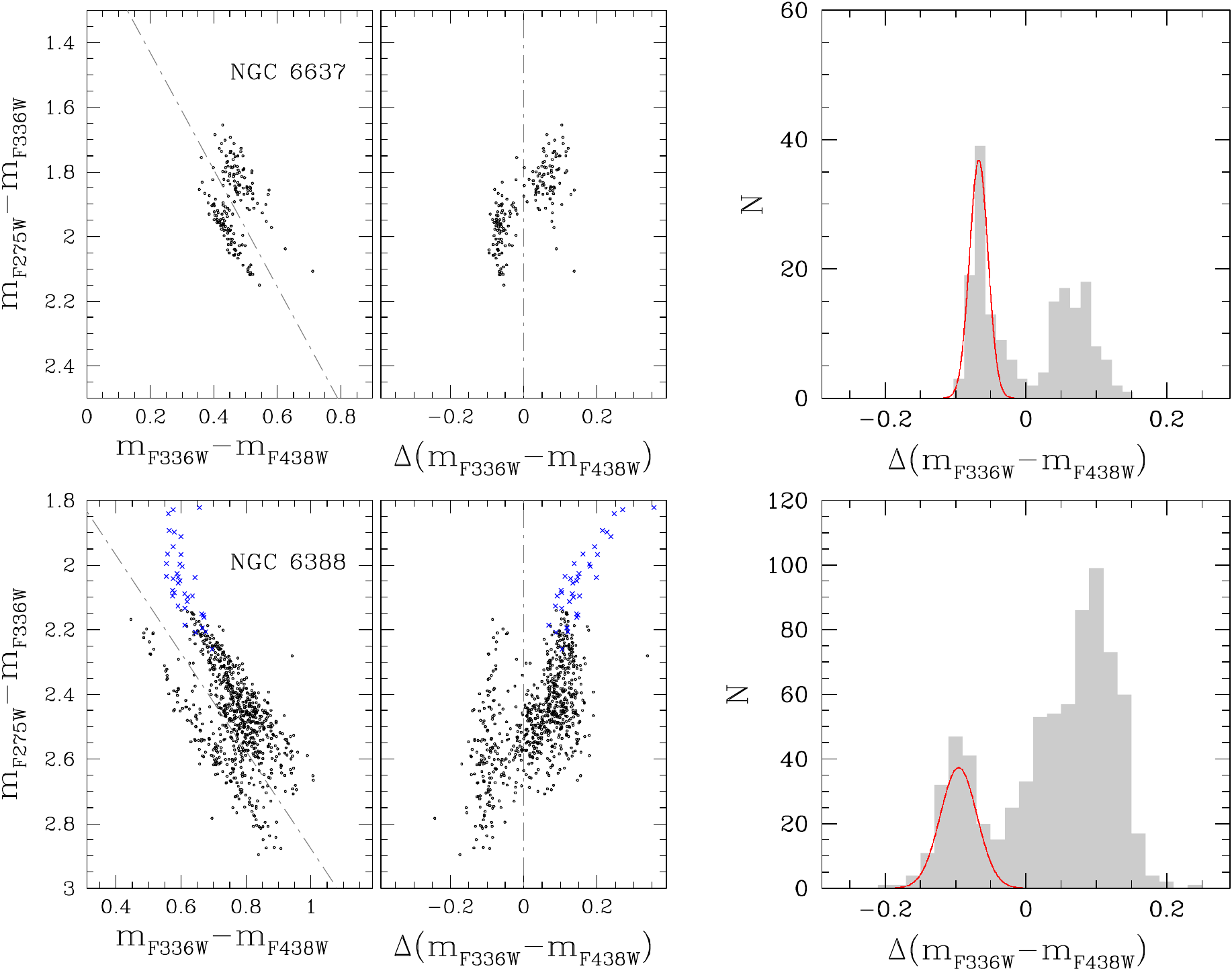}
  \caption{This figure summarises the main steps to identify 1G stars along the red HBs of NGC\,6637 (top panels) and NGC\,6388 (bottom panels). Left panels show the $m_{\rm F275W}-m_{\rm F336W}$ vs.\,$m_{\rm F336W}-m_{\rm F438W}$ two-color diagram of HB stars, blue crosses in the NGC\,6388 diagram are the blue HB stars. The gray dashed-dot lines separate the bulk of 1G stars from the remaining HB stars. 
  The verticalized $m_{\rm F275W}-m_{\rm F336W}$ vs.\,$\Delta$($m_{\rm F336W}-m_{\rm F438W}$) diagrams of HB stars are plotted in middle panels, whereas right panels show the $\Delta$($m_{\rm F336W}-m_{\rm F438W}$ histogram distributions. The Gaussian function that provides the least-squares best fit with the observed distribution is represented with the red solid line. See text for details.\\} 
 \label{fig:met} 
 \end{center} 
\end{figure*} 

\begin{figure*} 
 \begin{center} 
  \includegraphics[height=13.cm,clip]{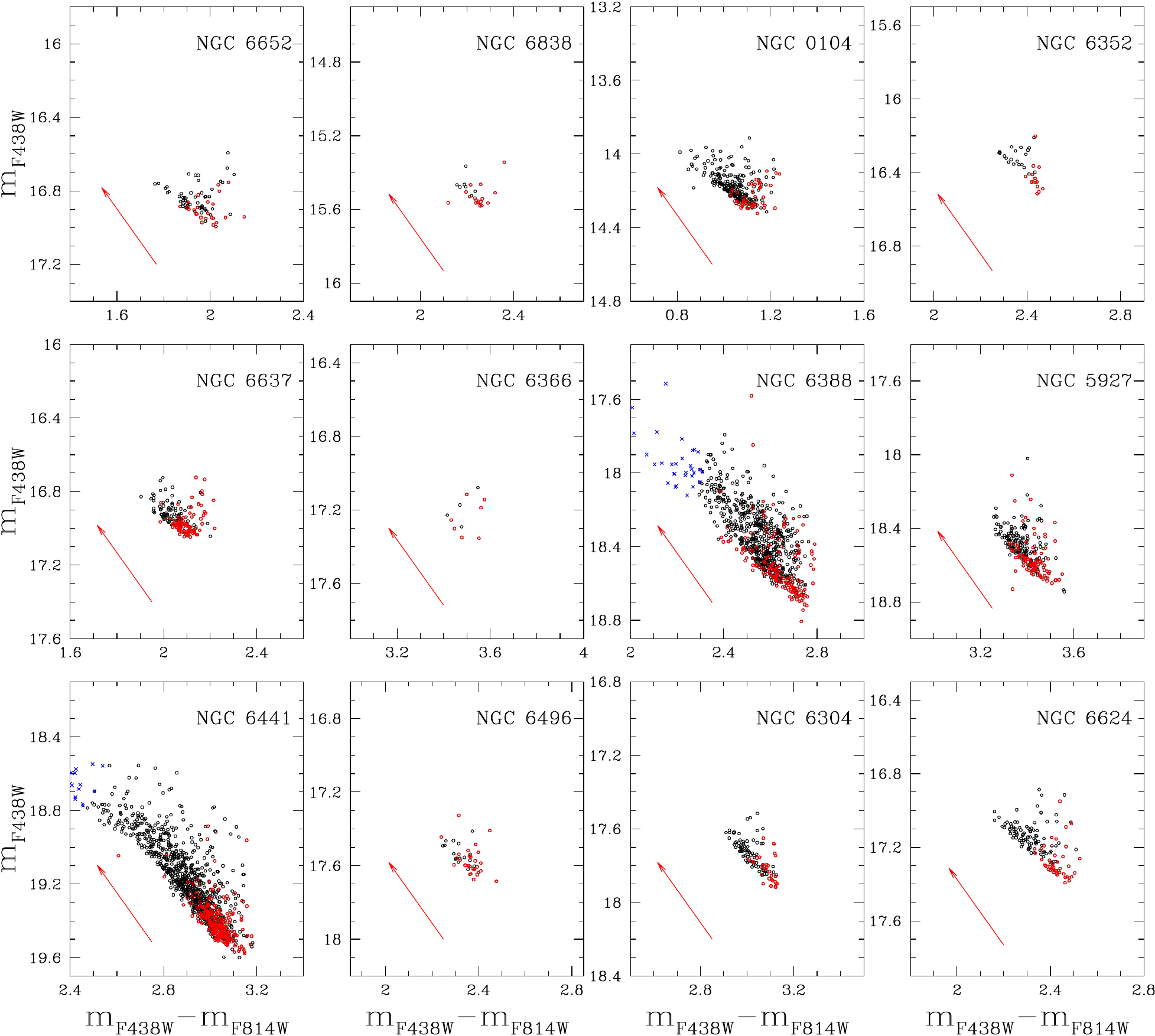}
  \caption{$m_{\rm F438W}$ vs.\,$m_{\rm F438W}-m_{\rm F814W}$ CMDs for the red-HB stars of the Galactic GCs studied in this paper. 1G and 2G stars are colored red and black, respectively. Blue crosses represent the blue HB stars.\\}
 \label{fig:cmd} 
  \end{center} 
\end{figure*} 

\subsection{The fraction of first-generation stars}\label{sec:pratio}

The procedure to identify the 1G stars and derive their fraction respect to the total number of HB stars is illustrated in Figure~\ref{fig:met} for NGC\,6637 and NGC\,6388 and is similar to the method introduced by \citet{milone2012a} to infer the fraction of 1G stars along the red HB of 47\,Tuc. 

Left panels of Figure~\ref{fig:met} show the $m_{\rm F275W}-m_{\rm F336W}$ vs.\,$m_{\rm F336W}-m_{\rm F438W}$ two-color diagrams of the cluster HB stars.
  The gray dashed-dot lines are used to derive the verticalized $m_{\rm F275W}-m_{\rm F336W}$ vs.\,$\Delta$($m_{\rm F336W}-m_{\rm F438W}$) diagrams, which are plotted in the middle panels of Figure~\ref{fig:met} and are obtained by subtracting from the  $m_{\rm F336W}-m_{\rm F438W}$ color of each star the color of the fiducial line with the same $m_{\rm F275W}-m_{\rm F336W}$.

The histogram distributions of $\Delta$($m_{\rm F336W}-m_{\rm F438W})$ are plotted in the right panels of Figure~\ref{fig:met} and clearly reveal two main peaks.
 The bimodal color distribution, which is evident from a visual inspection of Figure \ref{fig:2col}, is demonstrated by the large values of the Bimodality coefficient\footnote{The Bimodality coefficient is defined as \\ 
$BC = \frac{m_{3}^{2}+1}{m_{4} + 3\frac{(n-1)^{2}}{(n-2)(n-3)}}$ \\
 where $m_{3}$ and $m_{4}$ are the skewness of the distribution and its excess of kurtosis and $n$ is the number of points of a given distribution, respectively.  Bimodal distributions are characterized by values of BC that exceed the critical value $BC_{\rm crit} = 0.555$.}  (BC; \citealt{sasinc}), which is larger than BC$\sim$0.73 for all studied clusters.

We first selected a raw sample of 1G stars, including all stars with $\Delta$($m_{\rm F336W}-m_{\rm F438W})$ values lower than the value corresponding to the minimum of the histogram between the two peaks. We fitted these stars with a Gaussian function by means of least squares, plotted in red in the right panels of Figure~\ref{fig:met}.
 The fraction of 1G stars along the HB is derived as the ratio between the area below the red Gaussian and the area of the whole histogram. We verified that the results are not significantly affected by small changes in the slope of the dashed-gray lines.
To ensure a proper estimate of the fraction of 1G stars in NGC\,6388 and NGC\,6441 blue-HB stars have been included in the total number of 2G stars. 
 
The method described above for NGC\,6388 and NGC\,6637 has been extended to all Galactic GCs. The resulting fractions of 1G stars are listed in Table~\ref{tab:dati} and range from $\sim$15\% in NGC\,6388 to $\sim$68\% in NGC\,6838. 

For nine out twelve Galactic GCs, namely NGC\,104, NGC\,6352, NGC\,6366, NGC\,6388, NGC\,6496, NGC\,6652, NGC\,6624, NGC\,6637 and NGC\,6838 the fraction of 1G stars along the RGB has been derived by \citet{milone2017a} and in NGC\,6352 and NGC\,6838 the fraction of 1G stars along the MS are estimated by \citet{milone2020b}, by using the same dataset. Our measurements provide the first estimates of population ratios in NGC\,5927, NGC\,6304 and NGC\,6441. For each analyzed clusters, the fraction of 1G stars detected along the red HB is consistent within 1-$\sigma$ with that obtained for the RGB and MS by \citet[]{milone2017a, milone2020b}, with the exception of NGC\,6388, for which the fraction of 1G stars derived in this work, namely 0.183$\pm$0.120, is lower than that inferred by \citet{milone2017a}, with a difference significant at 3-$\sigma$ level.
We emphasize that the fractions of 1G stars provided in this section are derived from {\it HST} photometry and are representative of the central cluster regions alone. Although the field of view of the majority of studied clusters encloses the half-light radius  \citep[see Table~2 of this paper and Table~2 by][]{milone2017a}, the global fraction of 1G stars of some GC can differ from that observed in the central regions. Indeed, the 2G of some massive GCs is significantly more-centrally concentrated than the 1G \citep[e.g.][]{sollima2007a, bellini2009b, milone2012a, lee2019a, lee2020a}.


Figure~\ref{fig:cmd} shows a collection of $m_{\rm F438W}$ vs.\,$m_{\rm F438W} - m_{\rm F814W}$ CMDs for the studied clusters, in which we plotted in red the selected 1G stars. Colors made with optical magnitudes are strongly affected by $T_{\rm eff}$ variations, hence are sensitive to stellar populations with different helium abundances.  Optical magnitudes have low sensitivity to light-element abundance variations when compared with ultraviolet bands, although the F438W and F814W filters are significantly affected by carbon and nitrogen variations. As expected, the sample of 1G stars exhibit, on average, redder $m_{\rm F438W} - m_{\rm F814W}$ colors and fainter $m_{\rm F438W}$ magnitudes than the bulk of 2G stars. This fact is consistent with the previous findings that 2G stars are typically enhanced in helium  and depleted in carbon with respect to the 1G \citep[see e.g.\,][for determinations the chemical composition of 1G and 2G stars along the RGB and the HB]{lagioia2018a, milone2018a, tailo2020a}.

\subsection{Multiple populations in Magellanic Cloud clusters}\label{sec:mcpop}

In this section, we exploit multi-band photometry of the Galactic GC 47\,Tuc, where multiple populations along the HB have been extensively studied, to introduce new two-color diagrams that allow to disentangle 1G and 2G stars along the red HB. These diagrams will then be used as tools to identify for the first time 1G and 2G stars along the red HB of Small Magellanic Cloud cluster NGC\,416 and the red clump of the Large Magellanic Cloud cluster NGC\,1978.

\begin{figure*}
\begin{centering}
  \includegraphics[height=4.cm,clip]{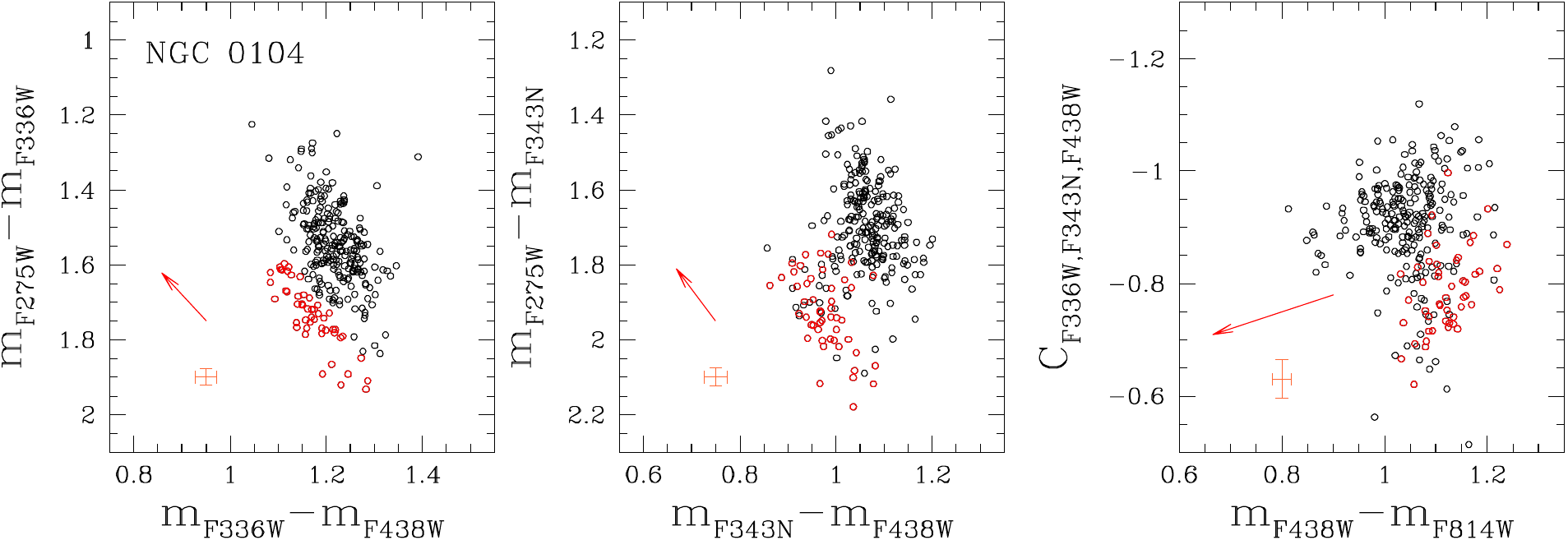}\\
  \includegraphics[height=4.cm,clip]{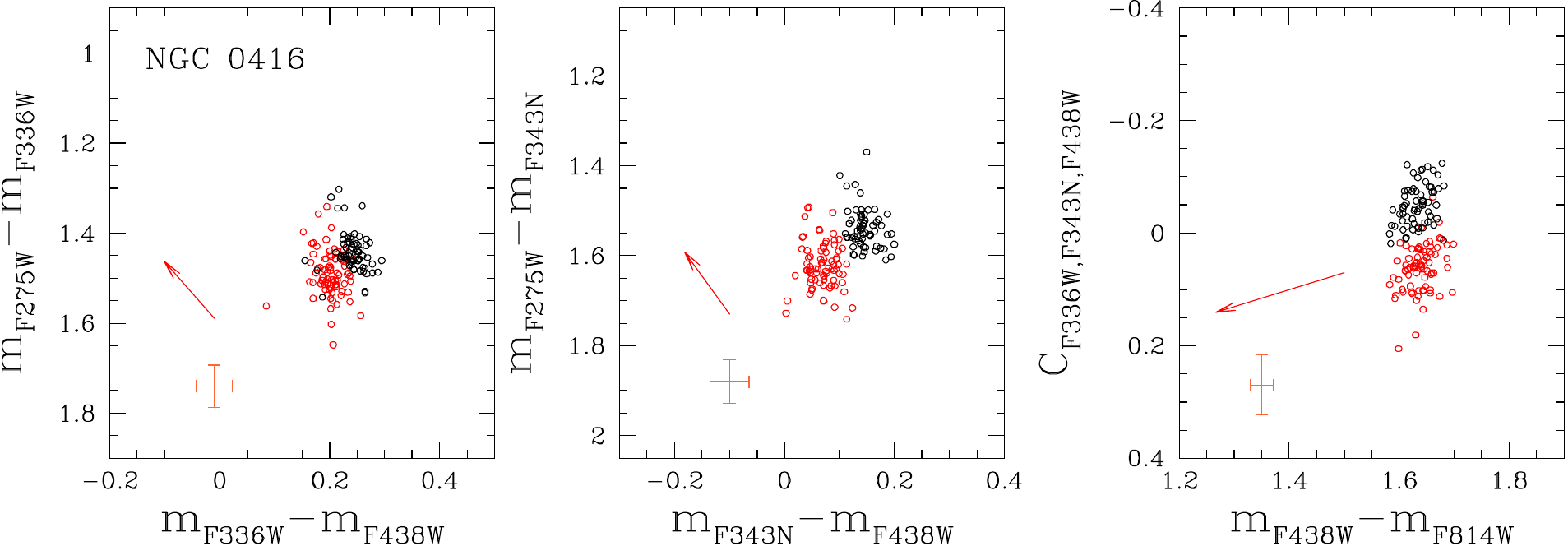}\\
  \includegraphics[height=4.cm,clip]{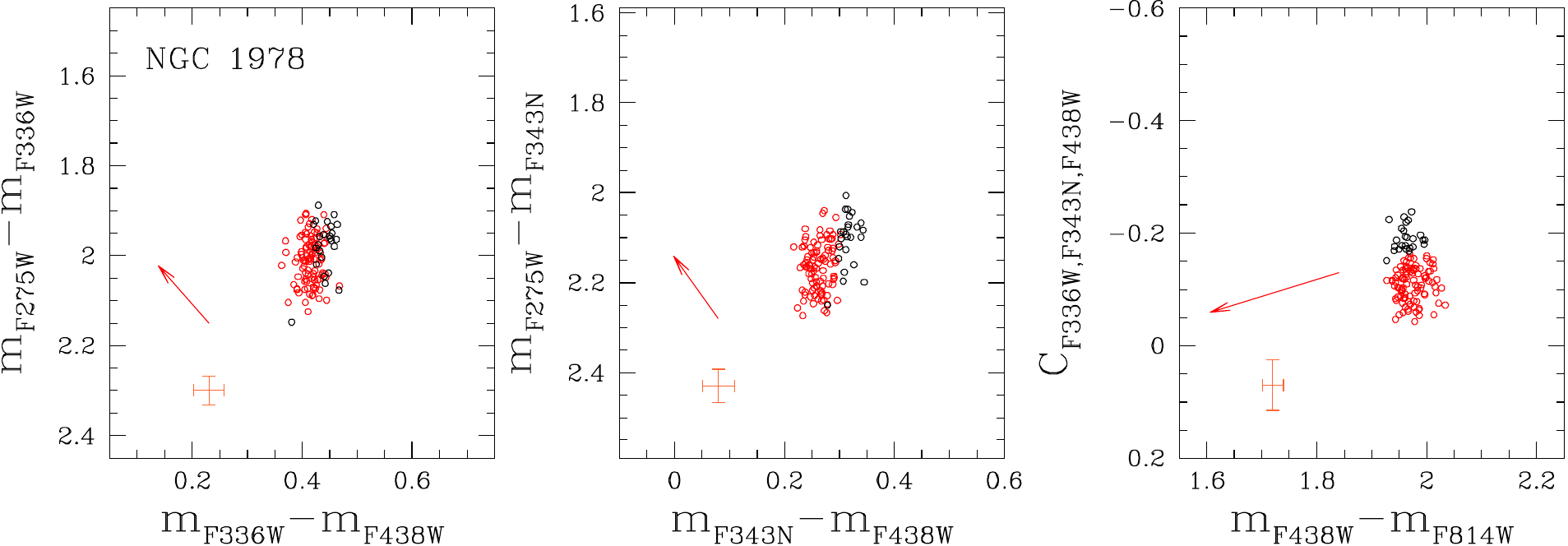}
  \caption{$(m_{\rm F275W}-m_{\rm F336W})$ vs.\,$(m_{\rm F336W}-m_{\rm F438W})$ (left panels),  $(m_{\rm F275W}-m_{\rm F343N})$ vs.\,$(m_{\rm F343N}-m_{\rm F438W})$ (central panels) and $C_{\rm F336W,F343N,F438W}$ vs.\,$m_{\rm F438W}-m_{\rm F814W}$ (right panels) two-colour diagrams for 47\,Tuc (top), NGC\,416 (middle) and NGC\,1978 (bottom). Selected 1G and 2G stars are colored red and black, respectively. Red arrows indicate the reddening vectors for $\Delta E(B-V) = 0.1$.}
 \label{fig:47tuc} 
\end{centering} 
\end{figure*}

Indeed, in addition to F275W, F336W and F438W data, images in F343N are available for 47\,Tuc, NGC\,1978 and NGC\,416. The F343N, which is a narrow filter that comprises the spectral region that includes various NH and molecular bands, is mostly sensitive to stellar populations with different nitrogen abundances. Hence, we exploited this filter to build the two-color $(m_{\rm F275W}-m_{\rm F343N})$ vs.\,$(m_{\rm F343N}-m_{\rm F438W})$ diagram and the $C_{\rm F336W,F343N,F438W}=(m_{\rm F336W}-m_{\rm F343N})-(m_{\rm F343N}-m_{\rm F438W})$ vs.\,$m_{F438W}-m_{F814W}$ pseudo two-color diagram for red HB and red clump stars. Results are illustrated in the upper panels of Figure~\ref{fig:47tuc} for 47 Tuc, where we compare the classic $(m_{\rm F275W}-m_{\rm F336W})$ vs.\,$(m_{\rm F336W}-m_{\rm F438W})$ two-color diagram with the diagrams introduced in this work. Clearly, the fact that the bulk of selected 1G and 2G stars (red and black points in Figure~\ref{fig:47tuc}) populate distinct regions in each diagram, demonstrates that the two-color and pseudo two-color diagrams that involve photometry in F343N are powerful tools to detect multiple populations along the red HB and the red clump. \footnote{We remind that the groups red-HB stars marked with red and blue colors include the majority of 1G and 2G stars, respectively, and that some contamination is expected. Such small contamination  is negligible for our purpose of showing that 1G and 2G stars populate distinct regions in the diagrams of Figure~\ref{fig:47tuc}.}.


\begin{figure*} 
\begin{centering} 
  \includegraphics[height=5.cm,clip]{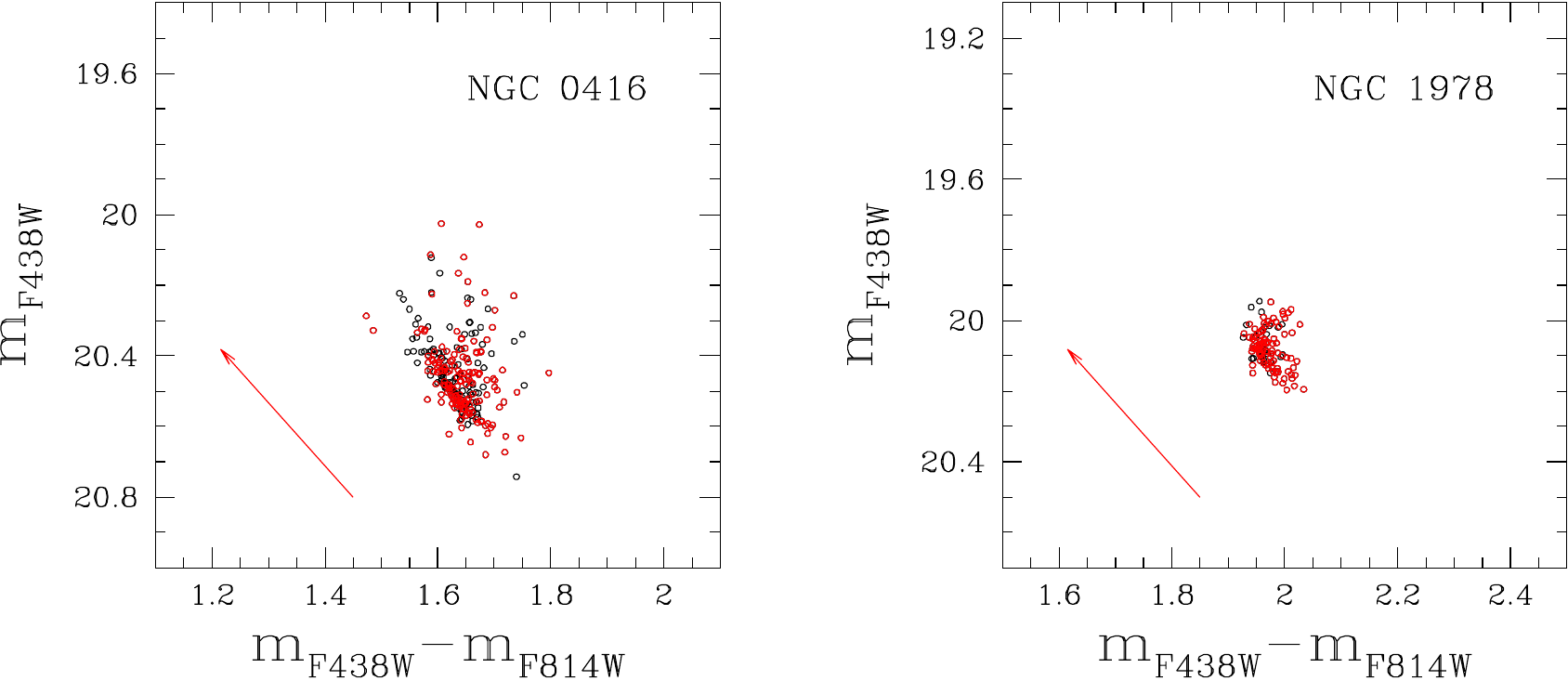}
  \caption{Comparison of the NGC\,416 red HB (left panel) and the NGC\,1978 red clump (right panel) in optical CMDs. 1G stars are colored in red.}
 \label{fig:cmde}
 \end{centering} 
\end{figure*}

These diagrams are plotted in the middle and lower panels of Figure~\ref{fig:47tuc} for NGC\,416 and NGC\,1978, respectively, and reveal that both clusters host multiple populations along the red HB and the red clump. 
 By using the method described in Section~\ref{sec:pratio}, we find that in contrast with what is observed in the majority of Galactic GCs, which are dominated by the 2G \citep{milone2017a}, the majority of red HB and red clump stars in both NGC\,416 and NGC\,1978 belong to the first generation ($54.2 \pm 4.4 \%$ and 84.6$\pm$2.9\%).
 Figure~\ref{fig:cmde} shows the optical CMDs of NGC\,416 and NGC\,1978, in which red and black points indicate respectively 1G and 2G stars. 
 While 2G stars are significantly brighter and bluer than the 1G in the optical CMDs of Galactic GCs (Figure~\ref{fig:cmd}), stars of both populations of NGC\,416 and NGC\,1978 are distributed along the whole red HB and red clump.
 

\subsection{Comparison with simulated multiple populations}\label{sec:teoria}
The behaviour of simple-population stars along the red clump and HB phases is well constrained from theory. In a nutshell, when a low-mass RGB star reaches the RGB tip, the temperature in its core becomes sufficiently high ($\sim10^8$ K) to allow the Helium ignition, starting the HB phase, during which stars burn Helium via $3\alpha$ reaction, producing Carbon and Oxygen.
At the start of Helium burning, the core mass is almost the same for all stars ($\sim0.45$ M$\odot$). Since the luminosity in this phase depends mainly on the core mass, stars in HB have approximately the same luminosity, so they are arranged along a horizontal sequence in CMDs.
During this evolutionary phase, the envelope mass can vary from star-to-star \citep[e.g.][]{iben1970a, dorman1992a}. Stars with lower initial mass or that have experienced a larger amount of mass loss during the RGB phase have smaller envelope masses, hence different colors. Indeed, the effective temperature ($T_{eff}$) depends on the envelope mass: lower envelope masses lead to lower $T_{eff}$, therefore to redder stars.
In the case of multiple-population GCs, the position of a star along the red HB or the red clump may also depend on its chemical composition \citep[e.g.][]{dantona2002a, salaris2008a}.

\begin{figure*} 
\begin{centering} 
  \includegraphics[height=8cm,clip, trim={0.0cm 4cm 0cm 4cm}]{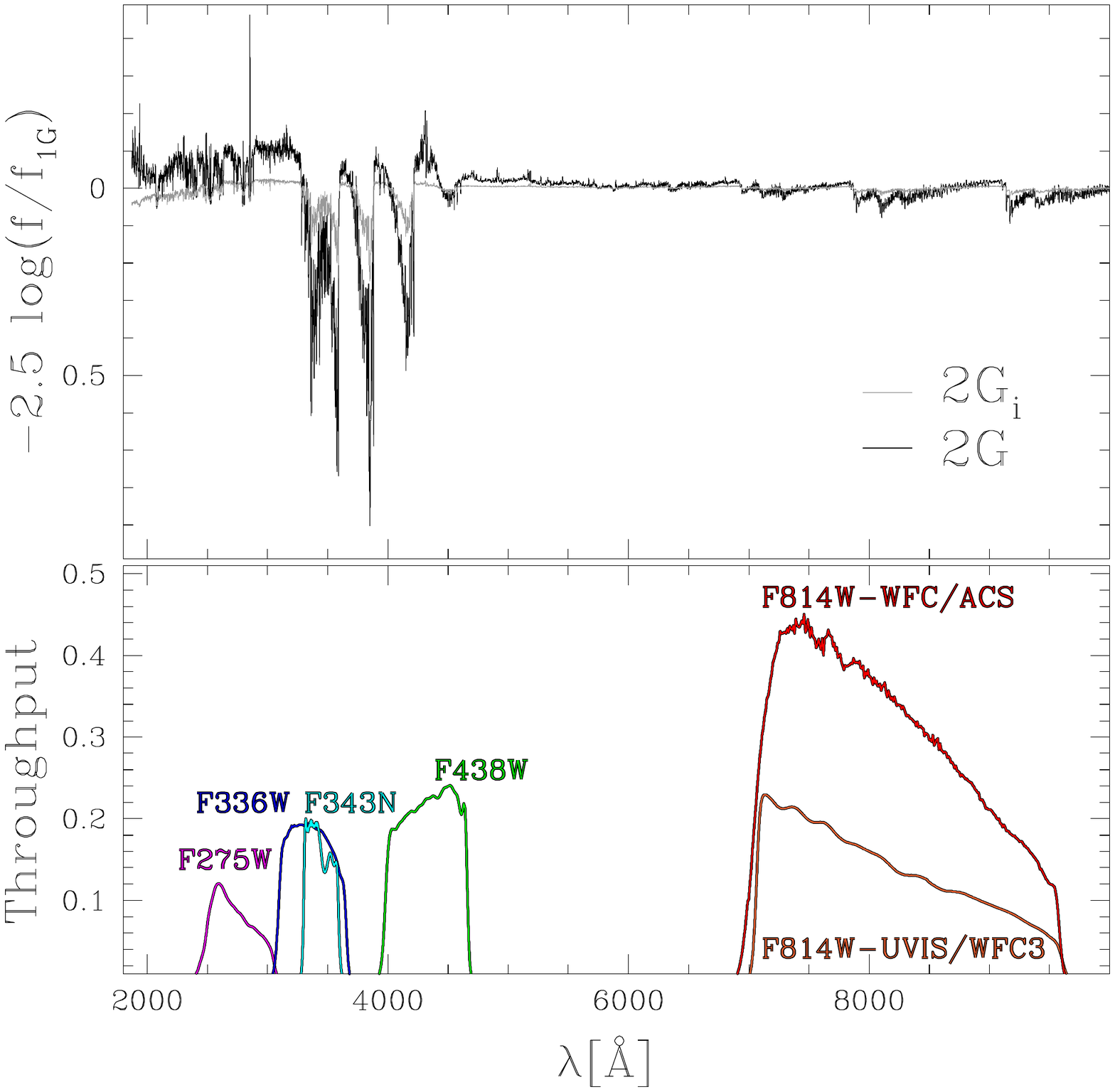}
  \includegraphics[height=8cm,clip, trim={0.0cm 4cm 0cm 4cm}]{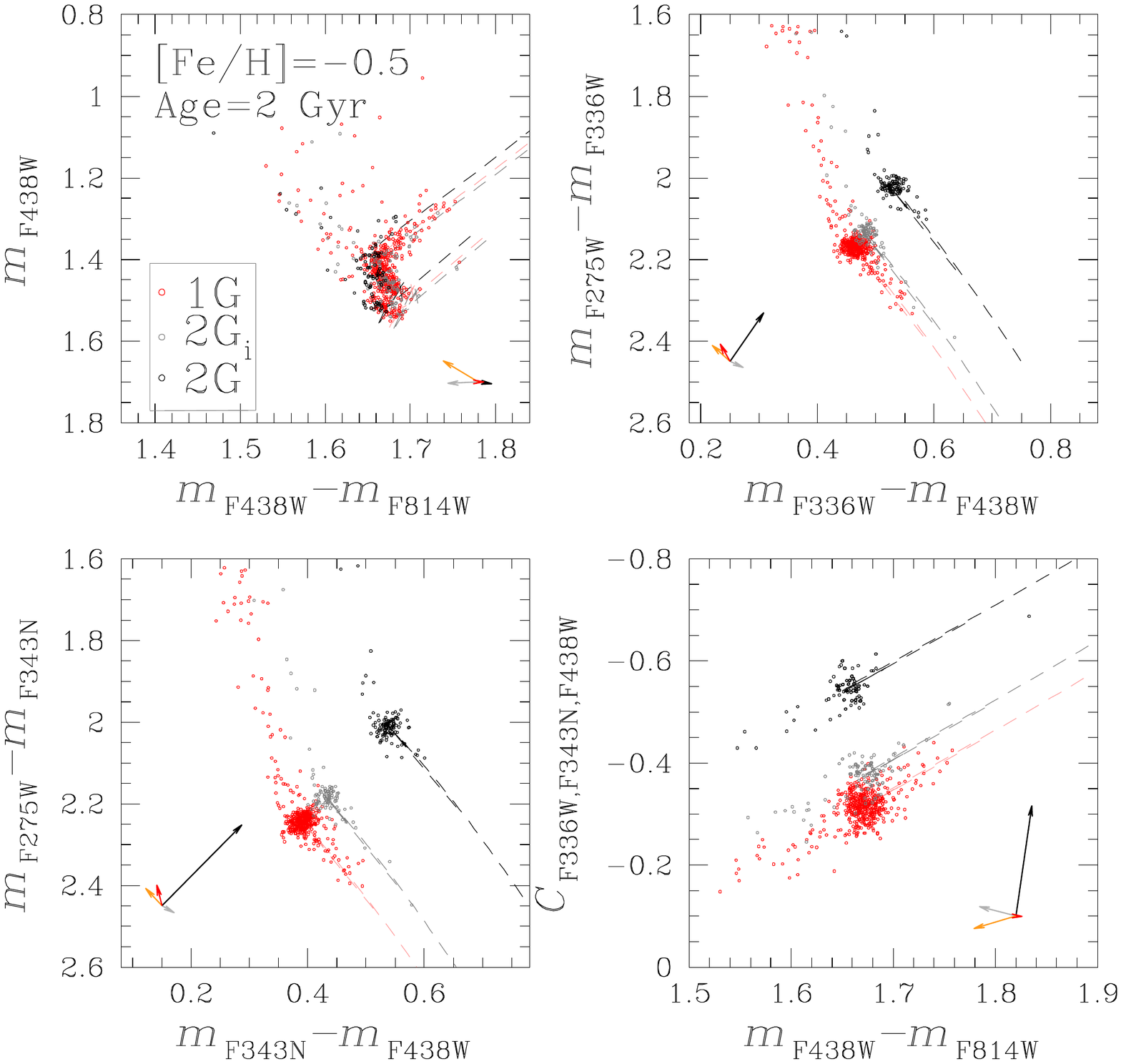}
  \caption{\textit{Left panels.} Flux ratio between the spectrum of 2G (black) or 2G$_{\rm i}$ star (gray) and  the spectrum of 1G star with $T_{\rm eff}=4,898$ K and $\log{g}$=2.46 (upper panel). The throughputs of the filters used in this paper are plotted in the bottom panel. \textit{Right panels.} Simulated diagrams of 2-Gyr old HB stars with the same iron abundance, [Fe/H]=$-$0.5. Red and black dots correspond to 1G and 2G stars, respectively, while gray dots have similar chemical composition as 2G stars of NGC\,1978, which are enhanced in nitrogen by 0.1 dex with respect to the 1G. The corresponding isochrones are represented with dashed lines. The gray, black, red and orange vectors indicate the effect of changing C, N, O and Fe, respectively, one at time, on the colors and magnitudes See text for details.}
 \label{fig:simu02} 
\end{centering} 
\end{figure*}

To further investigate the impact of light-element abundance variations on the colors and magnitudes of red HB and red clump stars, we qualitatively compared the observed photometric diagrams with simulations.
 To do this, we first extended to the red clump the method used in previous work from our group, which is based on synthetic spectra with different chemical compositions \citep[see e.g.][for details]{milone2012a, milone2018a}.
We first simulated the colors and magnitudes of red-clump stars in a stellar population with age 2.0 Gyr and [Fe/H]=$-$0.5, based on MESA isochrones \citep[][]{dotter2016a, choi2016a, paxton2011a, paxton2013a, paxton2015a}. To do this, we identified six points along the isochrone and extracted their effective temperatures and gravities. For each point, we computed a reference spectrum with solar-scaled light-element abundances that mimics 1G stars and two comparison spectra that correspond to 2G stars with different chemical compositions. Specifically, we simulated a spectrum of a 2G star enhanced in nitrogen by 0.6 dex and depleted in both C and O by 0.3 dex, and a spectrum of a stellar population (hereafter 2G$_{\rm i}$) with [C/Fe]=0.0, [N/Fe]=$+$0.1 and [O/Fe]=0.0 as inferred by \citet{milone2020a} for NGC\,1978. We assumed microturbolent velocity of 2 km s$^{-1}$
for all stars, which is higher than the values inferred for red-HB stars  \citep[e.g.][]{afsar2018a}. We verified that adopting microturbolent velocity of 2 km s$^{-1}$ has negligible impact on the relative colors of 1G and 2G stars and does not change our conclusions, thus confirming previous conclusion by \citet{sbordone2011a}.  
Atmosphere models are computed by using the computer program ATLAS12 \citep{kurucz1970a, kurucz1993a, sbordone2004a}, which is based on the opacity-sampling method and assumes local thermodynamic equilibrium. We derived synthetic spectra in the wavelength interval between 1,800 and 10,000 \AA~by using SYNTHE \citep{kurucz1981a, castelli2005a, kurucz2005a, sbordone2007a}.
As an example, in the upper-left panel of Figure~\ref{fig:simu02} we plot the wavelength against the fluxes  of a 2G star and a 2G$_{\rm i}$ star with $T_{\rm eff}=4,898$K and $\log{g}$=2.46, relative to the 1G star with the same atmosphere parameters. For completeness, we show the throughputs of the F275W, F336W, F343N, F438W and F814W UVIS/WFC3 filters and the F814W ACS/WFC filter used in this paper. 

Stellar magnitudes are calculated by integrating synthetic spectra over the bandpasses of the filters used in this paper, and are used to derive the magnitude difference, $\delta m_{\rm X}$, between the comparison and the reference spectrum. Hence, we derived the magnitudes of simulated 2G and 2G$_{\rm i}$ stars by adding to the 1G isochrones the corresponding values of $\delta m_{\rm X}$.

The isochrones of 2G and 2G$_{\rm i}$ stars are finally used the derive simulated diagrams illustrated in the right panels of Figure~\ref{fig:simu02}, where we adopted the mass function by \citet{salpeter1955a} and assumed that the 35\% of sources in the CMD are binary systems, which is the typical binary fraction inferred in intermediate-age Magellanic Cloud star clusters \citep{milone2009a}. We added to simulated photometry, typical uncertainties of our observations as inferred from NGC\,1978 by using artificial-star tests \citep[see e.g.][for details]{anderson2008a}. 

We find that 2G and 2G$_{\rm i}$ stars are almost indistinguishable from the 1G in the $m_{\rm F438W}$ vs.\,$m_{\rm F438W}-m_{\rm F814W}$ CMD and have higher $m_{\rm F336W}-m_{\rm F438W}$ and $m_{\rm F343N}-m_{\rm F438W}$ colors than the 1G. Moreover, 2G and 2G$_{\rm i}$ stars exhibit lower values of $m_{\rm F275W}-m_{\rm F336W}$, $m_{\rm F275W}-m_{\rm F343N}$ and $C_{\rm F336W,F343N,F438W}$ than the 1G. Although a quantitative comparison between the observed and the simulated diagrams is beyond the purposes of our work, we note that the observed behaviour of the selected stellar populations of NGC\,1978 qualitatively matches the simulated 1G and 2G$_{\rm i}$.  This fact further demonstrates that the red clump of NGC\,1978 is not consistent with a simple isochrone but hosts two stellar populations with different chemical composition. 

The color differences between the simulated stellar populations are mostly due to the NH molecular bands that affect the F336W and F343N filters. Since 2G and 2G$_{\rm i}$ stars are enhanced in nitrogen with respect to the 1G, they exhibit fainter magnitudes in the F336W and F343N than 1G stars with the same atmosphere parameters.
Moreover, 2G stars are O-poor and C-poor. Hence, they exhibit brighter F275W and F438W magnitudes as a consequence of the strengths of the OH and CN molecules that affect F275W and F438W fluxes, respectively.  
For completeness, we investigate the effect of C, N, O and Fe abundance variations on the diagrams Figure~\ref{fig:simu02}. The gray, black, red and orange vectors plotted on the bottom-left corner of each panel show the average effect of changing  [C/Fe], [N/Fe], [O/Fe] and [Fe/H] by $-$0.3, $+0.6$, $-0.3$ and $-0.1$ dex, respectively.

We also extended the analysis to 12-Gyr old HB stars with [Fe/H]=$-$0.5. In this case, we exploited $\alpha$-enhanced isochrones from the Roma database \citep[e.g.][and references therein]{tailo2019b}. We assumed that 1G stars have solar-scaled carbon and nitrogen abundances and [O/Fe]=0.4, while 2G stars are enhanced in N by 0.6 dex and depleted in both C and O by 0.3 dex, with respect to the 1G. We adopted helium content Y=0.25 for 1G stars and assumed that the 2G is  enhanced by 1\% in helium mass fraction, which is the typical helium difference between 2G and 1G stars in GCs \citep[e.g.][]{lagioia2018a, milone2018a}. We adopted a fraction of binaries of 0.10, which is consistent with results based on MS stars of GCs \citep[e.g.][]{milone2016a}. 

\begin{figure*} 
\begin{centering}
  \includegraphics[height=10cm,clip, trim={0.0cm 4cm 0cm 4cm}]{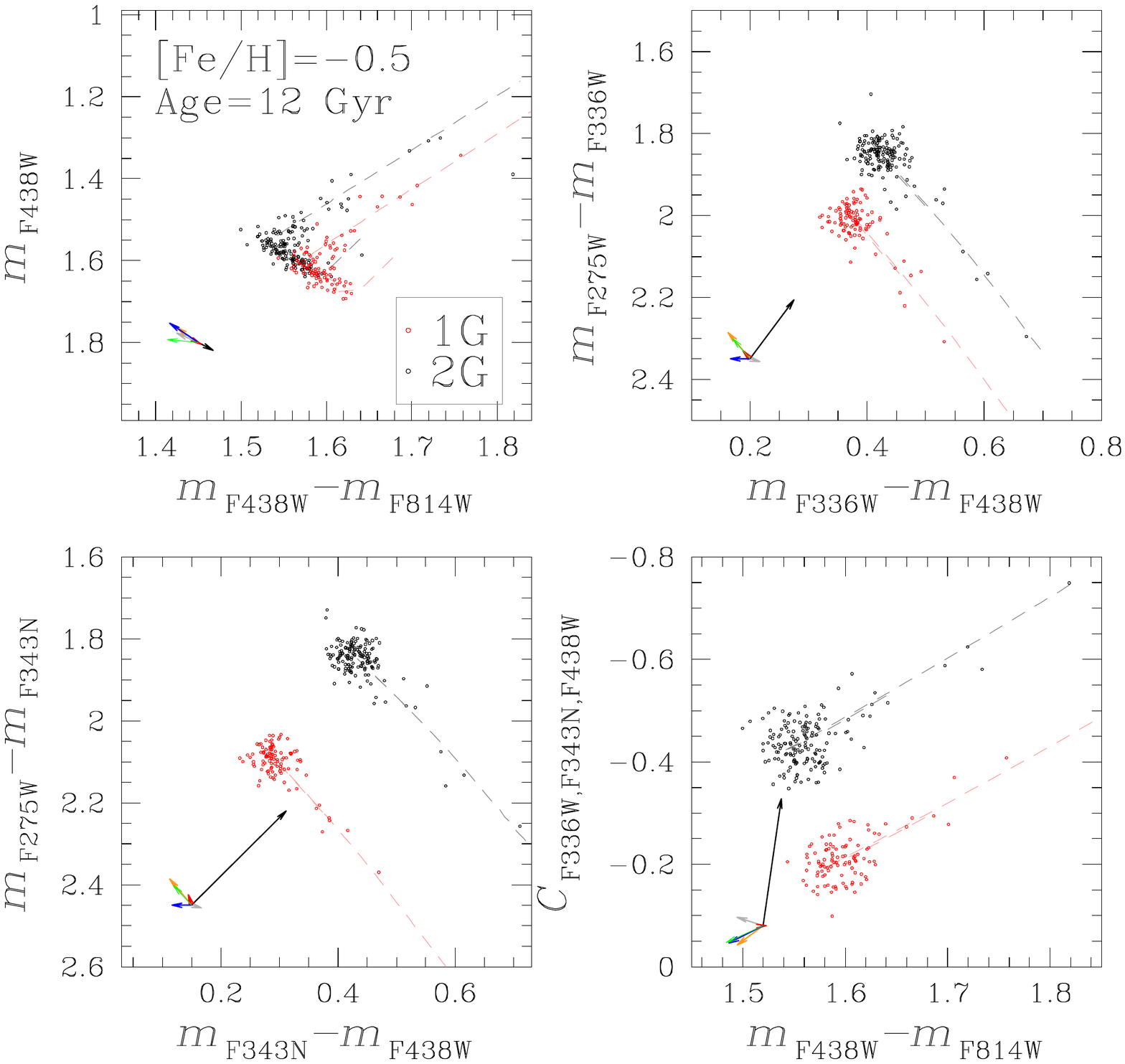}
  \caption{Simulated diagrams of 12-Gyr old stellar populations with [Fe/H]=$-$0.5. The blue, gray, black, red, orange and green vectors indicate the effect of changing He, C, N, O, Fe and mass loss, respectively, one at time, on the colors and magnitudes. See text for details.}
 \label{fig:simu12} 
\end{centering} 
\end{figure*}

Results are illustrated in Figure~\ref{fig:simu12}, and are qualitatively consistent with observations of Galactic GCs. Specifically, 2G stars exhibit bluer $m_{\rm F438W}-m_{\rm F814W}$ colors than the 1G. In this case, the color difference is mostly due to the hotter temperature of He-rich 2G HB stars.
 Similarly to what observed in the simulated red clumps, 2G stars have higher $m_{\rm F336W}-m_{\rm F438W}$ and $m_{\rm F343N}-m_{\rm F438W}$ colors and lower values of $m_{\rm F275W}-m_{\rm F336W}$, $m_{\rm F275W}-m_{\rm F343N}$ and $C_{\rm F336W,F343N,F438W}$ than the 1G. These color differences are mostly due to C, N, and O variations that impact the flux in the F275W, F336W, F343N and F438W bands mostly through CN, CH, OH and NH molecules. 
 The gray, black, red and orange vectors plotted on the bottom-left corner of each panel of Figure~\ref{fig:simu12} illustrate the effect of changing  [C/Fe], [N/Fe], [O/Fe] and [Fe/H], one at time, by $-$0.3, $+0.6$, $-0.3$ and $-0.1$ dex, respectively. Blue and green arrows correspond to helium mass fraction and RGB mass loss increase of $\Delta$Y=0.03 and $\Delta$ $\mathcal{M}$=0.02 $\mathcal{M}_{\odot}$, respectively.

\begin{table*}
\begin{centering}
\caption{Fractions of 1G stars of GCs measured in this paper along the HB and those derived in our previous work based on the RGB \citep{milone2017a, milone2018a, zennaro2019a, milone2020a} and the MS \citep{milone2020b}. $<$$N_{\rm 1G}/N_{\rm TOT}$$>$ provides the best estimates of the fractions of 1G stars, derived by combining all results. The last coloumn indicates the ratio between the maximum radius reached by the FoV and the half light radius.}
\begin{tabular}{cccccccc}
\label{tab:dati}
\\
\hline\hline
     \\
    ID & $N_{1G}/N_{TOT}$ (this work) & $<$$N_{\rm 1G}/N_{\rm TOT}$$>$ & $r_{\rm FoV}/r_{\rm hl}$ & ID & $N_{1G}/N_{TOT}$ (this work) & $<$$N_{\rm 1G}/N_{\rm TOT}$$>$ & $r_{\rm FoV}/r_{\rm hl}$ \\
    \\
    \hline
    \\
     NGC\,0104 & $0.218 \pm 0.025$ & $0.180 \pm 0.009$ & 0.56 & NGC\,6397 & $-$ & $0.345 \pm 0.036$ & 0.55\\
     NGC\,0288 & $-$ & $0.558 \pm 0.031$ & 0.89 & NGC\,6441 & $0.210 \pm 0.011$ & $0.210 \pm 0.011$ & 2.90\\
     NGC\,0362 & $-$ & $0.279 \pm 0.015$ & 2.01 & NGC\,6496 & $0.636 \pm 0.068$ & $0.666 \pm 0.035$ & 1.40\\
     NGC\,1261 & $-$ & $0.359 \pm 0.016$ & 2.35 & NGC\,6535 & $-$ & $0.536 \pm 0.081$ & 1.70\\
     NGC\,1851 & $-$ & $0.264 \pm 0.015$ & 3.00 & NGC\,6541 & $-$ & $0.396 \pm 0.020$ & 1.56\\
     NGC\,2298 & $-$ & $0.370 \pm 0.037$ & 1.64 & NGC\,6584 & $-$ & $0.451 \pm 0.026$ & 2.27\\
     NGC\,2419 & $-$ & $0.370 \pm 0.010$ & 2.18 & NGC\,6624 & $0.268 \pm 0.035$ & $0.276 \pm 0.020$ & 1.87\\
     NGC\,2808 & $-$ & $0.232 \pm 0.014$ & 2.32 & NGC\,6637 & $0.450 \pm 0.039$ & $0.426 \pm 0.017$ & 2.05\\
     NGC\,3201 & $-$ & $0.436 \pm 0.036$ & 0.52 & NGC\,6652 & $0.380 \pm 0.063$ & $0.349 \pm 0.026$ & 3.09\\
     NGC\,4590 & $-$ & $0.381 \pm 0.024$ & 1.13 & NGC\,6656 & $-$ & $0.274 \pm 0.020$ & 0.51\\
     NGC\,4833 & $-$ & $0.362 \pm 0.025$ & 0.73 & NGC\,6681 & $-$ & $0.234 \pm 0.019$ & 2.31\\
     NGC\,5024 & $-$ & $0.328 \pm 0.020$ & 1.35 & NGC\,6715 & $-$ & $0.267 \pm 0.012$ & 2.08\\
     NGC\,5053 & $-$ & $0.544 \pm 0.062$ & 0.53 & NGC\,6717 & $-$ & $0.637 \pm 0.039$ & 2.01\\
     NGC\,5139 & $-$ & $0.086 \pm 0.010$ & 0.50 & NGC\,6723 & $-$ & $0.363 \pm 0.017$ & 1.05\\
     NGC\,5272 & $-$ & $0.305 \pm 0.014$ & 0.83 & NGC\,6752 & $-$ & $0.294 \pm 0.023$ & 0.91\\
     NGC\,5286 & $-$ & $0.342 \pm 0.015$ & 2.25 & NGC\,6779 & $-$ & $0.469 \pm 0.041$ & 1.29\\
     NGC\,5466 & $-$ & $0.467 \pm 0.063$ & 0.67 & NGC\,6809 & $-$ & $0.311 \pm 0.029$ & 0.55\\
     NGC\,5897 & $-$ & $0.547 \pm 0.042$ & 0.79 & NGC\,6838 & $0.640 \pm 0.083$ & $0.630 \pm 0.035$ & 0.88\\
     NGC\,5904 & $-$ & $0.235 \pm 0.013$ & 0.90 & NGC\,6934 & $-$ & $0.326 \pm 0.020$ & 2.30\\
     NGC\,5927 & $0.373 \pm 0.033$ & $0.373 \pm 0.033$ & 1.52 & NGC\,6981 & $-$ & $0.542 \pm 0.027$ & 1.67\\
     NGC\,5986 & $-$ & $0.246 \pm 0.012$ & 1.81 & NGC\,7078 & $-$ & $0.399 \pm 0.019$ & 1.79\\
     NGC\,6093 & $-$ & $0.351 \pm 0.029$ & 2.52 & NGC\,7089 & $-$ & $0.224 \pm 0.014$ & 1.47\\
     NGC\,6101 & $-$ & $0.654 \pm 0.032$ & 1.48 & NGC\,7099 & $-$ & $0.380 \pm 0.028$ & 1.55\\
     NGC\,6121 & $-$ & $0.290 \pm 0.037$ & 0.39 & IC\,4499 & $-$ & $0.510 \pm 0.050$ & 1.18 \\
     NGC\,6144 & $-$ & $0.444 \pm 0.037$ & 0.45 & Lindsay\,1 & $-$ & $0.663 \pm 0.037$ & 0.65\\
     NGC\,6171 & $-$ & $0.397 \pm 0.031$ & 0.90 & Lindsay\,38 & $-$ & $1.000$ & 1.02\\
     NGC\,6205 & $-$ & $0.184 \pm 0.013$ & 1.05 & Lindsay\,113 & $-$ & $1.000$ & $-$\\
     NGC\,6218 & $-$ & $0.400 \pm 0.029$ & 0.93 & NGC\,0121 & $-$ & $0.517 \pm 0.026$ & 2.12\\
     NGC\,6254 & $-$ & $0.364 \pm 0.028$ & 0.86 & NGC\,0339 & $-$ & $0.883 \pm 0.022$ & 0.64\\
     NGC\,6304 & $0.330 \pm 0.046$ & $0.330 \pm 0.046$ & 1.13 & NGC\,0416 & $0.542 \pm 0.044$ & $0.500 \pm 0.025$ & 2.20\\
     NGC\,6341 & $-$ & $0.304 \pm 0.015$ & 1.63 & NGC\,0419 & $-$ & $1.000$ & 1.44\\
     NGC\,6352 & $0.417 \pm 0.083$ & $0.497 \pm 0.033$ & 0.76 & NGC\,1783 & $-$ & $1.000$ & 0.98\\
     NGC\,6362 & $-$ & $0.574 \pm 0.035$ & 0.81 & NGC\,1806 & $-$ & $1.000$ & 0.14\\
     NGC\,6366 & $0.636 \pm 0.182$ & $0.431 \pm 0.045$ & 0.51 & NGC\,1846 & $-$ & $1.000$ & 1.75\\
     NGC\,6388 & $0.183 \pm 0.0120$ & $-$ & 2.45 & NGC\,1978 & $0.846 \pm 0.029$ & $0.833 \pm 0.025$ & 0.82\\
\hline\hline
\end{tabular}\\
\footnotesize{{\it{Note.}} No estimate of the half light radius of Lindsay\,113 is currently available in literature.}
\end{centering}
\end{table*}

\section{Relations with the parameters of the host globular clusters}\label{sec:rela}
\begin{figure*} 
\begin{centering} 
  \includegraphics[width=16cm, height=7cm,clip]{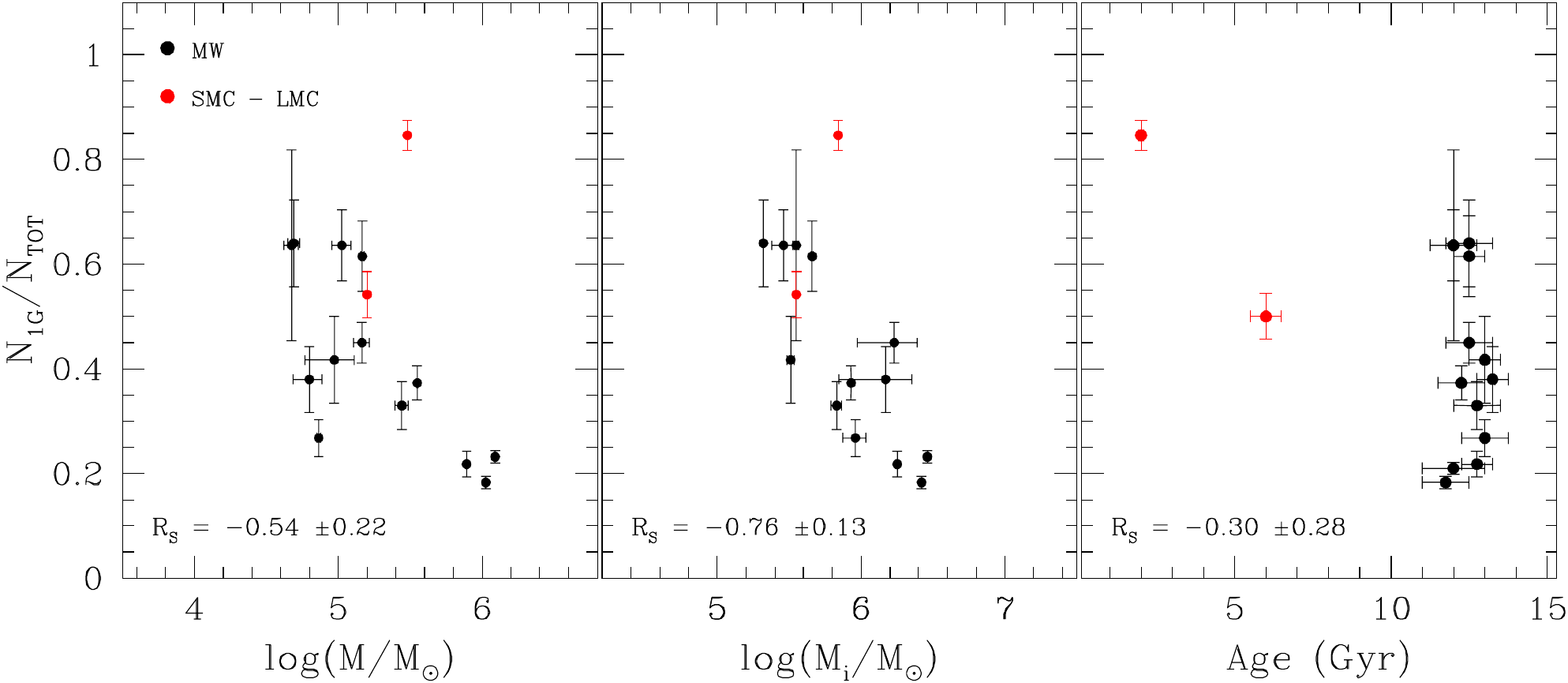}
  \caption{The fractions of 1G stars calculated in this work from the HB are plotted against the present-day mass (left) and the initial mass (middle) and cluster ages. Galactic GCs are shown in black, while red dots indicate the extragalactic clusters.\\}
 \label{fig:n1m} 
 \end{centering} 
\end{figure*}

In the following, we analyze the relation between the observed fraction of 1G stars and cluster mass, which represents the GC parameter that shows the strongest correlation with several MP indicators \citep[e.g.][]{milone2020a}, and with cluster age that is possibly associated with the MP phenomenon \citep[e.g.][]{martocchia2018a}.
In the left and middle panel of Figure~\ref{fig:n1m} we show that the fractions of 1G stars derived from the red HBs and the red clumps anti-correlate with both present-day mass, $M$, and the initial masses $M_{\rm i}$  of the host GCs \citep[from][]{glatt2011a, goudfrooij2014a, baumgardt2018a, baumgardt2019a, milone2020a}.\footnote{The values used in this paper are the state of the art for $M_{\rm i}$ of GCs. Nevertheless, these values are affected by various uncertainties, associated with our poor knowledge of the Galaxy the Magellanic Clouds and their tidal fields \citep[see Sect. 2.2 by ][]{milone2020a}. 
In addition, significant uncertainties may come from poorly-known processes occurred during the formation and early evolution of star clusters and of their MPs whose impacts are not taken into account in calculating $M_{\rm i}$ \citep[e.g.][]{renzini2015a}.}
 This fact is indicated by the Spearman's rank correlation coefficients, which are $R_{\rm s} = -0.54 \pm 0.22$ and $R_{\rm s} = -0.76 \pm 0.13$, respectively.
We do not find a significant correlation with cluster ages \citep[from][$\rm R_s$=-0.30$\pm$0.28]{milone2009a, dotter2010a, lagioia2019a, milone2014a} as illustrated in the right panel of Figure~\ref{fig:n1m}.

To increase the number of GCs we exploited results from this paper and from previous work based on RGB and MS stars from our group \citep{milone2017a, milone2018a, tailo2019a, zennaro2019a, milone2020a}.  
The fractions of 1G stars derived in in literature are consistent with those in our paper at 1-$\sigma$ level for all GCs, with the exception of NGC\,6388, which is consistent at $\sim$3-$\sigma$ level only.

\begin{figure*} 
\begin{centering} 
  \includegraphics[width=16cm, height=7cm,clip]{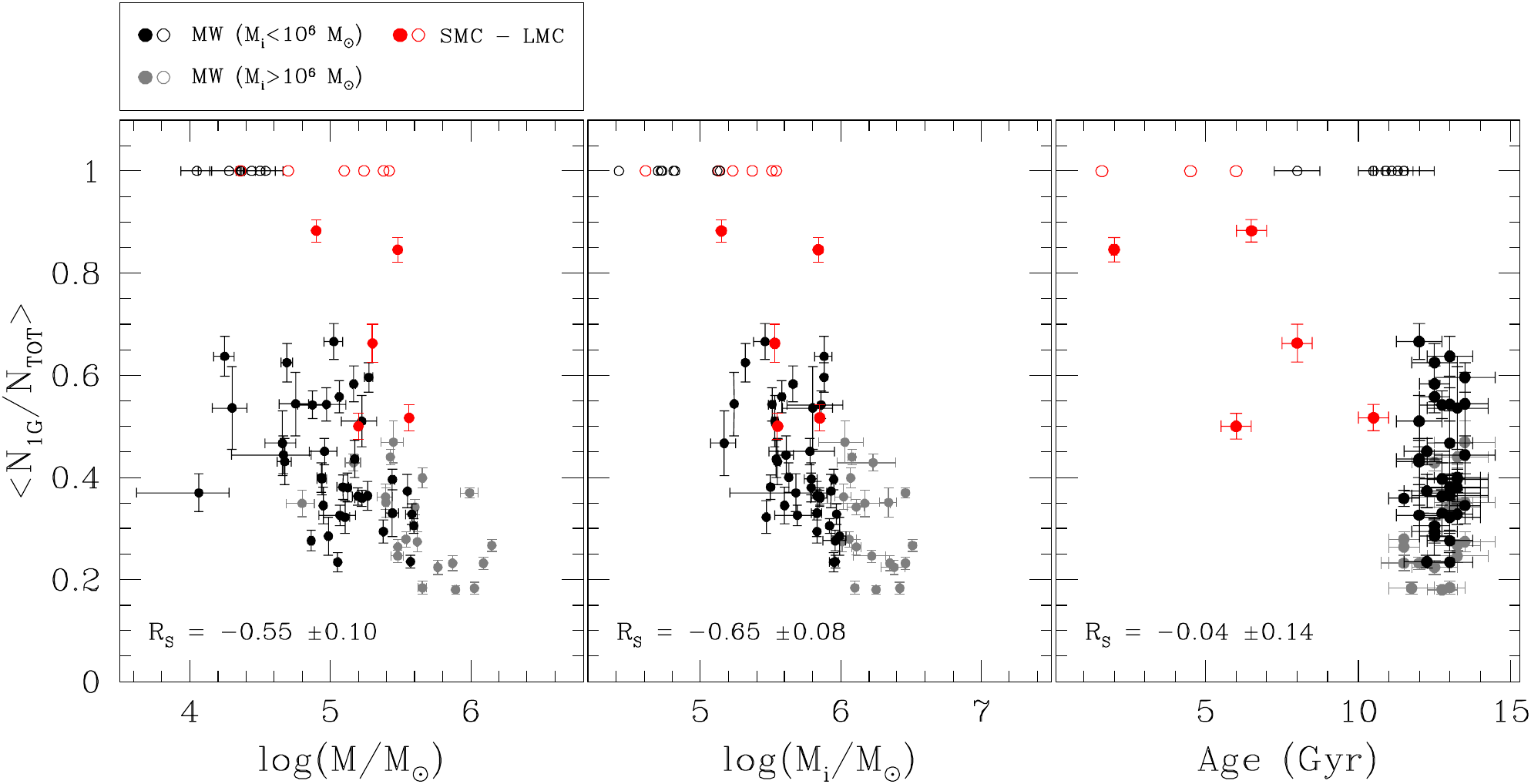}
  \caption{Weighted mean of the fraction of 1G stars versus the present-day mass (left) initial mass (middle) and the age of the host GC (right). Black and gray dots represent respectively Galactic GCs with $M_{i}<$10$^{6}~$M$_{\odot}$ and $M_{i}<$10$^{6}~$M$_{\odot}$, red dots represent extragalactic GCs. The cluster without MPs are represented with open circles.\\}
 \label{fig:masstot} 
\end{centering} 
\end{figure*}

We improved the determination of the fraction of 1G stars by calculating $<$$N_{\rm 1G}/N_{\rm TOT}$$>$, which is the weighted mean of the various estimates of $N_{\rm 1G}/N_{\rm TOT}$, when available. Results are listed in Table~\ref{tab:dati}. In Figure~\ref{fig:masstot}, the same diagrams of Figure~\ref{fig:n1m} are shown for this larger sample of clusters. 
Based on a large sample of GCs with multiple populations, we find correlations with present-day and initial masses ($R_{\rm s} = -0.55 \pm 0.10$ and $R_{\rm s} = -0.65 \pm 0.08$, respectively). Such correlations are confirmed when we extend the analysis to clusters without multiple populations (open symbols in Figure~\ref{fig:masstot}), which provides higher values of $R_{\rm s} = -0.65 \pm 0.08$ and $R_{\rm s} = -0.80 \pm 0.05$ for the anti-correlations with present day and initial masses, respectively.

When we consider GCs with multiple populations alone, we find no evidence between the fraction of 1G stars and cluster age ($R_{\rm s} = -0.04 \pm 0.14$) and we get the same conclusion when we consider only clusters with initial masses smaller than $10^{6} M_{\odot}$ (black and red filled dots in Figure~\ref{fig:masstot}). 
On the contrary, the analysis of all GCs, including simple population star clusters, provides a significant anticorrelation between the fraction of 1G stars and the cluster ages ($R_{\rm s} = -0.60 \pm 0.07$). This result reflects the evidence that all analyzed GCs older than $\sim$12 Gyr host multiple populations, while all studied clusters younger than $\sim$2 Gyr are consistent with simple populations. 
 
\section{The color and magnitude extensions of 1G stars}\label{sec:6388}

Figure~\ref{fig:2col} reveals that the color and magnitude extensions of 1G stars significantly change from one cluster to another. 

\begin{table}
\caption{Extension of the $F336W-F438W$ color of 1G red HB stars and whole red HB stars of Galactic and extragalactic clusters of our sample.}
\begin{tabular}{cccc}
\label{dati}
\\
\hline\hline
     \\
    CLUSTER & $W_{\rm F336W,F438W}^{\rm 1G, rHB}$ & $W_{\rm F336W,F438W}^{\rm rHB}$ \\
    \\
    \hline
    \\
     NGC\,0104 & $0.120 \pm 0.018$ & $0.138 \pm 0.009$ \\
     NGC\,5927 & $0.200 \pm 0.018$ & $0.180 \pm 0.013$ \\
     NGC\,6304 & $0.123 \pm 0.020$ & $0.100 \pm 0.012$ \\
     NGC\,6352 & $0.060 \pm 0.008$ & $0.095 \pm 0.016$ \\
     NGC\,6388 & $0.250 \pm 0.018$ & $0.210 \pm 0.008$ \\
     NGC\,6441 & $0.140 \pm 0.012$ & $0.200 \pm 0.008$ \\
     NGC\,6496 & $0.095 \pm 0.034$ & $0.107 \pm 0.023$ \\
     NGC\,6624 & $0.102 \pm 0.021$ & $0.120 \pm 0.011$ \\
     NGC\,6637 & $0.104 \pm 0.014$ & $0.110 \pm 0.006$ \\
     NGC\,6652 & $0.125 \pm 0.028$ & $0.100 \pm 0.014$ \\
     NGC\,6838 & $0.090 \pm 0.028$ & $0.094 \pm 0.023$ \\
     NGC\,1978 & $0.050 \pm 0.006$ & $0.051 \pm 0.007$ \\
     NGC\,0416 & $0.060 \pm 0.007$ & $0.080 \pm 0.006$ \\
\hline\hline
\end{tabular}\\
\label{tab:est}
\end{table}

NGC\,6388 and NGC\,6441, which are traditionally considered twin clusters, are remarkable examples \footnote{NGC\,6388 and NGC\,6441 share very similar masses, metallicities ($[Fe/H]\sim -0.50$) and they are both located in the Galactic bulge. The existence of bHB stars in these clusters was an unexpected feature given their relatively high metallicity (e.g.\,\citealt{rich1997a}) and was considered one of the earliest  signatures of stellar populations with extreme helium abundances in GCs (e.g.\,\citealt{dantona2008a, tailo2017a} and references therein).}.
 Indeed, NGC\,6388 shows a very extended 1G sequence in the two-color diagrams of Figure~\ref{fig:2col}, while 1G stars of the HB of NGC\,6441 span relatively-small $m_{\rm F275W}-m_{\rm F336W}$  and $m_{\rm F336W}-m_{\rm F438W}$ color intervals.
 
 \begin{figure*} 
 \begin{center}
  \includegraphics[height=13cm]{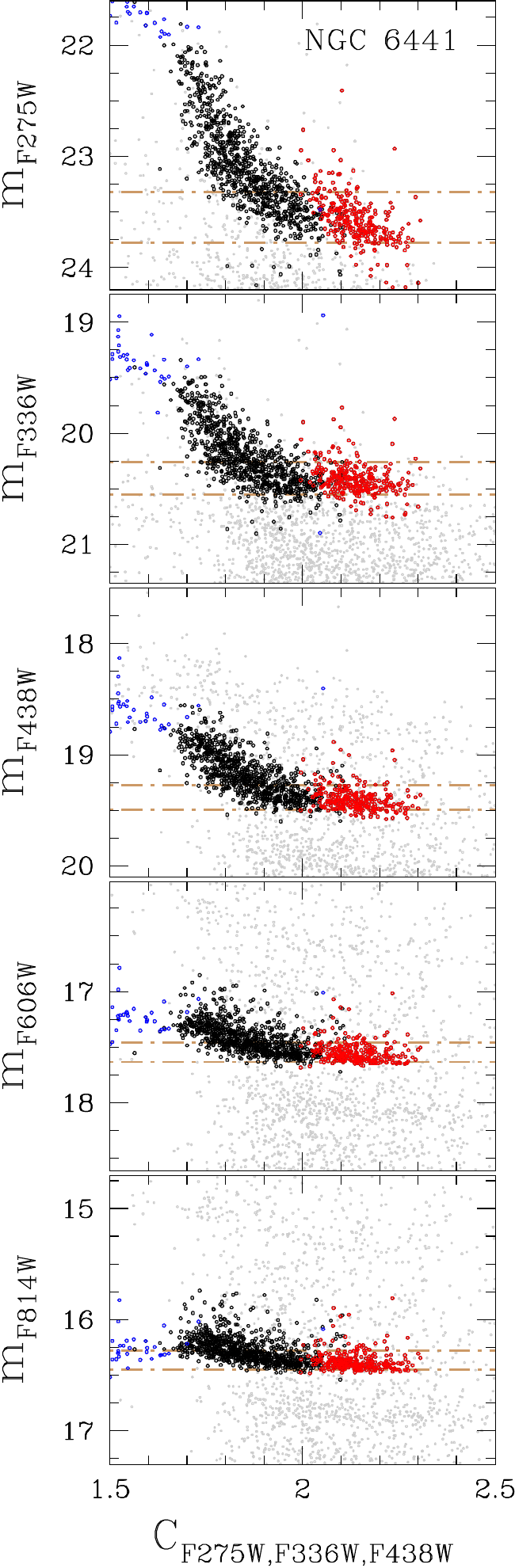}
  \includegraphics[height=13cm]{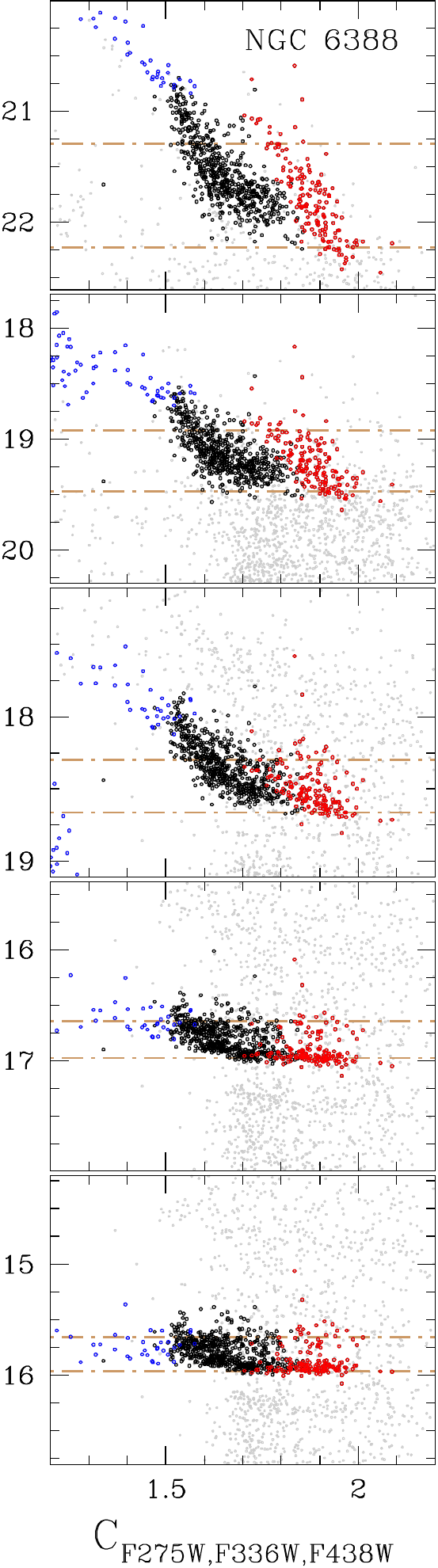}
  \raisebox{0.25\height}{\includegraphics[height=9cm]{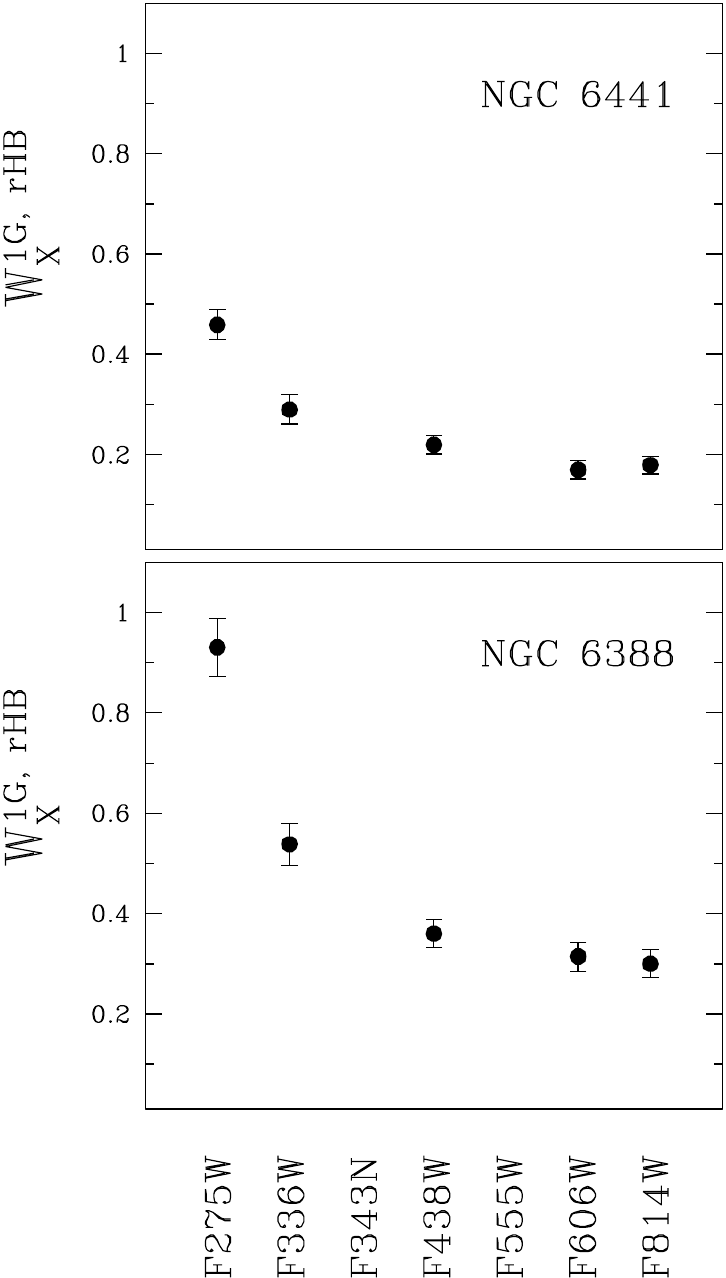}}
  \caption{Comparison between the red HBs of NGC~6441 (left panels) and NGC~6388 (middle panels) in the $m_{\rm X}$ vs.\,$C_{\rm F275W,F336W,F438W}$ planes, where X=F275W, F336W, F438W, F606W and F814W. 1G stars, 2G red-HB, and 2G blue-HB stars are colored red, black and blue, respectively, while the remaining clusters stars are represented with gray points. 
  The two brown horizontal dot-dashed lines mark the 10$^{\rm th}$ and 90$^{\rm th}$ percentile for the magnitude distribution of 1G stars. 
  Right panels show the magnitude extension of 1G stars,  $W_{\rm X}^{\rm 1G,rHB}$, for the available filters.\\}
 \label{fig:confr1} 
 \end{center} 
\end{figure*} 
 
To further compare the red HBs of NGC\,6388 and NGC\,6441, we plot in the left and middle panels of Figure~\ref{fig:confr1} the $C_{\rm F275W,F336W,F438W}$ vs.\,$m_{\rm X}$ diagram of their HB stars, where X=F275W, F336W, F438W, F606W  and F814W. In each panel, the two brown horizontal dot-dashed lines indicate the $10^{\rm th}$ (upper) and the $90^{\rm th}$ (lower) percentile of the magnitude distribution. The vertical distance observed between the two lines $W_{\rm X}^{\rm 1G,rHB}$, has been plotted in the right panels of the Figure as a function of the filter. From this figure we see that the red HB magnitude extension of NGC\,6388 is slightly larger than that of NGC\,6441 ($\sim$0.3 vs.\,$\sim$0.2 mag), but the difference suddenly diverges as we move to blue and UV wavelengths.
The most pronounced difference is in the F275W band, where the value of $W_{\rm F275W}^{\rm 1G,rHB}$ measured for NGC\,6388 is $\sim$0.45 mag larger than that of NGC\,6441.  

\begin{figure} 
 \begin{center}
    \includegraphics[height=9cm,clip]{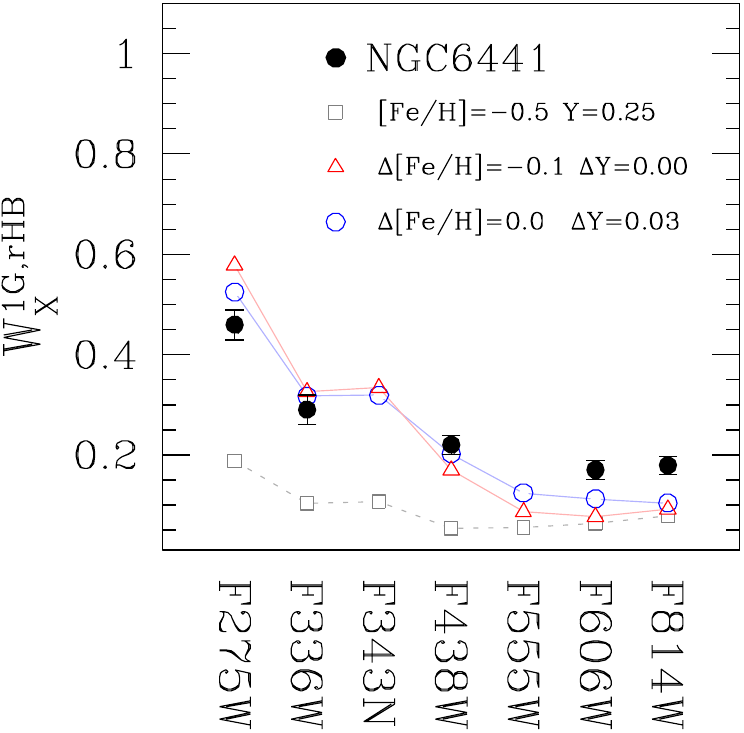}
  \caption{Comparison between the observed magnitude extension of 1G stars along the red HB of NGC\,6441 (black circles) and simulated HBs with different helium contents and metallicities. Gray squares correspond to a simple stellar population with [Fe/H]=$-$0.5 and pristine helium abundance (Y=0.25), red triangles represent a stellar system composed of two stellar populations with pristine helium content and [Fe/H]=$-$0.5 and [Fe/H]=$-$0.6, whereas blue circles correspond to a stellar system composed of two stellar populations with the same [Fe/H] and helium abundances Y=0.25 and Y=0.28.\\} 
 \label{fig:W1Gteo} 
 \end{center} 
\end{figure}

The 1G of NGC\,6441 is further investigated in Figure~\ref{fig:W1Gteo}, where we compare the observed values of $W_{\rm X}^{\rm 1G,rHB}$ with those derived from simulated HBs that account for observational errors and correspond to a simple population with pristine helium abundance, age of 12.0 Gyr and [Fe/H]=$-$0.5, suitable for NGC\,6441 \citep{harris1996a, dotter2010a}. 
Clearly, the observed red HB of 1G stars of NGC\,6441 spans wider magnitude ranges than the simulated HB (gray squares in Figure~\ref{fig:W1Gteo}), thus demonstrating that the 1G of this cluster is composed of stars with different chemical composition.

Recent work, based on MS and RGB stars, revealed that 1G stars of most studied clusters exhibit extended sequences in the ChM \citep{milone2015a, milone2017a}. The analysis of 1G stars through multi-band photometry shows that the extended sequence is consistent with either star-to-star variations of helium abundance \citep[e.g.\,][]{milone2015a, milone2018a} or with intrinsic metallicity spread \citep[e.g.][]{dantona2016a, tailo2019b}. The latter hypothesis is supported by direct spectroscopic measurements of iron abundances of NGC\,3201 \citep{marino2019b}.

To investigate the physical reasons that are responsible for the 1G extension of HB stars in NGC\,6441, we simulated the red HB of a stellar system composed of two stellar populations with pristine helium content and iron abundances [Fe/H]=$-$0.5 and $-$0.6, so with an iron variation of $\Delta$[Fe/H]=$-0.1$, and derived the corresponding values of $W_{\rm X}^{\rm 1G,rHB}$ (red triangles in Figure~\ref{fig:W1Gteo}). Moreover, we simulated the red HB of a stellar system hosting two stellar populations with [Fe/H]=$-$0.5 and different helium abundances of Y=0.25 and Y=0.28 (blue circles in Figure~\ref{fig:W1Gteo}).

The simulated HBs are derived from the stellar models by \citet[][2020 in preparation]{tailo2016}.
We determined the mass of the each HB star as $\rm M^{HB}=M^{Tip} - \Delta M(\mu,\delta)$, where $\rm M^{Tip}$ is the mass at the RGB tip derived from the best-fit isochrone and $\rm \Delta M$ is the mass lost by the star, during the RGB.
Specifically, we adopted a Gaussian profile for $\rm \Delta M$, with average mass loss of 0.25 $M_\odot$ and  mass loss dispersion of $0.006 M_\odot$, which is the average value inferred by \citep{tailo2020a} for the studied GCs.

Figure~\ref{fig:W1Gteo} reveals that the observed HB magnitude extensions of 1G stars in NGC\,6441 are consistent with two stellar populations with [Fe/H]=$-0.5$ that differ in helium mass fraction by $\Delta$Y$\sim$0.03. As an alternative, observations are well reproduced by stellar populations with the same helium content, Y=0.26, but different iron abundances at a level of $\sim$0.1~dex  ([Fe/H]=$-$0.5 and [Fe/H]=$-0.6$).  Hence, similarly to what observed on the ChM \citep{milone2017a}, from our dataset it is not possible to disentangle between internal helium and metallicity variations as responsible for the magnitude extension of 1G HB stars in NGC\,6441. Similar conclusion can be extended to the other studied GCs.
 
Since we verified that at the typical luminosities of the red-HB stars, both NGC\,6388 and NGC\,6441 show comparable magnitude errors in each band, we infer from the observation that in Figure~\ref{fig:confr1} the spread of the NGC\,6388 1G red-HB stars is larger than that of NGC\,6441 in all the five bands, that also the 1G population of NGC\,6388 is not consistent with being a simple stellar population.

\begin{figure*} 
\begin{center} 
  \includegraphics[height=10.cm,clip]{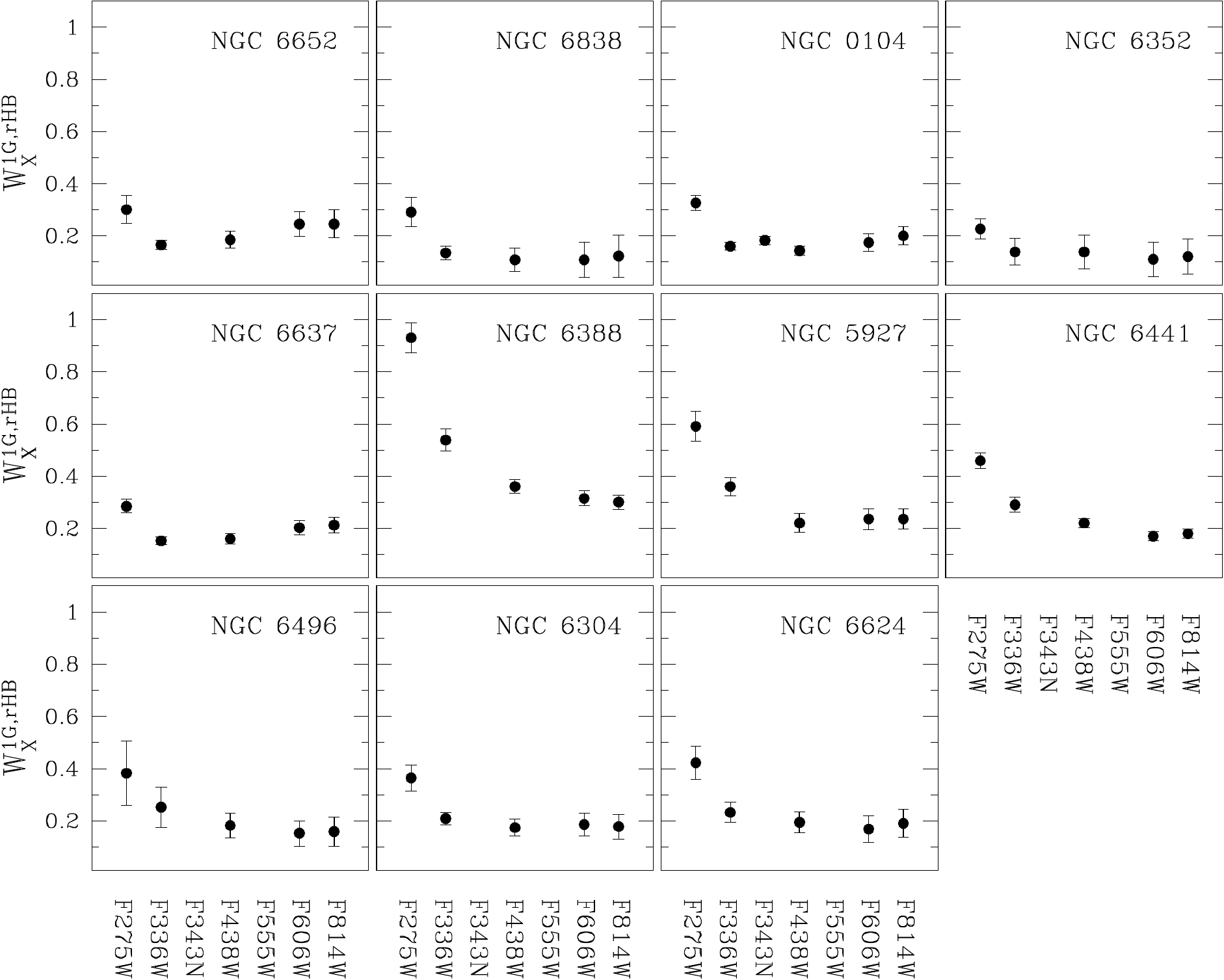}\\
  \includegraphics[height=4.cm,clip]{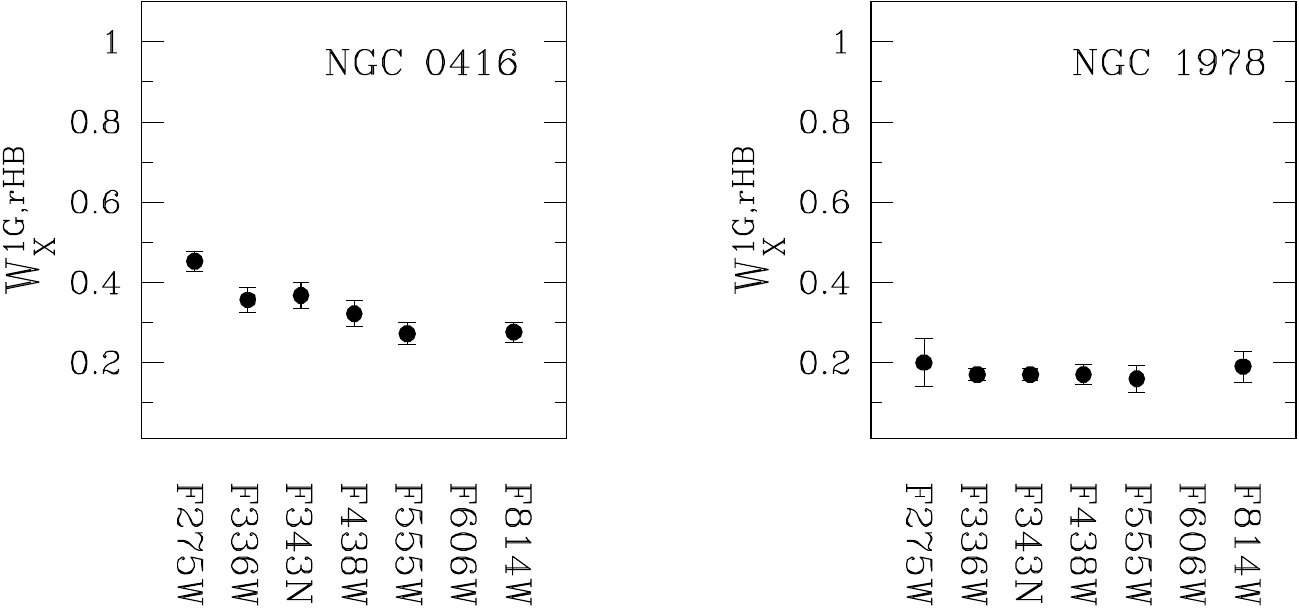}
  \caption{Width of 1G stars along the red HB in the X magnitude, $W_{\rm X}^{\rm 1G,rHB}$ for Galactic and extragalactic GCs, as a function of the various filters used in this work.}
 \label{fig:dmag}
\end{center}
\end{figure*}

Figure~\ref{fig:dmag} shows the magnitude extensions of 1G red HB and red clump stars in different filters for all studied GCs.
Similarly to what is observed in NGC\,6388 and NGC\,6441, the $W_{\rm X}^{\rm 1G,rHB}$ quantities values approach their maximum for X=F275W in all clusters but NGC\,1978, which exhibits nearly constant magnitude extension in all filters. The F336W extension is significantly wider than that measured in optical bands in  NGC\,416, NGC\,5927, NGC\,6388 and NGC\,6441,  whereas the other clusters share similar magnitude extension in F336W and optical bands. In NGC\,6637 and NGC\,6652 the F336W and F438W magnitude extensions are slightly narrower than those in F606W and F814W.

\begin{figure*} 
\begin{centering} 
\includegraphics[height=7cm,clip]{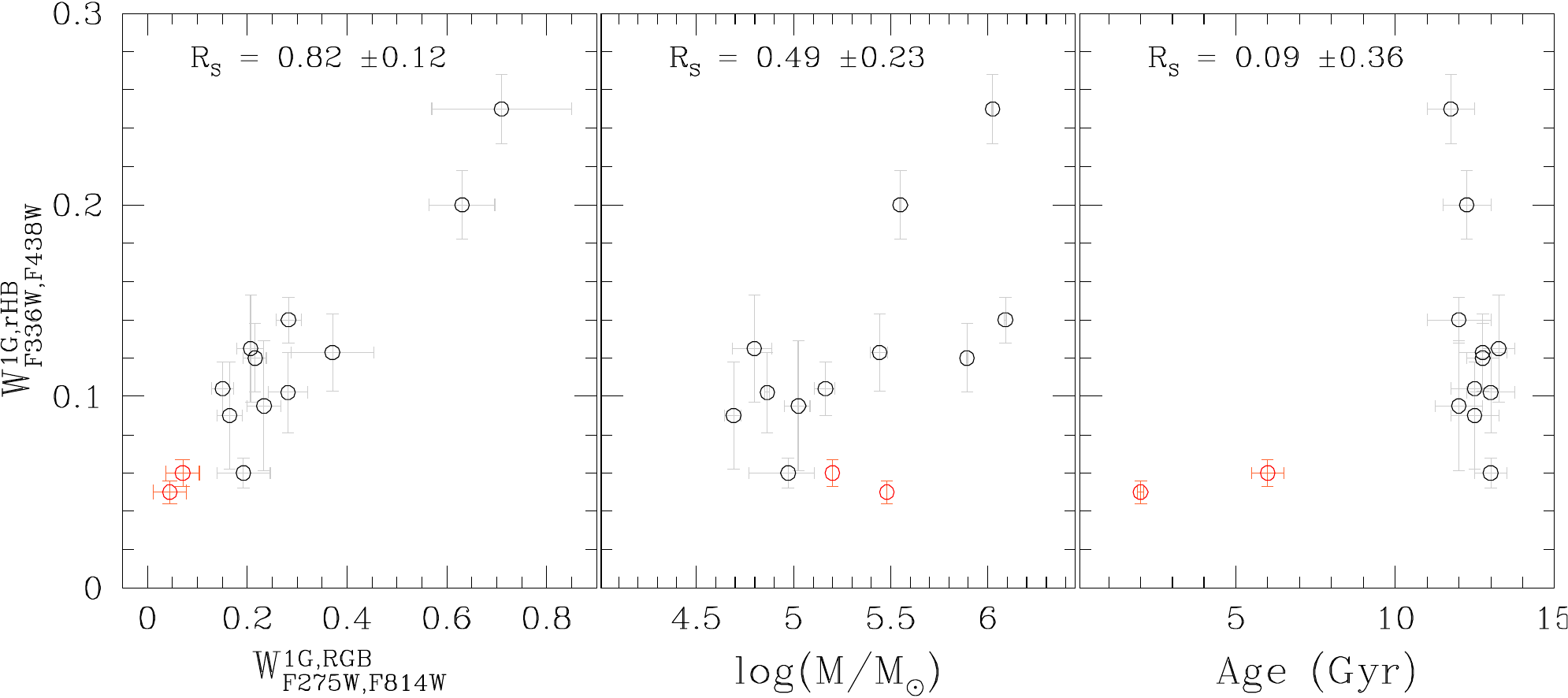}
  \caption{$m_{\rm F336W}-m_{\rm F438W}$ color extension of 1G stars along the HB against the width of 1G RGB stars along the ChM $W_{\rm F275W,F814W}^{\rm 1G, RGB}$ \citep[from][left panel]{milone2017a}, the mass of the host GC \citep[from][middle panel]{baumgardt2018a} and GC ages \citep[from][right panel]{dotter2010a}. The Spearman rank correlation coefficients are quoted on top of each panel.}
 \label{fig:cor}
 \end{centering} 
\end{figure*} 

\begin{figure*} 
\begin{centering} 
  \includegraphics[height=7cm,clip]{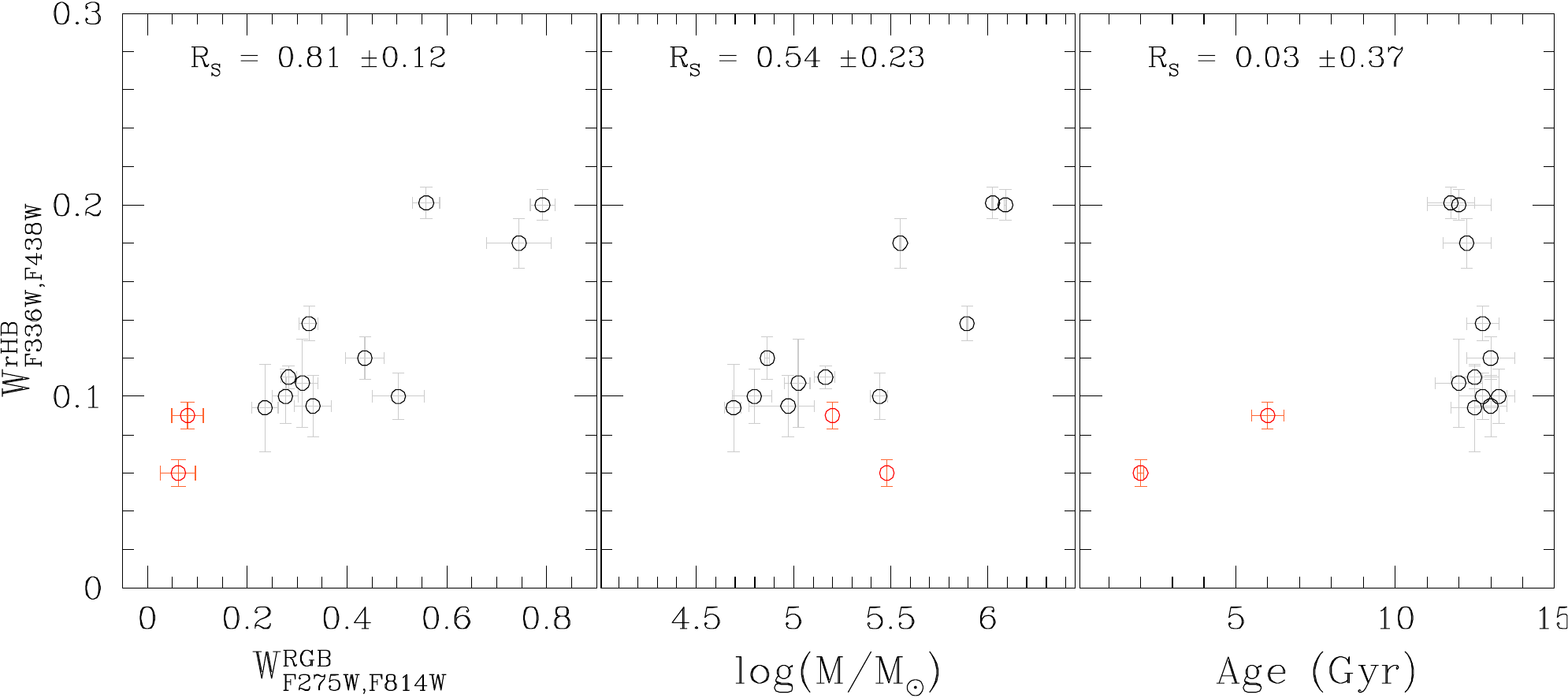}
  \caption{Same as Figure~\ref{fig:cor}, but for the $m_{\rm F336W}-m_{\rm F438W}$ color extension of stars along the whole red HB.}
 \label{fig:cor1}
\end{centering} 
\end{figure*} 
 
To investigate the relation between the color extension of the sequence formed by 1G stars along the red HB and the RGB, we exploit the red-HB width $W_{\rm F336W,F438W}^{\rm 1G, rHB}$, derived as the difference between the $90^{\rm th}$ and the $10^{\rm th}$ percentile of the red-HB 1G stars $m_{\rm F336W}-m_{\rm F438W}$ color distribution, listed in Table~\ref{tab:est}.
Figure~\ref{fig:cor} shows that $W_{\rm F336W,F438W}^{\rm 1G, rHB}$ correlates with the width of the ChM of 1G stars $W_{\rm F275W,F814W}^{\rm 1G, RGB}$, that is the difference between the $90^{\rm th}$ and the $10^{\rm th}$ percentile of the RGB 1G stars $\Delta_{\rm F275W,F814W}$ distribution \citep[from][]{milone2017a}, and with the present-day mass of the host GC \citep[from][]{baumgardt2018a}. There is no evidence between the color extension of 1G HB stars the age of the host GCs \citep[from][]{dotter2010a}.

For completeness, we extended the analysis to the whole HB and derived the corresponding color widths $W_{\rm F336W,F438W}^{\rm rHB}$ that are listed in Table~\ref{tab:est}. 
As shown in Figure~\ref{fig:cor1}, $W_{\rm F336W,F438W}^{\rm rHB}$ correlates with both the RGB width and the cluster mass. 
No significant correlations has been found with age, in close analogy with what is observed for the color extension of 1G HB stars.


\section{Radial distribution of multiple populations}\label{sec:RD}

To investigate the radial distributions of multiple populations in 47\,Tuc, NGC\,5927, NGC\,6366 and NGC\,6838, we combined information from the {\it HST} observations, that cover a central region of $\sim$2.7$\times$2.7 arcmin, with ground-based photometry catalogs by \citet{stetson2019a}, which are extended over a wider field of view.

Indeed, previous work has shown that the photometric diagrams including $U$ and $B$ photometry are efficient tools to identify 1G and 2G stars from ground-based photometry \citep[e.g.][]{marino2008a, milone2012a, monelli2013a}. 

\begin{figure*} 
\begin{center} 
  \includegraphics[height=10.cm, trim={3cm 5cm 3.cm 4cm}]{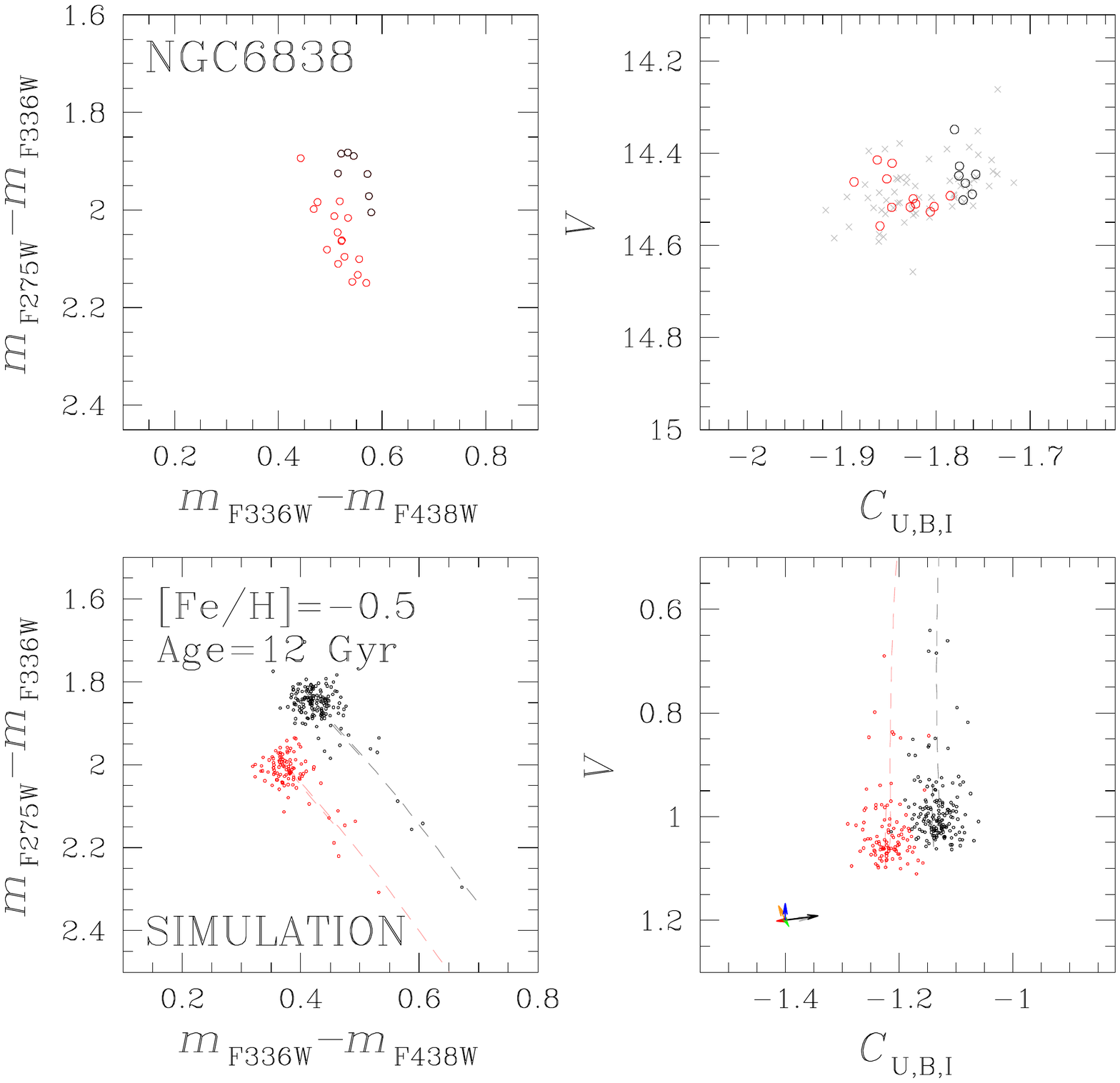}
  \caption{Comparison between $m_{\rm F275W}-m_{\rm F336W}$ vs.\,$m_{\rm F336W}-m_{\rm F438W}$ (left) and $V$ vs.\,$C_{\rm U,B,I}$ (right) diagrams for red-HB stars. 
  Upper panels show the observed diagrams of NGC\,6838 from {\it HST} and ground-based photometry. 1G and 2G stars, selected from the left-panel two-color diagram are plotted red and black, respectively, in both panels. 
 Lower panels illustrate results for simulated diagrams. The vectors plotted in the lower-right panel are defined as in Figure~\ref{fig:simu12}.}
 \label{fig:2CvsCUBI}
 \end{center}
\end{figure*}

To further demonstrate that the sequences of red-HB stars observed in the $m_{\rm F275W}-m_{\rm F336W}$ vs.\,$m_{\rm F336W}-m_{\rm F438W}$ and $V$ vs.\,$C_{\rm U,B,I}$ diagrams correspond to the 1G and 2G, we exploit both observations and simulated photometry. The upper panels of Figure~\ref{fig:2CvsCUBI} compare these two diagrams for NGC\,6838 and use red and black colors to represent 1G and 2G stars, respectively, with both ground-based and {\it HST} photometry. Clearly, the groups of 1G and 2G stars selected from the two-color diagram populate distinct sequences in the $V$ vs.\,$C_{\rm U,B,I}$ pseudo CMD and similar conclusion can be derived for the other analyzed GCs with available {\it HST} and ground-based photometry. 

This result is corroborated by the  simulations of 1G and 2G stars introduced in Section~\ref{sec:teoria} for 12 Gyr old stellar populations with [Fe/H]=$-$0.5.
As shown in the bottom panels of Figure~\ref{fig:2CvsCUBI} simulated 1G and 2G stars populate distinct regions of $m_{\rm F275W}-m_{\rm F336W}$ vs.\,$m_{\rm F336W}-m_{\rm F438W}$ and $V$ vs.\,$C_{\rm U,B,I}$ planes.

\begin{figure*} 
\begin{centering} 
  \includegraphics[height=8.3cm,trim={0.5cm 5cm 8.cm 12cm}]{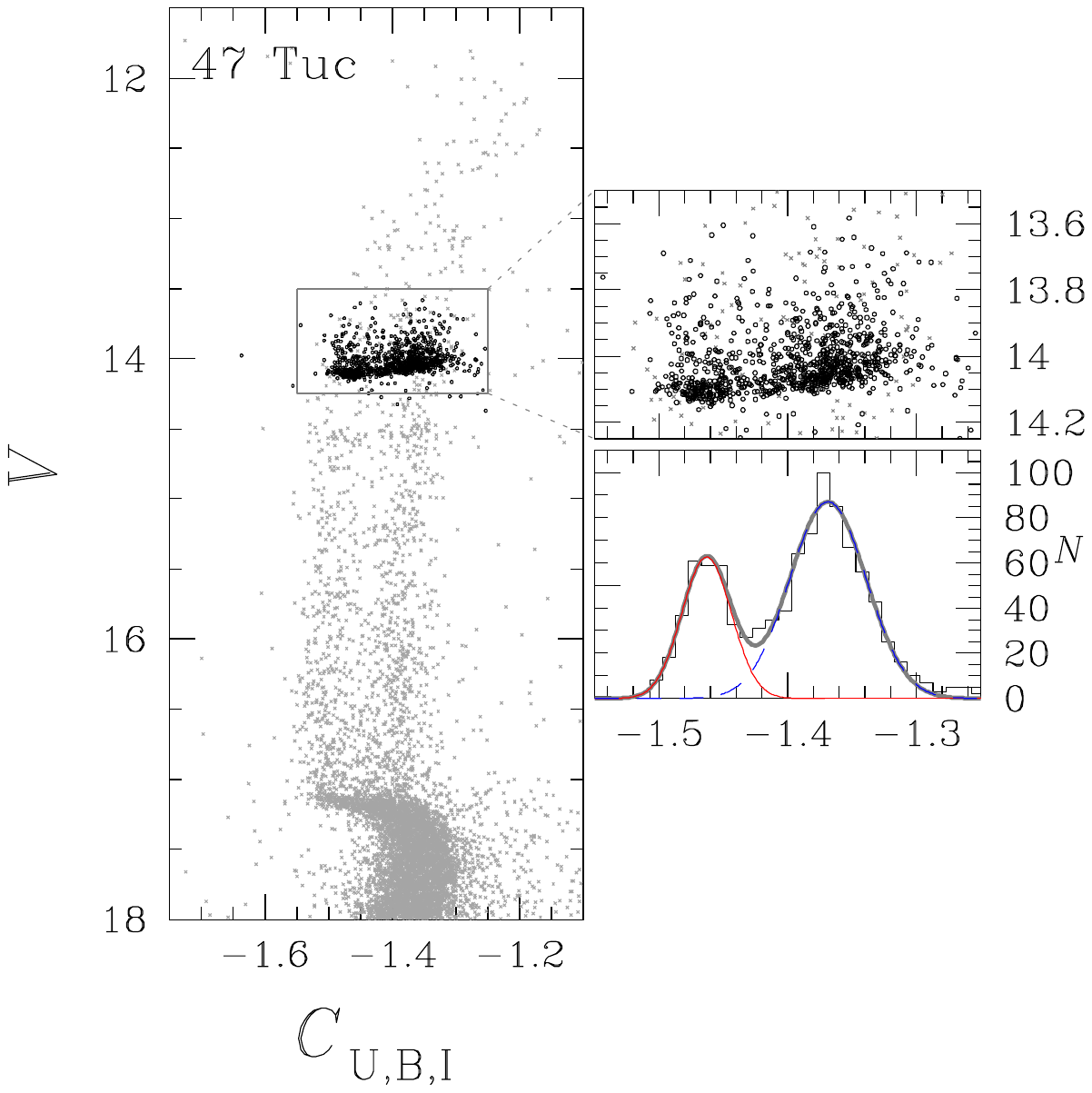}
  \includegraphics[height=8.3cm,trim={0.5cm 5cm 8.cm 12cm}]{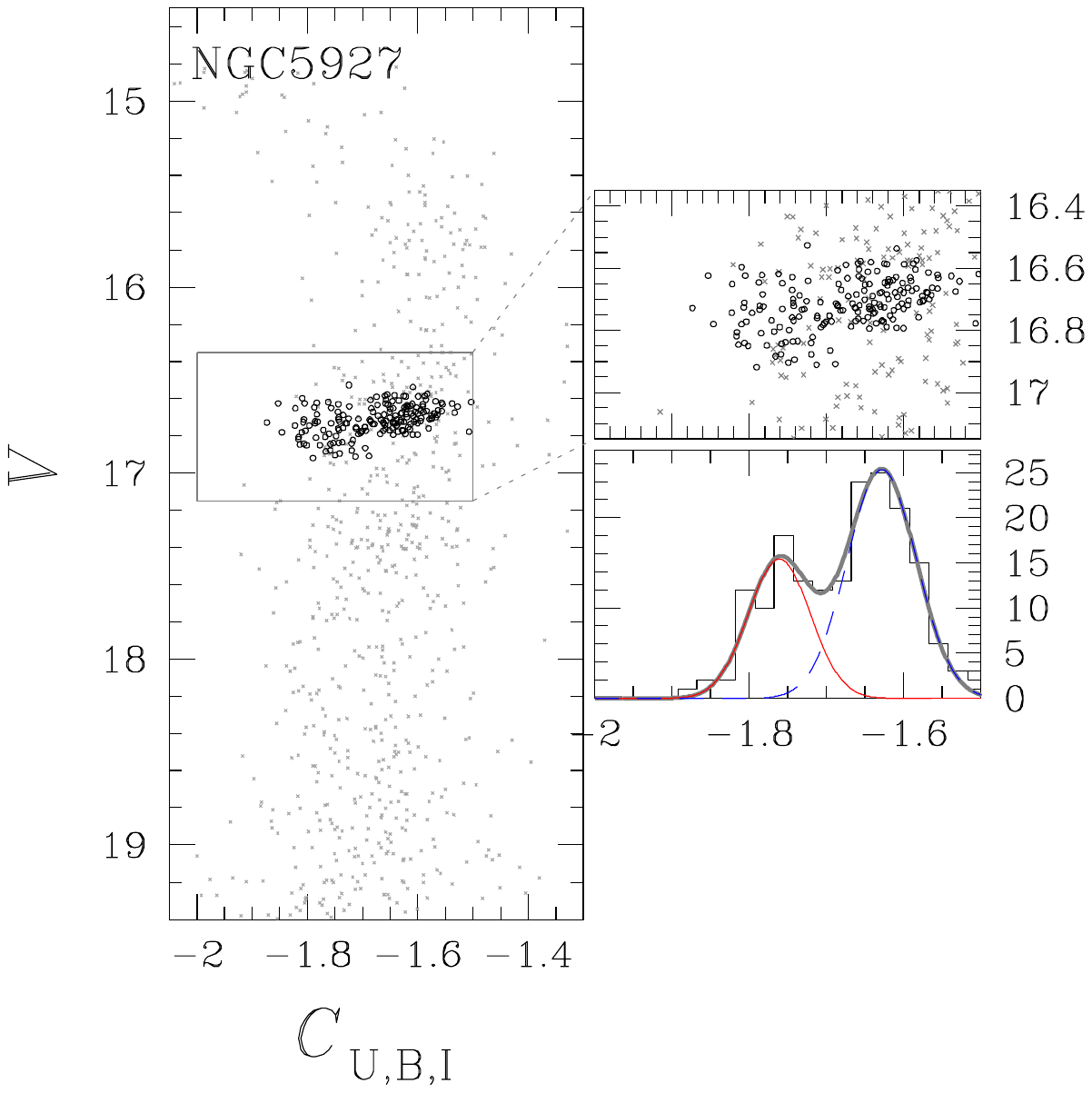}
  \includegraphics[height=8.3cm,trim={0.5cm 5cm 8.cm 12cm}]{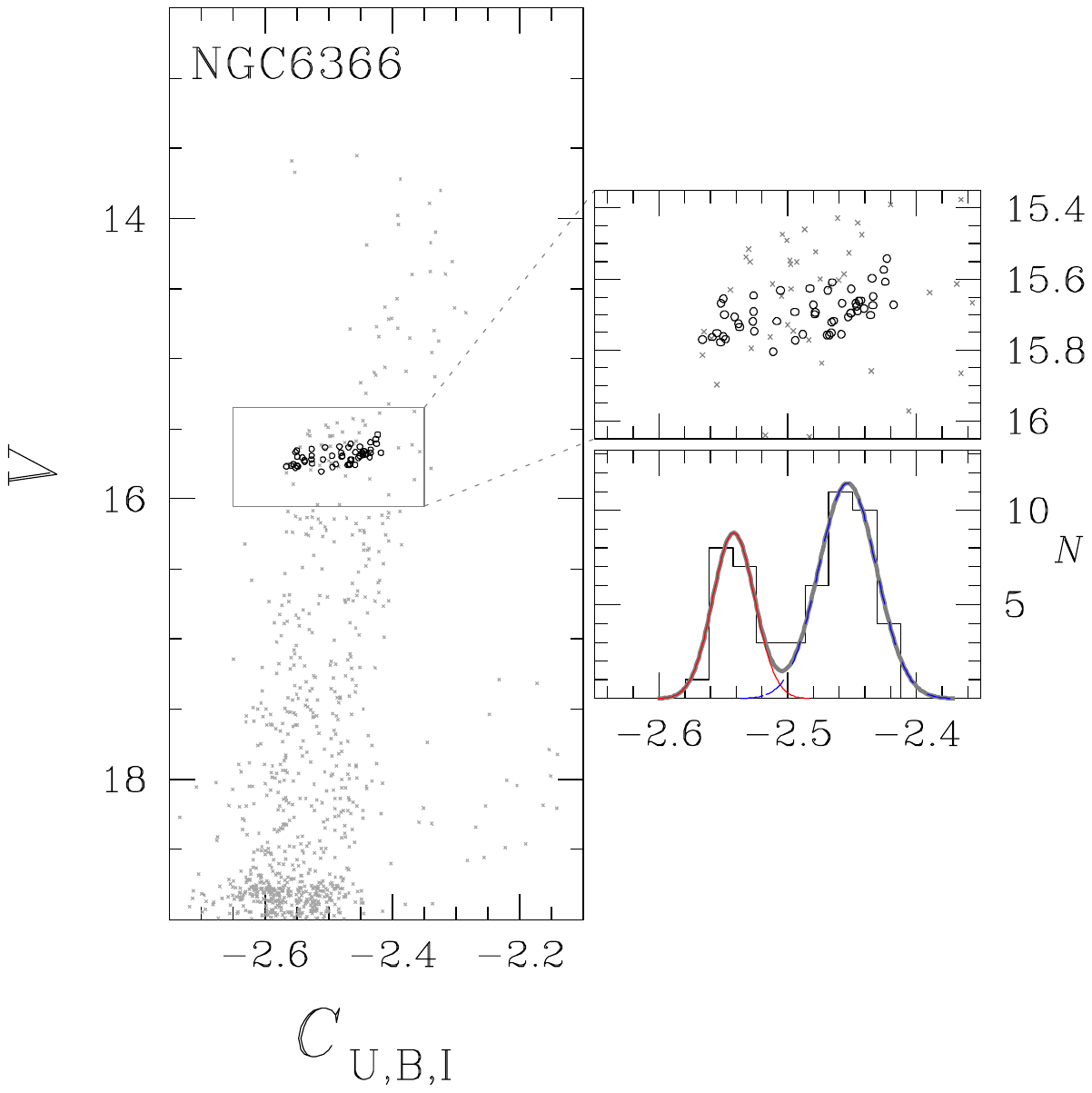}
  \includegraphics[height=8.3cm,trim={0.5cm 5cm 8.cm 12cm}]{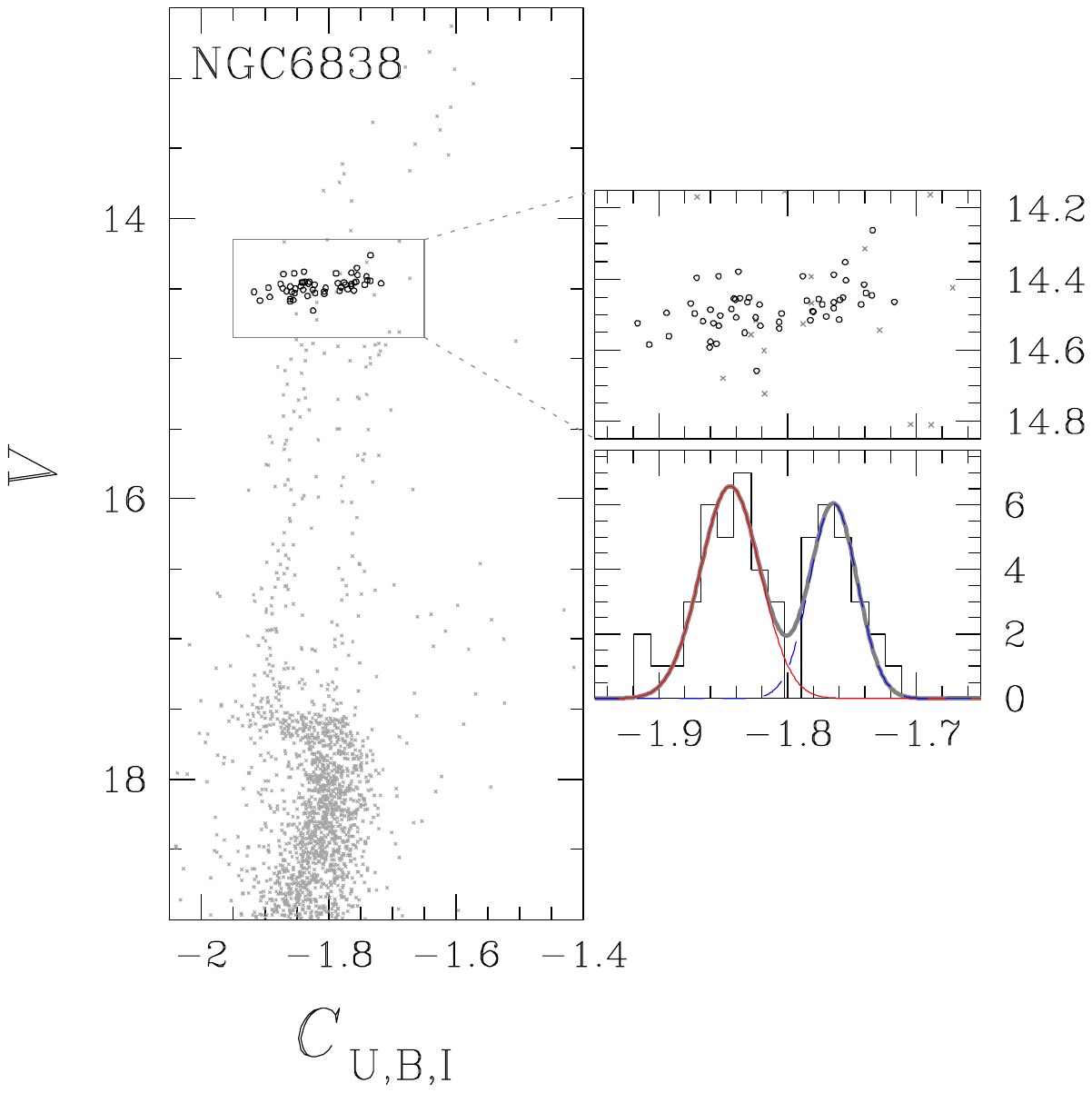}
  \caption{$V$ vs.\,$C_{\rm U,B,I}$ pseudo two-color diagrams of selected cluster members of 47\,Tuc, NGC\,5927, NGC\,6366 and NGC\,6838 from ground-based photometry \citep{stetson2019a}. Red-HB stars are marked with black dots, while the remaining stars are plotted with gray dots. A zoom of the CMD region around the HB is provided in the small panels on the right together with the histogram distributions of $C_{\rm U,B,I}$ for red-HB stars. The red and blue curves superimposed on the histogram represent the Gaussian functions that provide the best fit of the two peaks.\\}
 \label{fig:GB}
\end{centering}
\end{figure*}

The procedure to estimate the fraction of 1G and 2G stars along the red HB from ground-based photometry is summarized in the upper-left panels of Figure~\ref{fig:GB} for 47\,Tuc.
The $V$ vs.\,$C_{\rm U,B,I}$ diagram of this cluster exhibits split  HB as noticed in previous work \citep[][]{monelli2013a, milone2018b, cordoni2020a}. The $C_{\rm U,B,I}$ histogram distribution of HB stars is fitted by a bi-Gaussian function by means of least squares. The Gaussian components corresponding to the 1G and the 2G are colored red and blue, respectively, and the relative numbers of 1G and 2G stars are derived by comparing the area below the red and blue Gaussian.

We find that the fraction of 2G stars of 47\,Tuc derived from ground-based photometry between 1.5 and 24.0 arcmin is 0.67$\pm$0.02, and it is significantly smaller than that observed within $\sim$1 arcmin from the cluster center 0.78$\pm$0.03. 
Similarly, the fraction of 2G stars of NGC\,5927 in the region with radial distance between $\sim$0.5 and 6.0 arcmin from the cluster center, 0.58$\pm$0.03, is smaller than that derived from {\it HST} photometry (0.63$\pm$0.03). These results are consistent with clusters where the 2G is more centrally concentrated than the 1G.
On the contrary, the fraction of 2G stars of NGC\,6366 and NGC\,6838, measured within 8.0 and 9.0 arcmin from the center, are consistent with the corresponding values inferred from {\it HST} photometry in the central field. 

To further investigate the dependence of the relative numbers of 1G and 2G stars from the radial distance, we divided the field of view of ground-based photometry into different radial bins, each containing almost the same number of HB stars. We estimated the fractions of 1G and 2G stars in each radial bin by using the procedure described in Figure~\ref{fig:GB}. 

\begin{figure*} 
\begin{center} 
  \includegraphics[height=10.cm,trim={0.5cm 6.5cm 5.cm 5cm}]{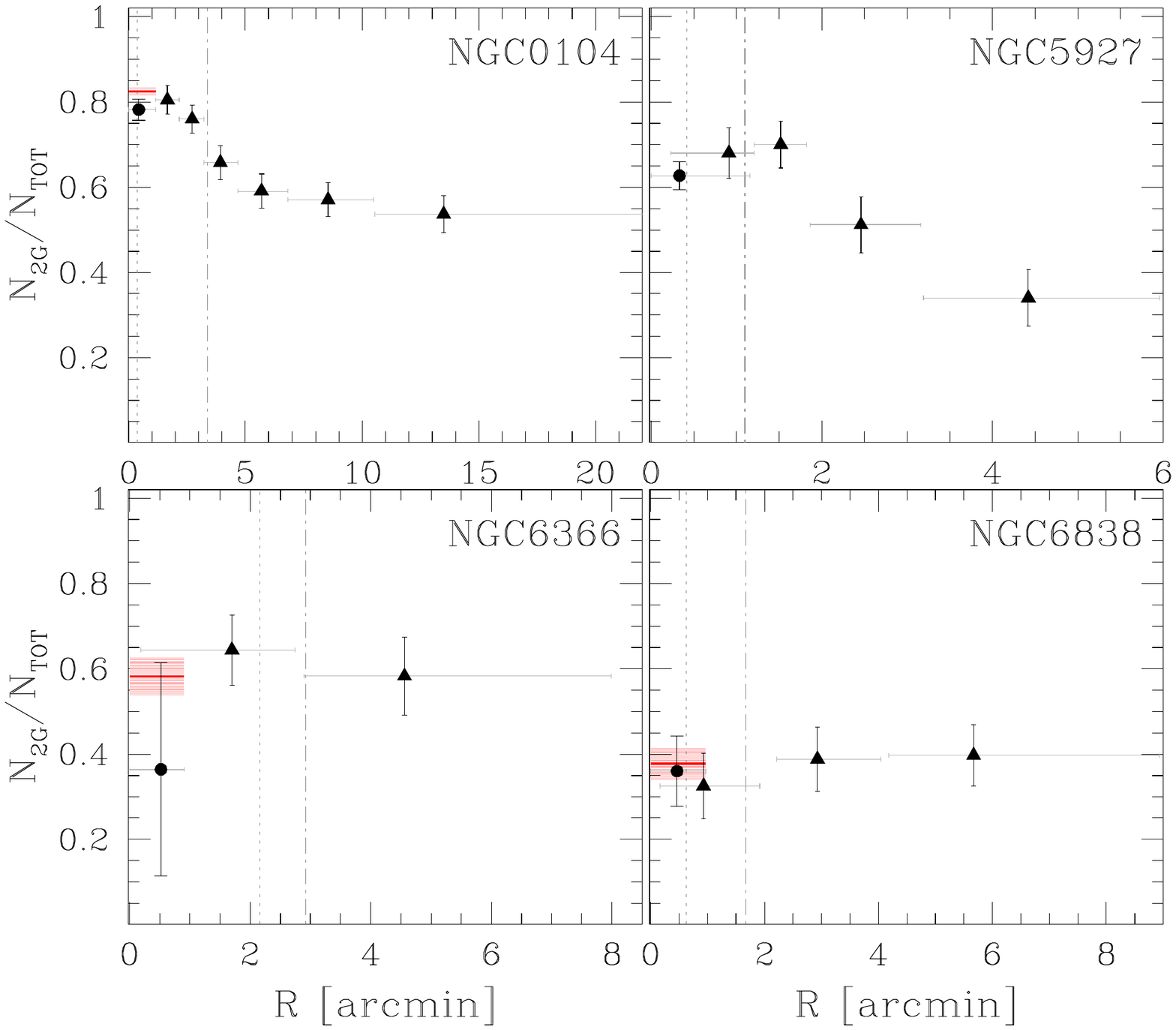}
  \caption{Fraction of 2G stars as a function of radial distance for 47\,Tuc, NGC\,5927, NGC\,6366 and NGC\,6838. Black circles and triangles mark the results derived from {\it HST} and ground-based photometry, respectively. Gray horizontal lines mark the extension of each radial intervals, while the red segments indicate results from \citet{milone2017a} and \citet{milone2020a} based on RGB stars. 
  The vertical dotted and dashed-dotted lines indicate the core and the half-light radius.}
 \label{fig:rad}
 \end{center}
\end{figure*}

Results are illustrated in Figure~\ref{fig:rad}, where we plot the fraction of 2G stars as a function of the radial distance from the cluster center. The dashed and dot-dashed gray lines correspond respectively to the values of the core radius and the half light radius (from \citealt{harris1996a} 2010 edition for Galactic GCs, from \citealt{mclaughlin2005a} for NGC\,416 and from \citealt{fischer1992} for NGC\,1978).
In 47\,Tuc the fraction of 2G stars is maximum near the cluster center and is consistent with a flat distribution within $\sim$0.7 half-light radii. 
 The fraction of 2G stars drops from $\sim$0.8 to $\sim$0.65 at about 1.0 half light radius and then it slightly decreases to $\sim$0.55 at larger radial distances.
 Similarly, the fraction of 2G stars of NGC\,5927 is close to $\sim$0.7 within $\sim$1.5 half-light radii and decreases to $\sim$0.4 in the region around three half-light radii.
 
 On the contrary, in NGC\,6366 and NGC\,6838 there is no evidence for radial gradient. Noticeably, the fractions of 2G stars  in the internal regions inferred in this paper from the HB are in agreement with those derived from RGB and MS stars by \citep{milone2017a} and \citep{milone2020b}.
 
The radial distributions of multiple populations in GCs with no available ground-based photometry have been derived by using {\it HST} photometry alone. Hence, the investigation is limited to the innermost $\sim$1 arcmin from the cluster center. 
We divided the {\it HST} field of view of each cluster in various radial intervals comprising similar numbers of HB stars and derived the fraction of 1G and 2G by following the procedure of Section~\ref{sec:pratio}. 

\begin{figure*} 
\begin{center} 
  \includegraphics[height=11.cm,clip]{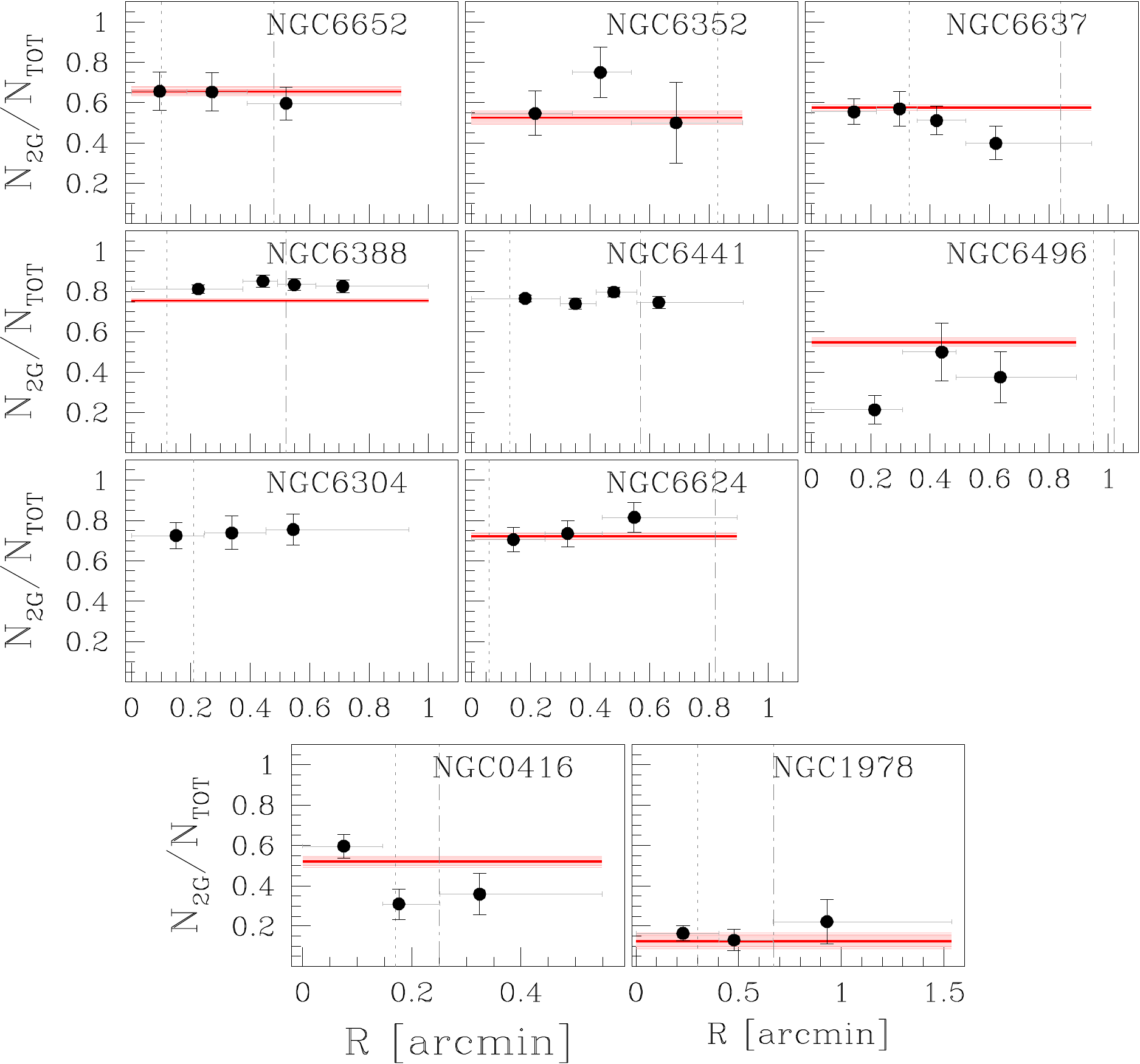} 
  \caption{Same as Figure~\ref{fig:rad}, but for the Galactic GCs NGC\,6304, NGC\,6352, NGC\,6388, NGC\,6441, NGC\,6496, NGC\,6624, NGC\,6652 and for the Magellanic-Cloud clusters NGC\,416 and NGC\,1978.}
 \label{fig:rade}
 \end{center}
\end{figure*} 
 

We find that the majority of Galactic clusters are consistent with a flat radial distribution of 2G stars in the innermost arcmin.
NGC\,416 is a possible exception. Indeed its 2G seems to be more centrally-concentrated than the 1G.

The statistical significance of the observed 1G and 2G radial distributions has been estimated by running 10,000 simulations with a flat radial distribution of stars, under the null hypothesis that the observed distribution profiles are produced by statistical fluctuations. For each simulation, the radial distribution has been obtained starting from the the observed $N_{\rm 2G}/N_{\rm TOT}$ weighted-average ratios across the covered radial distance and then adding up a random radial scatter based on the observed errors.
By using a chi-square test, we measured the deviation from flatness, represented by the quantity $\chi ^{2}_{\rm sim}$. We finally compared this value with $\chi ^{2}_{\rm obs}$ and determined the number of times for which $\chi^{2}_{\rm sim}>\chi^{2}_{\rm obs}$. This number divided by the total number of simulations, gives an estimate of the p-value, which is the probability that the chi-square is equal or higher than the one measured. The p-value corresponding to $X^2_{\rm obs}$ value of each cluster is reported in Table~\ref{tab:prob}. We see that $p\leq0.05$ only for three out of 14 GCs (47\,Tuc, NGC\,5927 and NGC\,416), which means in turn that the observed scatter can be truly associated to a different population radial distribution.



Our results corroborate previous evidence that the 2G of 47\,Tuc is more centrally concentrated than the 1G \citep[e.g.][]{milone2012a, cordero2014a}.

\citet{dalessandro2018a} investigated the radial distribution of multiple populations in 20 GCs and quantified the radial difference between 1G and 2G stars by using the area enclosed between their cumulative radial distributions, $A^{+}$, within two half-light radii from the center. The sample of GCs studied by \citet{dalessandro2018a} comprises three GCs, namely NGC\,1978, NGC\,6624 and NGC\,6637, also studied in this paper.

Although it is not possible to quantitatively compare our results with those by Dalessandro and collaborators, due to the different methods adopted, we notice that NGC\,6624 and NGC\,6637 exhibit values of $A^{+}$ that are close to zero, thus indicating that 1G and 2G stars exhibit similar radial distributions as found in this paper. On the contrary, results on NGC\,1978 from \citet{dalessandro2018a} are in disagreement with our conclusion of mixed 1G and 2G stars in this cluster. Indeed, the large and negative value of $A^{+}=-0.081$, indicates that its 2G is significantly more centrally concentrated than the 1G. 

The comparison between our results on NGC\,6441 and those by \citet{bellini2013a} is even more puzzling. Based on $m_{\rm F390W}$ vs.\,$m_{\rm F390W}-m_{\rm F606W}$ CMD,   Bellini and collaborators identified split MS and RGB in NGC\,6441. Both the blue MS and the blue RGB are more centrally concentrated than their red counterparts. Specifically, the fraction of blue-MS stars ranges from $\sim$0.4 for radial distance from the cluster center, R$\sim 0.9$ arcmin, to $\sim0.35$ at R$\sim$2.5 arcmin. The fraction of blue-RGB stars varies from $\sim$0.6, near the cluster center, to $\sim$0.5 at R$\sim$2.5 arcmin. Clearly, the fractions of blue-RGB and blue-MS stars are smaller then the fraction of 2G stars derived in this paper, thus indicating that the blue sequences identified by Bellini and collaborators encloses only part of the 2G of NGC\,6441.
This fact, together with the larger radial interval covered by the observations analyzed by \citet{bellini2013a}, are likely the reasons for the different conclusions from these two papers. 

\begin{table}
\centering
\caption{Probability that the observed radial distribution of $N_{\rm 2G}/N_{\rm TOT}$ is produced by a flat distribution.}
\begin{tabular}{cc}
\\
\hline\hline
     \\
    CLUSTER & p-value \\
    \\
    \hline
    \\
     NGC\,0104 & $<$ 0.01 \\
     NGC\,5927 & $<$ 0.01 \\
     NGC\,6304 & 0.99 \\
     NGC\,6352 & 0.60 \\
     NGC\,6366 & 0.75 \\
     NGC\,6388 & 0.87 \\
     NGC\,6441 & 0.59 \\
     NGC\,6496 & 0.29 \\
     NGC\,6624 & 0.71 \\
     NGC\,6637 & 0.60 \\
     NGC\,6652 & 0.96 \\
     NGC\,6838 & 0.97 \\
     NGC\,1978 & 0.89 \\
     NGC\,0416 & 0.02 \\
\hline\hline
\end{tabular}\\

\label{tab:prob}
\end{table}

\section{Summary and conclusions}\label{sec:conclusions}

In this work we exploited the red HB to investigate the phenomenon of multiple stellar populations in 14 GCs, based on the distribution of red HB stars in UV-optical two-colour diagrams. This allowed us to identify and characterize, for the first time, the MPs along the red HB in a large sample of $\sim 13$-Gyr old Galactic GCs and in the  extragalactic GCs NGC\,416 and NGC\,1978, which have ages of $\sim$2 and $\sim$6 Gyr, respectively. 

To do this, we exploited multi-band photometry obtained from images collected with the ACS/WFC and WFC3/UVIS cameras on board {\it HST}. In particular, we derived high-precision astrometry and  multi-band photometry of stars in NGC\,416 and NGC\,1978. 
The main results can be summarized as follows:
\begin{itemize}
    \item We identified distinct sequences of 1G and 2G stars along the red HB of twelve Milky-Way GCs, NGC\,416 in the SMC and NGC\,1978 in the LMC.
    These results confirm that MPs are a common feature of both Galactic and extragalactic GCs. 
    \item  MPs along the red HB exhibit high degree of variety, with the extension of the 1G and 2G sequences, the number of sub-populations and the relative numbers of stars in each population, changing from one cluster to another. 
    \item 
    We measured the fraction of 1G stars of fourteen GCs. This is the first time that this quantity is inferred in a large sample of clusters, homogeneously analyzed by using the red HB and the red clump. The fraction of 1G stars in Milky Way clusters ranges from $\sim$18\% in the massive GC NGC\,6388 ($\sim$1.1$\times$10$^{6}~$M$_{\odot}$) to $\sim$68\% in the low-mass cluster NGC\,6838 ($\sim$4.9$\times$10$^{4}~$M$_{\odot}$).
    Noticeably, by using HB stars, it was possible for the first time to measure the population ratios in NGC\,5927, NGC\,6304 and NGC\,6441.
    We combined our results based on HB stars with previous findings based on MS and RGB stars thus deriving improved estimates of the fractions of 1G and 2G stars in GCs.
    \item 
    The 1G fractions derived from red HB stars correlate with the present day and the initial mass of the host cluster, with massive GCs having larger fraction of 2G stars. The conclusion is confirmed also when we extend the number of clusters by including literature determination of the fraction of 1G star inferred from the RGB and the MS.
    
    Similarly, the $m_{\rm F336W} - m_{\rm F438W}$ color extension of red HB correlates with cluster mass.
    These facts confirm that the incidence and complexity of the MPs phenomenon depend on GC mass \citep{milone2020a}.
    There is no correlation between the fraction of 1G stars and the age of the host GC.

        \item We combined results from {\it HST} photometry of stars in the innermost $\sim$2.7$\times$2.7 arcmin region and from ground-based wide-field photometry from \citet{stetson2019a} to investigate the radial distributions of 1G and 2G stars identified along the red HBs of NGC\,104, NGC\,5927, NGC\,6366 and NGC\,6838. 
    We find that 2G stars of NGC\,5927 and NGC\,104 are more centrally concentrated than the 1G, whereas the two stellar populations of NGC\,6366 and NGC\,6838 share similar radial distributions.
    1G and 2G stars of the remaining clusters are consistent with the same radial distribution within the {\it HST} field of view, except for NGC\,416, which is consistent with a more centrally concentrated 2G.
    \item We discovered that GCs typically exhibit extended sequences of 1G stars along the red HB.
        NGC\,6388 is the most extreme case and shows a more extended 1G sequence in the $m_{\rm F336W}-m_{\rm F438W}$ vs $m_{\rm F275W}-m_{\rm F336W}$ two color diagram, when compared with NGC\,6441, which is considered its `twin' cluster. By comparing the observed magnitude width of 1G stars along the red HB with simulated HBs with appropriate chemical composition, we find that the extension of the HB sequence of 1G stars is consistent with an internal spread in either helium or metallicity. 
        
        Furthermore, the color extensions of the 1G sequence along the red HB and the RGB 
         correlate with each other, thus suggesting that the extended 1G sequences in the ChM of RGB stars and in the two-color diagrams of red HB stars are associated with the same physical process. 
         
        \item The fraction of 1G stars in the extragalactic clusters NGC\,416 and NGC\,1978 are $\sim$0.55 and $\sim$0.85, respectively, and are larger than those of Galactic GCs with similar masses, which are typically dominated by the second generation. This finding corroborates previous evidence that the environment may affect the MPs phenomenon.

These results are consistent with a scenario in which the GCs are dominated by the 1G at formation and most 1G stars are stripped away from the cluster due to interaction with the host galaxy. Indeed, in this scenario, we expect that the studied LMC and SMC clusters, which are younger than Milky-Way GCs, are still dominated by the 1G.
Moreover, due to their relatively small masses, the Magellanic Clouds would be less efficient than the Milky Way in stripping 1G stars from their GCs.

    \end{itemize}

\section*{acknowledgments} 
\small
We thank H. Baumgardt and M. Hilker for providing initial masses of GCs.
This work has received funding from the European Research Council (ERC) under the European Union's Horizon 2020 research innovation programme (Grant Agreement ERC-StG 2016, No 716082 'GALFOR', PI: Milone, http://progetti.dfa.unipd.it/GALFOR), and the European Union's Horizon 2020 research and innovation programme under the Marie Sklodowska-Curie (Grant Agreement No 797100). ED, APM and MT  have been supported by MIUR under PRIN program 2017Z2HSMF (PI: Bedin). APM acknowledges support from MIUR through the FARE project R164RM93XW SEMPLICE (PI: Milone).

\bibliography{ms}{}

\begin{thebibliography}{}
\expandafter\ifx\csname natexlab\endcsname\relax\def\natexlab#1{#1}\fi
\providecommand{\url}[1]{\href{#1}{#1}}
\providecommand{\dodoi}[1]{doi:~\href{http://doi.org/#1}{\nolinkurl{#1}}}
\providecommand{\doeprint}[1]{\href{http://ascl.net/#1}{\nolinkurl{http://ascl.net/#1}}}
\providecommand{\doarXiv}[1]{\href{https://arxiv.org/abs/#1}{\nolinkurl{https://arxiv.org/abs/#1}}}

\bibitem[{{Af{\textcommabelow s}ar} {et~al.}(2018){Af{\textcommabelow s}ar},
  {Sneden}, {Wood}, {Lawler}, {Bozkurt}, {B{\"o}cek Topcu}, {Mace}, {Kim}, \&
  {Jaffe}}]{afsar2018a}
{Af{\textcommabelow s}ar}, M., {Sneden}, C., {Wood}, M.~P., {et~al.} 2018,
  \apj, 865, 44, \dodoi{10.3847/1538-4357/aada0c}

\bibitem[{{Anderson} {et~al.}(2008){Anderson}, {Sarajedini}, {Bedin}, {King},
  {Piotto}, {Reid}, {Siegel}, {Majewski}, {Paust}, \&
  {Aparicio}}]{anderson2008a}
{Anderson}, J., {Sarajedini}, A., {Bedin}, L.~R., {et~al.} 2008, \aj, 135,
  2055, \dodoi{10.1088/0004-6256/135/6/2055}

\bibitem[{{Baumgardt} \& {Hilker}(2018)}]{baumgardt2018a}
{Baumgardt}, H., \& {Hilker}, M. 2018, \mnras, 478, 1520,
  \dodoi{10.1093/mnras/sty1057}

\bibitem[{{Baumgardt} {et~al.}(2019){Baumgardt}, {Hilker}, {Sollima}, \&
  {Bellini}}]{baumgardt2019a}
{Baumgardt}, H., {Hilker}, M., {Sollima}, A., \& {Bellini}, A. 2019, \mnras,
  482, 5138, \dodoi{10.1093/mnras/sty2997}

\bibitem[{{Bedin} {et~al.}(2005){Bedin}, {Cassisi}, {Castelli}, {Piotto},
  {Anderson}, {Salaris}, {Momany}, \& {Pietrinferni}}]{bedin2005a}
{Bedin}, L.~R., {Cassisi}, S., {Castelli}, F., {et~al.} 2005, \mnras, 357,
  1038, \dodoi{10.1111/j.1365-2966.2005.08735.x}

\bibitem[{{Bedin} {et~al.}(2009){Bedin}, {Salaris}, {Piotto}, {Anderson},
  {King}, \& {Cassisi}}]{bedin2009a}
{Bedin}, L.~R., {Salaris}, M., {Piotto}, G., {et~al.} 2009, \apj, 697, 965,
  \dodoi{10.1088/0004-637X/697/2/965}

\bibitem[{{Bellini} {et~al.}(2011){Bellini}, {Anderson}, \&
  {Bedin}}]{bellini2011a}
{Bellini}, A., {Anderson}, J., \& {Bedin}, L.~R. 2011, \pasp, 123, 622,
  \dodoi{10.1086/659878}

\bibitem[{{Bellini} \& {Bedin}(2009)}]{bellini2009a}
{Bellini}, A., \& {Bedin}, L.~R. 2009, \pasp, 121, 1419, \dodoi{10.1086/649061}

\bibitem[{{Bellini} {et~al.}(2009){Bellini}, {Piotto}, {Bedin}, {King},
  {Anderson}, {Milone}, \& {Momany}}]{bellini2009b}
{Bellini}, A., {Piotto}, G., {Bedin}, L.~R., {et~al.} 2009, \aap, 507, 1393,
  \dodoi{10.1051/0004-6361/200912757}

\bibitem[{{Bellini} {et~al.}(2013){Bellini}, {Piotto}, {Milone}, {King},
  {Renzini}, {Cassisi}, {Anderson}, {Bedin}, {Nardiello}, {Pietrinferni}, \&
  {Sarajedini}}]{bellini2013a}
{Bellini}, A., {Piotto}, G., {Milone}, A.~P., {et~al.} 2013, \apj, 765, 32,
  \dodoi{10.1088/0004-637X/765/1/32}

\bibitem[{{Carretta} {et~al.}(2009){Carretta}, {Bragaglia}, {Gratton}, \&
  {Lucatello}}]{carretta2009a}
{Carretta}, E., {Bragaglia}, A., {Gratton}, R., \& {Lucatello}, S. 2009, \aap,
  505, 139, \dodoi{10.1051/0004-6361/200912097}

\bibitem[{{Castelli}(2005)}]{castelli2005a}
{Castelli}, F. 2005, Memorie della Societa Astronomica Italiana Supplementi, 8,
  25

\bibitem[{{Chantereau} {et~al.}(2019){Chantereau}, {Salaris}, {Bastian}, \&
  {Martocchia}}]{chantereau2019a}
{Chantereau}, W., {Salaris}, M., {Bastian}, N., \& {Martocchia}, S. 2019,
  \mnras, 484, 5236, \dodoi{10.1093/mnras/stz378}

\bibitem[{{Choi} {et~al.}(2016){Choi}, {Dotter}, {Conroy}, {Cantiello},
  {Paxton}, \& {Johnson}}]{choi2016a}
{Choi}, J., {Dotter}, A., {Conroy}, C., {et~al.} 2016, \apj, 823, 102,
  \dodoi{10.3847/0004-637X/823/2/102}

\bibitem[{{Cordero} {et~al.}(2014){Cordero}, {Pilachowski}, {Johnson},
  {McDonald}, {Zijlstra}, \& {Simmerer}}]{cordero2014a}
{Cordero}, M.~J., {Pilachowski}, C.~A., {Johnson}, C.~I., {et~al.} 2014, \apj,
  780, 94, \dodoi{10.1088/0004-637X/780/1/94}

\bibitem[{{Cordoni} {et~al.}(2018){Cordoni}, {Milone}, {Marino}, {Di
  Criscienzo}, {D'Antona}, {Dotter}, {Lagioia}, \& {Tailo}}]{cordoni2018a}
{Cordoni}, G., {Milone}, A.~P., {Marino}, A.~F., {et~al.} 2018, \apj, 869, 139,
  \dodoi{10.3847/1538-4357/aaedc1}

\bibitem[{{Cordoni} {et~al.}(2020){Cordoni}, {Milone}, {Mastrobuono-Battisti},
  {Marino}, {Lagioia}, {Tailo}, {Baumgardt}, \& {Hilker}}]{cordoni2020a}
{Cordoni}, G., {Milone}, A.~P., {Mastrobuono-Battisti}, A., {et~al.} 2020,
  \apj, 889, 18, \dodoi{10.3847/1538-4357/ab5aee}

\bibitem[{{Dalessandro} {et~al.}(2018){Dalessandro}, {Mucciarelli},
  {Bellazzini}, {Sollima}, {Vesperini}, {Hong}, {H{\'e}nault-Brunet},
  {Ferraro}, {Ibata}, {Lanzoni}, {Massari}, \& {Salaris}}]{dalessandro2018a}
{Dalessandro}, E., {Mucciarelli}, A., {Bellazzini}, M., {et~al.} 2018, \apj,
  864, 33, \dodoi{10.3847/1538-4357/aad4b3}

\bibitem[{{D'Antona} \& {Caloi}(2008)}]{dantona2008a}
{D'Antona}, F., \& {Caloi}, V. 2008, \mnras, 390, 693,
  \dodoi{10.1111/j.1365-2966.2008.13760.x}

\bibitem[{{D'Antona} {et~al.}(2002){D'Antona}, {Caloi}, {Montalb{\'a}n},
  {Ventura}, \& {Gratton}}]{dantona2002a}
{D'Antona}, F., {Caloi}, V., {Montalb{\'a}n}, J., {Ventura}, P., \& {Gratton},
  R. 2002, \aap, 395, 69, \dodoi{10.1051/0004-6361:20021220}

\bibitem[{{D'Antona} {et~al.}(2016){D'Antona}, {Vesperini}, {D'Ercole},
  {Ventura}, {Milone}, {Marino}, \& {Tailo}}]{dantona2016a}
{D'Antona}, F., {Vesperini}, E., {D'Ercole}, A., {et~al.} 2016, \mnras, 458,
  2122, \dodoi{10.1093/mnras/stw387}

\bibitem[{{Dorman}(1992)}]{dorman1992a}
{Dorman}, B. 1992, \apjs, 80, 701, \dodoi{10.1086/191678}

\bibitem[{{Dotter}(2016)}]{dotter2016a}
{Dotter}, A. 2016, \apjs, 222, 8, \dodoi{10.3847/0067-0049/222/1/8}

\bibitem[{{Dotter} {et~al.}(2010){Dotter}, {Sarajedini}, {Anderson},
  {Aparicio}, {Bedin}, {Chaboyer}, {Majewski}, {Mar{\'\i}n-Franch}, {Milone},
  {Paust}, {Piotto}, {Reid}, {Rosenberg}, \& {Siegel}}]{dotter2010a}
{Dotter}, A., {Sarajedini}, A., {Anderson}, J., {et~al.} 2010, \apj, 708, 698,
  \dodoi{10.1088/0004-637X/708/1/698}

\bibitem[{{Fischer} {et~al.}(1992){Fischer}, {Welch}, \& {Mateo}}]{fischer1992}
{Fischer}, P., {Welch}, D.~L., \& {Mateo}, M. 1992, \aj, 104, 1086,
  \dodoi{10.1086/116299}

\bibitem[{{Gaia Collaboration} {et~al.}(2018){Gaia Collaboration}, {Brown},
  {Vallenari}, {Prusti}, {de Bruijne}, {Babusiaux}, {Bailer-Jones}, {Biermann},
  {Evans}, {Eyer}, {Jansen}, {Jordi}, {Klioner}, {Lammers}, {Lindegren},
  {Luri}, {Mignard}, {Panem}, {Pourbaix}, {Randich}, {Sartoretti}, {Siddiqui},
  {Soubiran}, {van Leeuwen}, {Walton}, {Arenou}, {Bastian}, {Cropper},
  {Drimmel}, {Katz}, {Lattanzi}, {Bakker}, {Cacciari}, {Casta{\~n}eda},
  {Chaoul}, {Cheek}, {De Angeli}, {Fabricius}, {Guerra}, {Holl}, {Masana},
  {Messineo}, {Mowlavi}, {Nienartowicz}, {Panuzzo}, {Portell}, {Riello},
  {Seabroke}, {Tanga}, {Th{\'e}venin}, {Gracia-Abril}, {Comoretto},
  {Garcia-Reinaldos}, {Teyssier}, {Altmann}, {Andrae}, {Audard},
  {Bellas-Velidis}, {Benson}, {Berthier}, {Blomme}, {Burgess}, {Busso},
  {Carry}, {Cellino}, {Clementini}, {Clotet}, {Creevey}, {Davidson}, {De
  Ridder}, {Delchambre}, {Dell'Oro}, {Ducourant},
  {Fern{\'a}ndez-Hern{\'a}ndez}, {Fouesneau}, {Fr{\'e}mat}, {Galluccio},
  {Garc{\'\i}a-Torres}, {Gonz{\'a}lez-N{\'u}{\~n}ez}, {Gonz{\'a}lez-Vidal},
  {Gosset}, {Guy}, {Halbwachs}, {Hambly}, {Harrison}, {Hern{\'a}ndez},
  {Hestroffer}, {Hodgkin}, {Hutton}, {Jasniewicz}, {Jean-Antoine-Piccolo},
  {Jordan}, {Korn}, {Krone-Martins}, {Lanzafame}, {Lebzelter}, {L{\"o}ffler},
  {Manteiga}, {Marrese}, {Mart{\'\i}n-Fleitas}, {Moitinho}, {Mora}, {Muinonen},
  {Osinde}, {Pancino}, {Pauwels}, {Petit}, {Recio-Blanco}, {Richards},
  {Rimoldini}, {Robin}, {Sarro}, {Siopis}, {Smith}, {Sozzetti}, {S{\"u}veges},
  {Torra}, {van Reeven}, {Abbas}, {Abreu Aramburu}, {Accart}, {Aerts},
  {Altavilla}, {{\'A}lvarez}, {Alvarez}, {Alves}, {Anderson}, {Andrei},
  {Anglada Varela}, {Antiche}, {Antoja}, {Arcay}, {Astraatmadja}, {Bach},
  {Baker}, {Balaguer-N{\'u}{\~n}ez}, {Balm}, {Barache}, {Barata}, {Barbato},
  {Barblan}, {Barklem}, {Barrado}, {Barros}, {Barstow}, {Bartholom{\'e}
  Mu{\~n}oz}, {Bassilana}, {Becciani}, {Bellazzini}, {Berihuete}, {Bertone},
  {Bianchi}, {Bienaym{\'e}}, {Blanco-Cuaresma}, {Boch}, {Boeche}, {Bombrun},
  {Borrachero}, {Bossini}, {Bouquillon}, {Bourda}, {Bragaglia}, {Bramante},
  {Breddels}, {Bressan}, {Brouillet}, {Br{\"u}semeister}, {Brugaletta},
  {Bucciarelli}, {Burlacu}, {Busonero}, {Butkevich}, {Buzzi}, {Caffau},
  {Cancelliere}, {Cannizzaro}, {Cantat-Gaudin}, {Carballo}, {Carlucci},
  {Carrasco}, {Casamiquela}, {Castellani}, {Castro-Ginard}, {Charlot},
  {Chemin}, {Chiavassa}, {Cocozza}, {Costigan}, {Cowell}, {Crifo}, {Crosta},
  {Crowley}, {Cuypers}, {Dafonte}, {Damerdji}, {Dapergolas}, {David}, {David},
  {de Laverny}, {De Luise}, {De March}, {de Martino}, {de Souza}, {de Torres},
  {Debosscher}, {del Pozo}, {Delbo}, {Delgado}, {Delgado}, {Di Matteo},
  {Diakite}, {Diener}, {Distefano}, {Dolding}, {Drazinos}, {Dur{\'a}n},
  {Edvardsson}, {Enke}, {Eriksson}, {Esquej}, {Eynard Bontemps}, {Fabre},
  {Fabrizio}, {Faigler}, {Falc{\~a}o}, {Farr{\`a}s Casas}, {Federici},
  {Fedorets}, {Fernique}, {Figueras}, {Filippi}, {Findeisen}, {Fonti},
  {Fraile}, {Fraser}, {Fr{\'e}zouls}, {Gai}, {Galleti}, {Garabato},
  {Garc{\'\i}a-Sedano}, {Garofalo}, {Garralda}, {Gavel}, {Gavras}, {Gerssen},
  {Geyer}, {Giacobbe}, {Gilmore}, {Girona}, {Giuffrida}, {Glass}, {Gomes},
  {Granvik}, {Gueguen}, {Guerrier}, {Guiraud}, {Guti{\'e}rrez-S{\'a}nchez},
  {Haigron}, {Hatzidimitriou}, {Hauser}, {Haywood}, {Heiter}, {Helmi}, {Heu},
  {Hilger}, {Hobbs}, {Hofmann}, {Holland}, {Huckle}, {Hypki}, {Icardi},
  {Jan{\ss}en}, {Jevardat de Fombelle}, {Jonker}, {Juh{\'a}sz}, {Julbe},
  {Karampelas}, {Kewley}, {Klar}, {Kochoska}, {Kohley}, {Kolenberg},
  {Kontizas}, {Kontizas}, {Koposov}, {Kordopatis}, {Kostrzewa-Rutkowska},
  {Koubsky}, {Lambert}, {Lanza}, {Lasne}, {Lavigne}, {Le Fustec}, {Le
  Poncin-Lafitte}, {Lebreton}, {Leccia}, {Leclerc}, {Lecoeur-Taibi},
  {Lenhardt}, {Leroux}, {Liao}, {Licata}, {Lindstr{\o}m}, {Lister}, {Livanou},
  {Lobel}, {L{\'o}pez}, {Managau}, {Mann}, {Mantelet}, {Marchal}, {Marchant},
  {Marconi}, {Marinoni}, {Marschalk{\'o}}, {Marshall}, {Martino}, {Marton},
  {Mary}, {Massari}, {Matijevi{\v{c}}}, {Mazeh}, {McMillan}, {Messina},
  {Michalik}, {Millar}, {Molina}, {Molinaro}, {Moln{\'a}r}, {Montegriffo},
  {Mor}, {Morbidelli}, {Morel}, {Morris}, {Mulone}, {Muraveva}, {Musella},
  {Nelemans}, {Nicastro}, {Noval}, {O'Mullane}, {Ord{\'e}novic},
  {Ord{\'o}{\~n}ez-Blanco}, {Osborne}, {Pagani}, {Pagano}, {Pailler},
  {Palacin}, {Palaversa}, {Panahi}, {Pawlak}, {Piersimoni}, {Pineau}, {Plachy},
  {Plum}, {Poggio}, {Poujoulet}, {Pr{\v{s}}a}, {Pulone}, {Racero}, {Ragaini},
  {Rambaux}, {Ramos-Lerate}, {Regibo}, {Reyl{\'e}}, {Riclet}, {Ripepi}, {Riva},
  {Rivard}, {Rixon}, {Roegiers}, {Roelens}, {Romero-G{\'o}mez}, {Rowell},
  {Royer}, {Ruiz-Dern}, {Sadowski}, {Sagrist{\`a} Sell{\'e}s}, {Sahlmann},
  {Salgado}, {Salguero}, {Sanna}, {Santana-Ros}, {Sarasso}, {Savietto},
  {Schultheis}, {Sciacca}, {Segol}, {Segovia}, {S{\'e}gransan}, {Shih},
  {Siltala}, {Silva}, {Smart}, {Smith}, {Solano}, {Solitro}, {Sordo}, {Soria
  Nieto}, {Souchay}, {Spagna}, {Spoto}, {Stampa}, {Steele},
  {Steidelm{\"u}ller}, {Stephenson}, {Stoev}, {Suess}, {Surdej}, {Szabados},
  {Szegedi-Elek}, {Tapiador}, {Taris}, {Tauran}, {Taylor}, {Teixeira},
  {Terrett}, {Teyssand ier}, {Thuillot}, {Titarenko}, {Torra Clotet}, {Turon},
  {Ulla}, {Utrilla}, {Uzzi}, {Vaillant}, {Valentini}, {Valette}, {van Elteren},
  {Van Hemelryck}, {van Leeuwen}, {Vaschetto}, {Vecchiato}, {Veljanoski},
  {Viala}, {Vicente}, {Vogt}, {von Essen}, {Voss}, {Votruba}, {Voutsinas},
  {Walmsley}, {Weiler}, {Wertz}, {Wevers}, {Wyrzykowski}, {Yoldas},
  {{\v{Z}}erjal}, {Ziaeepour}, {Zorec}, {Zschocke}, {Zucker}, {Zurbach}, \&
  {Zwitter}}]{gaia2018a}
{Gaia Collaboration}, {Brown}, A.~G.~A., {Vallenari}, A., {et~al.} 2018, \aap,
  616, A1, \dodoi{10.1051/0004-6361/201833051}

\bibitem[{{Glatt} {et~al.}(2009){Glatt}, {Grebel}, {Gallagher}, {Nota},
  {Sabbi}, {Sirianni}, {Clementini}, {Da Costa}, {Tosi}, {Harbeck}, {Koch}, \&
  {Kayser}}]{glatt2009a}
{Glatt}, K., {Grebel}, E.~K., {Gallagher}, John~S., I., {et~al.} 2009, \aj,
  138, 1403, \dodoi{10.1088/0004-6256/138/5/1403}

\bibitem[{{Glatt} {et~al.}(2011){Glatt}, {Grebel}, {Jordi}, {Gallagher}, {Da
  Costa}, {Clementini}, {Tosi}, {Harbeck}, {Nota}, {Sabbi}, \&
  {Sirianni}}]{glatt2011a}
{Glatt}, K., {Grebel}, E.~K., {Jordi}, K., {et~al.} 2011, \aj, 142, 36,
  \dodoi{10.1088/0004-6256/142/2/36}

\bibitem[{{Goudfrooij} {et~al.}(2014){Goudfrooij}, {Girardi},
  {Kozhurina-Platais}, {Kalirai}, {Platais}, {Puzia}, {Correnti}, {Bressan},
  {Chandar}, {Kerber}, {Marigo}, \& {Rubele}}]{goudfrooij2014a}
{Goudfrooij}, P., {Girardi}, L., {Kozhurina-Platais}, V., {et~al.} 2014, \apj,
  797, 35, \dodoi{10.1088/0004-637X/797/1/35}

\bibitem[{{Gratton} {et~al.}(2011){Gratton}, {Lucatello}, {Carretta},
  {Bragaglia}, {D'Orazi}, \& {Momany}}]{gratton2011a}
{Gratton}, R.~G., {Lucatello}, S., {Carretta}, E., {et~al.} 2011, \aap, 534,
  A123, \dodoi{10.1051/0004-6361/201117690}

\bibitem[{{Harris}(1996)}]{harris1996a}
{Harris}, W.~E. 1996, \aj, 112, 1487, \dodoi{10.1086/118116}

\bibitem[{{Iben} \& {Rood}(1970)}]{iben1970a}
{Iben}, Icko, J., \& {Rood}, R.~T. 1970, \apj, 161, 587, \dodoi{10.1086/150563}

\bibitem[{{Kraft}(1994)}]{kraft1994a}
{Kraft}, R.~P. 1994, \pasp, 106, 553, \dodoi{10.1086/133416}

\bibitem[{{Kurucz}(1993)}]{kurucz1993a}
{Kurucz}, R. 1993, SYNTHE Spectrum Synthesis Programs and Line Data. Kurucz
  CD-ROM No. 18. Cambridge, 18

\bibitem[{Kurucz(2005)}]{kurucz2005a}
Kurucz, R. 2005, Memorie della Societa Astronomica Italiana Supplementi, 8, 14

\bibitem[{{Kurucz}(1970)}]{kurucz1970a}
{Kurucz}, R.~L. 1970, SAO Special Report, 309

\bibitem[{{Kurucz} \& {Avrett}(1981)}]{kurucz1981a}
{Kurucz}, R.~L., \& {Avrett}, E.~H. 1981, SAO Special Report, 391

\bibitem[{{Lagioia} {et~al.}(2019{\natexlab{a}}){Lagioia}, {Milone}, {Marino},
  {Cordoni}, \& {Tailo}}]{lagioia2019b}
{Lagioia}, E.~P., {Milone}, A.~P., {Marino}, A.~F., {Cordoni}, G., \& {Tailo},
  M. 2019{\natexlab{a}}, \aj, 158, 202, \dodoi{10.3847/1538-3881/ab45f2}

\bibitem[{{Lagioia} {et~al.}(2019{\natexlab{b}}){Lagioia}, {Milone}, {Marino},
  \& {Dotter}}]{lagioia2019a}
{Lagioia}, E.~P., {Milone}, A.~P., {Marino}, A.~F., \& {Dotter}, A.
  2019{\natexlab{b}}, \apj, 871, 140, \dodoi{10.3847/1538-4357/aaf729}

\bibitem[{{Lagioia} {et~al.}(2018){Lagioia}, {Milone}, {Marino}, {Cassisi},
  {Aparicio}, {Piotto}, {Anderson}, {Barbuy}, {Bedin}, {Bellini}, {Brown},
  {D'Antona}, {Nardiello}, {Ortolani}, {Pietrinferni}, {Renzini}, {Salaris},
  {Sarajedini}, {van der Marel}, \& {Vesperini}}]{lagioia2018a}
{Lagioia}, E.~P., {Milone}, A.~P., {Marino}, A.~F., {et~al.} 2018, \mnras, 475,
  4088, \dodoi{10.1093/mnras/sty083}

\bibitem[{{Lee}(2017)}]{lee2017a}
{Lee}, J.-W. 2017, \apj, 844, 77, \dodoi{10.3847/1538-4357/aa7b8c}

\bibitem[{Lee(2018)}]{lee2018a}
Lee, J.-W. 2018, \apjs, 238, 24, \dodoi{10.3847/1538-4365/aadcad}

\bibitem[{{Lee}(2019)}]{lee2019a}
{Lee}, J.-W. 2019, \apjl, 875, L27, \dodoi{10.3847/2041-8213/ab1594}

\bibitem[{{Lee} \& {Sneden}(2020)}]{lee2020a}
{Lee}, J.-W., \& {Sneden}, C. 2020, arXiv e-prints, arXiv:2006.01274.
\newblock \doarXiv{2006.01274}

\bibitem[{{Lindegren} {et~al.}(2018){Lindegren}, {Hern{\'a}ndez}, {Bombrun},
  {Klioner}, {Bastian}, {Ramos-Lerate}, {de Torres}, {Steidelm{\"u}ller},
  {Stephenson}, {Hobbs}, {Lammers}, {Biermann}, {Geyer}, {Hilger}, {Michalik},
  {Stampa}, {McMillan}, {Casta{\~n}eda}, {Clotet}, {Comoretto}, {Davidson},
  {Fabricius}, {Gracia}, {Hambly}, {Hutton}, {Mora}, {Portell}, {van Leeuwen},
  {Abbas}, {Abreu}, {Altmann}, {Andrei}, {Anglada}, {Balaguer-N{\'u}{\~n}ez},
  {Barache}, {Becciani}, {Bertone}, {Bianchi}, {Bouquillon}, {Bourda},
  {Br{\"u}semeister}, {Bucciarelli}, {Busonero}, {Buzzi}, {Cancelliere},
  {Carlucci}, {Charlot}, {Cheek}, {Crosta}, {Crowley}, {de Bruijne}, {de
  Felice}, {Drimmel}, {Esquej}, {Fienga}, {Fraile}, {Gai}, {Garralda},
  {Gonz{\'a}lez-Vidal}, {Guerra}, {Hauser}, {Hofmann}, {Holl}, {Jordan},
  {Lattanzi}, {Lenhardt}, {Liao}, {Licata}, {Lister}, {L{\"o}ffler},
  {Marchant}, {Martin-Fleitas}, {Messineo}, {Mignard}, {Morbidelli}, {Poggio},
  {Riva}, {Rowell}, {Salguero}, {Sarasso}, {Sciacca}, {Siddiqui}, {Smart},
  {Spagna}, {Steele}, {Taris}, {Torra}, {van Elteren}, {van Reeven}, \&
  {Vecchiato}}]{lindegren2018a}
{Lindegren}, L., {Hern{\'a}ndez}, J., {Bombrun}, A., {et~al.} 2018, \aap, 616,
  A2, \dodoi{10.1051/0004-6361/201832727}

\bibitem[{{Marino} {et~al.}(2011){Marino}, {Villanova}, {Milone}, {Piotto},
  {Lind}, {Geisler}, \& {Stetson}}]{marino2011a}
{Marino}, A.~F., {Villanova}, S., {Milone}, A.~P., {et~al.} 2011, \apjl, 730,
  L16, \dodoi{10.1088/2041-8205/730/2/L16}

\bibitem[{{Marino} {et~al.}(2008){Marino}, {Villanova}, {Piotto}, {Milone},
  {Momany}, {Bedin}, \& {Medling}}]{marino2008a}
{Marino}, A.~F., {Villanova}, S., {Piotto}, G., {et~al.} 2008, \aap, 490, 625,
  \dodoi{10.1051/0004-6361:200810389}

\bibitem[{{Marino} {et~al.}(2014){Marino}, {Milone}, {Przybilla}, {Bergemann},
  {Lind}, {Asplund}, {Cassisi}, {Catelan}, {Casagrande}, {Valcarce}, {Bedin},
  {Cort{\'e}s}, {D'Antona}, {Jerjen}, {Piotto}, {Schlesinger}, {Zoccali}, \&
  {Angeloni}}]{marino2014a}
{Marino}, A.~F., {Milone}, A.~P., {Przybilla}, N., {et~al.} 2014, \mnras, 437,
  1609, \dodoi{10.1093/mnras/stt1993}

\bibitem[{{Marino} {et~al.}(2019{\natexlab{a}}){Marino}, {Milone}, {Renzini},
  {D'Antona}, {Anderson}, {Bedin}, {Bellini}, {Cordoni}, {Lagioia}, {Piotto},
  \& {Tailo}}]{marino2019a}
{Marino}, A.~F., {Milone}, A.~P., {Renzini}, A., {et~al.} 2019{\natexlab{a}},
  \mnras, 487, 3815, \dodoi{10.1093/mnras/stz1415}

\bibitem[{{Marino} {et~al.}(2019{\natexlab{b}}){Marino}, {Milone}, {Sills},
  {Yong}, {Renzini}, {Bedin}, {Cordoni}, {D{\textquoteright}Antona}, {Jerjen},
  {Karakas}, {Lagioia}, {Piotto}, \& {Tailo}}]{marino2019b}
{Marino}, A.~F., {Milone}, A.~P., {Sills}, A., {et~al.} 2019{\natexlab{b}},
  \apj, 887, 91, \dodoi{10.3847/1538-4357/ab53d9}

\bibitem[{{Martocchia} {et~al.}(2018{\natexlab{a}}){Martocchia}, {Niederhofer},
  {Dalessandro}, {Bastian}, {Kacharov}, {Usher}, {Cabrera-Ziri}, {Lardo},
  {Cassisi}, {Geisler}, {Hilker}, {Hollyhead}, {Kozhurina-Platais}, {Larsen},
  {Mackey}, {Mucciarelli}, {Platais}, \& {Salaris}}]{martocchia2018a}
{Martocchia}, S., {Niederhofer}, F., {Dalessandro}, E., {et~al.}
  2018{\natexlab{a}}, \mnras, 477, 4696, \dodoi{10.1093/mnras/sty916}

\bibitem[{{Martocchia} {et~al.}(2018{\natexlab{b}}){Martocchia},
  {Cabrera-Ziri}, {Lardo}, {Dalessand ro}, {Bastian}, {Kozhurina-Platais},
  {Usher}, {Niederhofer}, {Cordero}, {Geisler}, {Hollyhead}, {Kacharov},
  {Larsen}, {Li}, {Mackey}, {Hilker}, {Mucciarelli}, {Platais}, \&
  {Salaris}}]{martocchia2018b}
{Martocchia}, S., {Cabrera-Ziri}, I., {Lardo}, C., {et~al.} 2018{\natexlab{b}},
  \mnras, 473, 2688, \dodoi{10.1093/mnras/stx2556}

\bibitem[{{Mateo}(1987)}]{mateo1987a}
{Mateo}, M. 1987, \apjl, 323, L41, \dodoi{10.1086/185053}

\bibitem[{{McLaughlin} \& {van der Marel}(2005)}]{mclaughlin2005a}
{McLaughlin}, D.~E., \& {van der Marel}, R.~P. 2005, \apjs, 161, 304,
  \dodoi{10.1086/497429}

\bibitem[{{Milone} {et~al.}(2009){Milone}, {Bedin}, {Piotto}, \&
  {Anderson}}]{milone2009a}
{Milone}, A.~P., {Bedin}, L.~R., {Piotto}, G., \& {Anderson}, J. 2009, \aap,
  497, 755, \dodoi{10.1051/0004-6361/200810870}

\bibitem[{{Milone} {et~al.}(2018{\natexlab{a}}){Milone}, {Marino},
  {Mastrobuono-Battisti}, \& {Lagioia}}]{milone2018b}
{Milone}, A.~P., {Marino}, A.~F., {Mastrobuono-Battisti}, A., \& {Lagioia},
  E.~P. 2018{\natexlab{a}}, \mnras, 479, 5005, \dodoi{10.1093/mnras/sty1873}

\bibitem[{Milone {et~al.}(2012{\natexlab{a}})Milone, {Piotto}, {Bedin}, {King},
  {Anderson}, {Marino}, {Bellini}, {Gratton}, {Renzini}, {Stetson}, {Cassisi},
  {Aparicio}, {Bragaglia}, {Carretta}, {D'Antona}, {Di Criscienzo},
  {Lucatello}, {Monelli}, \& {Pietrinferni}}]{milone2012a}
Milone, A.~P., {Piotto}, G., {Bedin}, L.~R., {et~al.} 2012{\natexlab{a}}, \apj,
  744, 58, \dodoi{10.1088/0004-637X/744/1/58}

\bibitem[{Milone {et~al.}(2012{\natexlab{b}})Milone, Piotto, {Bedin},
  {Aparicio}, {Anderson}, {Sarajedini}, {Marino}, {Moretti}, {Davies},
  {Chaboyer}, {Dotter}, {Hempel}, {Mar{\'\i}n-Franch}, {Majewski}, {Paust},
  {Reid}, {Rosenberg}, \& {Siegel}}]{milone2012b}
Milone, A.~P., Piotto, G., {Bedin}, L.~R., {et~al.} 2012{\natexlab{b}}, \aap,
  540, A16, \dodoi{10.1051/0004-6361/201016384}

\bibitem[{{Milone} {et~al.}(2013){Milone}, {Marino}, {Piotto}, {Bedin},
  {Anderson}, {Aparicio}, {Bellini}, {Cassisi}, {D'Antona}, {Grundahl},
  {Monelli}, \& {Yong}}]{milone2013a}
{Milone}, A.~P., {Marino}, A.~F., {Piotto}, G., {et~al.} 2013, \apj, 767, 120,
  \dodoi{10.1088/0004-637X/767/2/120}

\bibitem[{{Milone} {et~al.}(2014){Milone}, {Marino}, {Dotter}, {Norris},
  {Jerjen}, {Piotto}, {Cassisi}, {Bedin}, {Recio Blanco}, {Sarajedini},
  {Asplund}, {Monelli}, \& {Aparicio}}]{milone2014a}
{Milone}, A.~P., {Marino}, A.~F., {Dotter}, A., {et~al.} 2014, \apj, 785, 21,
  \dodoi{10.1088/0004-637X/785/1/21}

\bibitem[{{Milone} {et~al.}(2015){Milone}, {Marino}, {Piotto}, {Renzini},
  {Bedin}, {Anderson}, {Cassisi}, {D'Antona}, {Bellini}, {Jerjen},
  {Pietrinferni}, \& {Ventura}}]{milone2015a}
{Milone}, A.~P., {Marino}, A.~F., {Piotto}, G., {et~al.} 2015, \apj, 808, 51,
  \dodoi{10.1088/0004-637X/808/1/51}

\bibitem[{{Milone} {et~al.}(2016){Milone}, {Marino}, {Bedin}, {Dotter},
  {Jerjen}, {Kim}, {Nardiello}, {Piotto}, \& {Cong}}]{milone2016a}
{Milone}, A.~P., {Marino}, A.~F., {Bedin}, L.~R., {et~al.} 2016, \mnras, 455,
  3009, \dodoi{10.1093/mnras/stv2415}

\bibitem[{{Milone} {et~al.}(2017){Milone}, {Piotto}, {Renzini}, {Marino},
  {Bedin}, {Vesperini}, {D'Antona}, {Nardiello}, {Anderson}, {King}, {Yong},
  {Bellini}, {Aparicio}, {Barbuy}, {Brown}, {Cassisi}, {Ortolani}, {Salaris},
  {Sarajedini}, \& {van der Marel}}]{milone2017a}
{Milone}, A.~P., {Piotto}, G., {Renzini}, A., {et~al.} 2017, \mnras, 464, 3636,
  \dodoi{10.1093/mnras/stw2531}

\bibitem[{{Milone} {et~al.}(2018{\natexlab{b}}){Milone}, {Marino}, {Renzini},
  {D'Antona}, {Anderson}, {Barbuy}, {Bedin}, {Bellini}, {Brown}, {Cassisi},
  {Cordoni}, {Lagioia}, {Nardiello}, {Ortolani}, {Piotto}, {Sarajedini},
  {Tailo}, {van der Marel}, \& {Vesperini}}]{milone2018a}
{Milone}, A.~P., {Marino}, A.~F., {Renzini}, A., {et~al.} 2018{\natexlab{b}},
  \mnras, 481, 5098, \dodoi{10.1093/mnras/sty2573}

\bibitem[{{Milone} {et~al.}(2020{\natexlab{a}}){Milone}, {Marino}, {Da Costa},
  {Lagioia}, {D'Antona}, {Goudfrooij}, {Jerjen}, {Massari}, {Renzini}, {Yong},
  {Baumgardt}, {Cordoni}, {Dondoglio}, {Li}, {Tailo}, {Asa'd}, \&
  {Ventura}}]{milone2020a}
{Milone}, A.~P., {Marino}, A.~F., {Da Costa}, G.~S., {et~al.}
  2020{\natexlab{a}}, \mnras, 491, 515, \dodoi{10.1093/mnras/stz2999}

\bibitem[{{Milone} {et~al.}(2020{\natexlab{b}}){Milone}, {Vesperini}, {Marino},
  {Hong}, {van der Marel}, {Anderson}, {Renzini}, {Cordoni}, {Bedin},
  {Bellini}, {Brown}, {D'Antona}, {Lagioia}, {Libralato}, {Nardiello},
  {Piotto}, {Tailo}, {Cool}, {Salaris}, \& {Sarajedini}}]{milone2020b}
{Milone}, A.~P., {Vesperini}, E., {Marino}, A.~F., {et~al.} 2020{\natexlab{b}},
  \mnras, 492, 5457, \dodoi{10.1093/mnras/stz3629}

\bibitem[{{Monelli} {et~al.}(2013){Monelli}, {Milone}, {Stetson}, {Marino},
  {Cassisi}, {del Pino Molina}, {Salaris}, {Aparicio}, {Asplund}, {Grundahl},
  {Piotto}, {Weiss}, {Carrera}, {Cebri{\'a}n}, {Murabito}, {Pietrinferni}, \&
  {Sbordone}}]{monelli2013a}
{Monelli}, M., {Milone}, A.~P., {Stetson}, P.~B., {et~al.} 2013, \mnras, 431,
  2126, \dodoi{10.1093/mnras/stt273}

\bibitem[{{Niederhofer} {et~al.}(2017){Niederhofer}, {Bastian},
  {Kozhurina-Platais}, {Larsen}, {Hollyhead}, {Lardo}, {Cabrera-Ziri},
  {Kacharov}, {Platais}, {Salaris}, {Cordero}, {Dalessandro}, {Geisler},
  {Hilker}, {Li}, {Mackey}, \& {Mucciarelli}}]{niederhofer2017a}
{Niederhofer}, F., {Bastian}, N., {Kozhurina-Platais}, V., {et~al.} 2017,
  \mnras, 465, 4159, \dodoi{10.1093/mnras/stw3084}

\bibitem[{{Norris} \& {Freeman}(1982)}]{norris1982a}
{Norris}, J., \& {Freeman}, K.~C. 1982, \apj, 254, 143, \dodoi{10.1086/159717}

\bibitem[{{Paxton} {et~al.}(2011){Paxton}, {Bildsten}, {Dotter}, {Herwig},
  {Lesaffre}, \& {Timmes}}]{paxton2011a}
{Paxton}, B., {Bildsten}, L., {Dotter}, A., {et~al.} 2011, \apjs, 192, 3,
  \dodoi{10.1088/0067-0049/192/1/3}

\bibitem[{{Paxton} {et~al.}(2013){Paxton}, {Cantiello}, {Arras}, {Bildsten},
  {Brown}, {Dotter}, {Mankovich}, {Montgomery}, {Stello}, {Timmes}, \&
  {Townsend}}]{paxton2013a}
{Paxton}, B., {Cantiello}, M., {Arras}, P., {et~al.} 2013, \apjs, 208, 4,
  \dodoi{10.1088/0067-0049/208/1/4}

\bibitem[{{Paxton} {et~al.}(2015){Paxton}, {Marchant}, {Schwab}, {Bauer},
  {Bildsten}, {Cantiello}, {Dessart}, {Farmer}, {Hu}, {Langer}, {Townsend},
  {Townsley}, \& {Timmes}}]{paxton2015a}
{Paxton}, B., {Marchant}, P., {Schwab}, J., {et~al.} 2015, \apjs, 220, 15,
  \dodoi{10.1088/0067-0049/220/1/15}

\bibitem[{{Renzini} {et~al.}(2015){Renzini}, {D'Antona}, {Cassisi}, {King},
  {Milone}, {Ventura}, {Anderson}, {Bedin}, {Bellini}, {Brown}, {Piotto}, {van
  der Marel}, {Barbuy}, {Dalessandro}, {Hidalgo}, {Marino}, {Ortolani},
  {Salaris}, \& {Sarajedini}}]{renzini2015a}
{Renzini}, A., {D'Antona}, F., {Cassisi}, S., {et~al.} 2015, \mnras, 454, 4197,
  \dodoi{10.1093/mnras/stv2268}

\bibitem[{{Rich} {et~al.}(1997){Rich}, {Sosin}, {Djorgovski}, {Piotto}, {King},
  {Renzini}, {Phinney}, {Dorman}, {Liebert}, \& {Meylan}}]{rich1997a}
{Rich}, R.~M., {Sosin}, C., {Djorgovski}, S.~G., {et~al.} 1997, \apjl, 484,
  L25, \dodoi{10.1086/310758}

\bibitem[{{Sabbi} {et~al.}(2016){Sabbi}, {Lennon}, {Anderson}, {Cignoni}, {van
  der Marel}, {Zaritsky}, {De Marchi}, {Panagia}, {Gouliermis}, \&
  {Grebel}}]{sabbi2016a}
{Sabbi}, E., {Lennon}, D.~J., {Anderson}, J., {et~al.} 2016, \apjs, 222, 11,
  \dodoi{10.3847/0067-0049/222/1/11}

\bibitem[{{Salaris} {et~al.}(2008){Salaris}, {Cassisi}, \&
  {Pietrinferni}}]{salaris2008a}
{Salaris}, M., {Cassisi}, S., \& {Pietrinferni}, A. 2008, \apjl, 678, L25,
  \dodoi{10.1086/588467}

\bibitem[{{Salpeter}(1955)}]{salpeter1955a}
{Salpeter}, E.~E. 1955, \apj, 121, 161, \dodoi{10.1086/145971}

\bibitem[{SAS Institute Inc.~Staff(1988)}]{sasinc}
SAS Institute Inc.~Staff, C. 1988, SAS-STAT User's Guide: Release 6.03 Edition
  (USA: SAS Institute Inc.)

\bibitem[{{Sbordone} {et~al.}(2007){Sbordone}, {Bonifacio}, \&
  {Castelli}}]{sbordone2007a}
{Sbordone}, L., {Bonifacio}, P., \& {Castelli}, F. 2007, in IAU Symposium, Vol.
  239, Convection in Astrophysics, ed. F.~{Kupka}, I.~{Roxburgh}, \& K.~L.
  {Chan}, 71--73

\bibitem[{{Sbordone} {et~al.}(2004){Sbordone}, {Bonifacio}, {Castelli}, \&
  {Kurucz}}]{sbordone2004a}
{Sbordone}, L., {Bonifacio}, P., {Castelli}, F., \& {Kurucz}, R.~L. 2004,
  Memorie della Societa Astronomica Italiana Supplementi, 5, 93.
\newblock \doarXiv{astro-ph/0406268}

\bibitem[{{Sbordone} {et~al.}(2011){Sbordone}, {Salaris}, {Weiss}, \&
  {Cassisi}}]{sbordone2011a}
{Sbordone}, L., {Salaris}, M., {Weiss}, A., \& {Cassisi}, S. 2011, \aap, 534,
  A9, \dodoi{10.1051/0004-6361/201116714}

\bibitem[{{Smith} \& {Penny}(1989)}]{smith1989a}
{Smith}, G.~H., \& {Penny}, A.~J. 1989, \aj, 97, 1397, \dodoi{10.1086/115080}

\bibitem[{{Sollima} {et~al.}(2007){Sollima}, {Ferraro}, {Bellazzini},
  {Origlia}, {Straniero}, \& {Pancino}}]{sollima2007a}
{Sollima}, A., {Ferraro}, F.~R., {Bellazzini}, M., {et~al.} 2007, \apj, 654,
  915, \dodoi{10.1086/509711}

\bibitem[{{Stetson}(2005)}]{stetson2005a}
{Stetson}, P.~B. 2005, \pasp, 117, 563, \dodoi{10.1086/430281}

\bibitem[{{Stetson} {et~al.}(2019){Stetson}, {Pancino}, {Zocchi}, {Sanna}, \&
  {Monelli}}]{stetson2019a}
{Stetson}, P.~B., {Pancino}, E., {Zocchi}, A., {Sanna}, N., \& {Monelli}, M.
  2019, \mnras, 485, 3042, \dodoi{10.1093/mnras/stz585}

\bibitem[{{Tailo} {et~al.}(2019{\natexlab{a}}){Tailo}, {D'Antona}, {Caloi},
  {Milone}, {Marino}, {Lagioia}, \& {Cordoni}}]{tailo2019b}
{Tailo}, M., {D'Antona}, F., {Caloi}, V., {et~al.} 2019{\natexlab{a}}, \mnras,
  486, 5895, \dodoi{10.1093/mnras/stz1273}

\bibitem[{{Tailo} {et~al.}(2016){Tailo}, {Di Criscienzo}, {D'Antona}, {Caloi},
  \& {Ventura}}]{tailo2016}
{Tailo}, M., {Di Criscienzo}, M., {D'Antona}, F., {Caloi}, V., \& {Ventura}, P.
  2016, \mnras, 457, 4525, \dodoi{10.1093/mnras/stw319}

\bibitem[{{Tailo} {et~al.}(2019{\natexlab{b}}){Tailo}, {Milone}, {Marino},
  {D'Antona}, {Lagioia}, \& {Cordoni}}]{tailo2019a}
{Tailo}, M., {Milone}, A.~P., {Marino}, A.~F., {et~al.} 2019{\natexlab{b}},
  \apj, 873, 123, \dodoi{10.3847/1538-4357/ab05cc}

\bibitem[{{Tailo} {et~al.}(2017){Tailo}, {D'Antona}, {Milone}, {Bellini},
  {Ventura}, {Di Criscienzo}, {Cassisi}, {Piotto}, {Salaris}, {Brown},
  {Vesperini}, {Bedin}, {Marino}, {Nardiello}, \& {Anderson}}]{tailo2017a}
{Tailo}, M., {D'Antona}, F., {Milone}, A.~P., {et~al.} 2017, \mnras, 465, 1046,
  \dodoi{10.1093/mnras/stw2790}

\bibitem[{{Tailo} {et~al.}(2020){Tailo}, {Milone}, {Lagioia}, {D'Antona},
  {Marino}, {Vesperini}, {Caloi}, {Ventura}, {Dondoglio}, \&
  {Cordoni}}]{tailo2020a}
{Tailo}, M., {Milone}, A.~P., {Lagioia}, E.~P., {et~al.} 2020, \mnras,
  \dodoi{10.1093/mnras/staa2639}

\bibitem[{{Zennaro} {et~al.}(2019){Zennaro}, {Milone}, {Marino}, {Cordoni},
  {Lagioia}, \& {Tailo}}]{zennaro2019a}
{Zennaro}, M., {Milone}, A.~P., {Marino}, A.~F., {et~al.} 2019, \mnras, 487,
  3239, \dodoi{10.1093/mnras/stz1477}

\end{thebibliography}
\bibliographystyle{aasjournal}

\end{document}